\documentclass[paper=A4,11pt]{scrartcl}
\usepackage[bookmarks=false]{hyperref}
\usepackage{tabularx}
\usepackage{array}
\usepackage{graphicx}
\usepackage{amsmath}
\usepackage{amssymb}
\usepackage{color}
\usepackage{longtable}
\usepackage{verbatim}
\usepackage{cite}
\usepackage{bbm}
\usepackage[normalem]{ulem}
\addtokomafont{disposition}{\rmfamily\boldmath}

\numberwithin{equation}{section}

\newcommand{\be}{\begin{equation}}
\newcommand{\ee}{\end{equation}}
\newcommand{\bea}{\begin{eqnarray}}
\newcommand{\eea}{\end{eqnarray}}
\newcommand{\ta}{\tilde\alpha}
\newcommand{\tb}{\tilde\beta}

\newcommand{\bm}{\bar\mu}
\newcommand{\bn}{\bar\nu}
\newcommand{\da}{\dot\alpha}

\newcommand{\la}{\lambda}
\newcommand{\ga}{\gamma}
\newcommand{\al}{\alpha}
\newcommand{\bet}{\beta}
\newcommand{\e}{\eta}
\newcommand{\nnu}{\nonumber\\}
\newcommand{\GSM}{$SU(3)\times SU(2)_L \times U(1)_Y\quad $}

\newcommand{\at}{\tilde a}
\newcommand{\pt}{\tilde p}
\newcommand{\omt}{\tilde \omega}
\newcommand{\om}{\omega}
\newcommand{\sgt}{\tilde \sigma}

\newcommand{\sgbt}{\tilde {\overline{ \sigma}}}

\newcommand{\sss}{\sigma}
\newcommand{\ssb}{{\overline{ \sigma}}}
\newcommand{\sq}{\sqrt{2}}
\newcommand{\sqs}{\sqrt{6}}
\newcommand{\sqt}{\sqrt{3}}
\newcommand{\sqf}{\sqrt{5}}
\newcommand{\sqtt}{\sqrt{3\over 2}}
\newcommand{\os}{\overline\Sigma}
\newcommand{\s}{\Sigma}
\newcommand{\Sigb}{{\overline\Sigma}}
\newcommand{\oot}{\overline {126}}
\newcommand{\bt}{\bar t}

\newcommand{\ovl}{\overline}
\newcommand{\boot}{${\bf{\oot}}$ }
\newcommand{\bten}{${\bf{10}}$ }
\begin{document}
%-------------------------------------------------------------------

\thispagestyle{empty}
\setcounter{page}{0}

\vskip1.5cm

\begin{center}
{\LARGE \bf \boldmath The New Minimal Supersymmetric GUT :\\ ${}$ \\ Spectra, RG analysis and Fermion Fits}\\[0.8 cm]
{\large Charanjit S. Aulakh$^{a}$ and  Sumit K. Garg$^{b}$  } \\[0.5 cm]

\bigskip
\small $^a$ {\em  Dept. of Physics, Panjab University,  Chandigarh, India}\\[0.1cm]
\small$^b${\em Center for High Energy Physics, IISc, Bangalore, India  } \\[0.1cm]
 \end{center}

\bigskip
%%%%%%%%%%%%%%%%%%%%%%%%%%%%%%%%%%%%%%%%%%%%%%%%%%%%%%%%%%%%%%%%%%%%
\begin{abstract}
The supersymmetric SO(10) GUT based on the ${\bf{210\oplus
10\oplus 120\oplus 126\oplus {\overline {126}} }}$ Higgs system
provides a minimal framework   for the emergence of the  R-parity
exact MSSM at low energies and a viable supersymmetric seesaw
explanation for the observed neutrino masses and mixing angles. We
present  formulae for MSSM decomposition of the superpotential
invariants, tree level light charged fermion effective Yukawa
couplings, Weinberg neutrino mass generation operator, and the
$d=5,\Delta B=\Delta L \neq 0$ effective  superpotential
     in terms of GUT  parameters.
 We use them     to
determine  fits   of the 18 available fermion mass-mixing data in
terms of the superpotential parameters of the NMSGUT and
SUGRY(NUHM) type  soft supersymmetry breaking parameters
($\{m_{\tilde f},m_{1/2},A_0,M^2_{H,\bar H}\} $)
 specified at the MSSM one loop unification
scale $M_X^0=10^{16.33} $ GeV. Our fits are compatible with
electroweak symmetry breaking and Unification constraints and
yield right-handed neutrino masses in the leptogenesis relevant
range : $10^8-10^{13} $ GeV. Matching the SM   data requires
lowering the  strange and down quark Yukawas in the MSSM via large
$\tan\beta$ driven threshold corrections and characteristic soft
Susy breaking spectra.  The
 Susy spectra  have   light pure  Bino LSP,  heavy exotic Higgs(inos) and large $
\mu,A_0,M_{H,\bar{H}}$ parameters $\sim 100$  TeV. Typically third
generation sfermions are much \emph{ heavier}  than the first two
generations. The smuon is often the lightest charged sfermion thus
offering a Bino-CDM co-annihilation channel. The parameter sets
obtained are used to calculate B violation rates which are found
to be generically much faster($\sim 10^{-28}\, yr^{-1}$) than the
current experimental limits. Improvements which may allow
    acceptable B violation rates are identified.

\end{abstract}
%%%%%%%%%%%%%%%%%%%%%%%%%%%%%%%%%%%%%%%%%%%%%%%%%%%%%%%%%%%%%%%%%%%%
\vfill

\newpage
\pagenumbering{arabic}
\section{Introduction}
 The discovery of neutrino mass was both
preceded by\cite{rpar2} and itself
provoked\cite{rpar3,ag1,bmsv,ag2,fuku04,allferm} intensive
investigation of unifying models   that naturally incorporate
supersymmetry and the seesaw mechanisms\cite{seesaw} : in
particular models with the left-right gauge group as a part of the
gauge symmetry and $B-L$ broken at a high scale and R/M-parity
preserved to low energies\cite{rpar2}. The close contiguity of the
seesaw scale and the Grand Unified scale pushed SO(10) GUTs, which
are the natural GUT home of both Type I and Type II seesaw
mechanisms, but were long relegated as baroque cousins of the
-seemingly- more elegant  minimal  SU(5) GUT, into centre stage.
The understanding that the Susy SO(10) GUT based on the
${\bf{210\oplus 10 \oplus 126\oplus {\overline {126}} }}$ Higgs
  system   proposed\cite{aulmoh,ckn}  long  ago was an optimal  candidate for the
  Minimal Supersymmetric GUT (MSGUT)  crystallized after the
   demonstration of its minimality on parameter counting
 grounds and an elegant reduction of its spontaneous symmetry
 breaking problem to a single  cubic equation with just one unknown parameter\cite{abmsv}.
  Careful computations of the symmetry breaking\cite{aulmoh,ckn,abmsv} and
   mass spectra\cite{ag1,bmsv,ag2,fuku04} became available for the  MSGUT. These theories naturally maintain a
structural distinction between Higgs and matter fields and
therefore naturally preserve R-parity down to low
energies\cite{rpar1,rpar2,rpar3}.

 The initial euphoria\cite{allferm} that the version utilizing only
$\mathbf{10,\oot}$  Higgs representations might prove
sufficient\cite{babmoh} to fit    all low energy fermion data in
an elegant and predictive  way ran aground when the successful
generic fits of fermion mass data were shown to be
unrealizable\cite{gmblm,blmdm,bert3} in the context of the actual
seesaw mechanisms(both Type I and Type II)  available in the
MSGUT. Both types of seesaw yielded neutrino masses that were too
small and Type I was shown to generically dominate Type II. Faced
with this impasse it is natural to have recourse to the third
allowed type of Fermion Mass (FM) Higgs, i.e the ${\bf{120}}$-plet
of SO(10). The ${\bf{120}}$-plet had previously played a
relatively minor role in fitting the fermion mass
data\cite{bert,dattmim}.

  In view of our no-go result in the MSGUT, however,
  we proposed\cite{blmdm}  a re-allocation of roles among
the three types of FM Higgs representation by suppressing the
 \boot  Yukawa couplings relative to those of ${\bf{10,120}}$.
Since the Type I seesaw neutrino masses are inversely proportional
to the  \boot  Yukawa coupling  this   would   enhance  the Type I
seesaw masses to viable   levels(Type II contributions get further
suppressed) while perhaps still allowing sufficient freedom to fit
all the fermion mass and mixing data. This also has the
interesting consequence that right-handed neutrino masses would be
significantly lowered into a range ($10^8-10^{13}$) GeV compatible
with leptogenesis.

Related subsequent work\cite{grimus1,grimus2,grimus3}   gave mixed
signals  regarding the viability of our proposal to use
 ${\bf{10}}+{\bf{120}}$ Higgs to fit charged fermion masses  while
 small \boot Yukawa couplings to   raise Type I neutrino  masses by
lowering \boot vev generated   right-handed neutrino masses.
However we find accurate NMSGUT specific  fits using ultra small
\boot couplings but a somewhat enlarged fitting scenario that
takes recourse to the strong influence (at the large $\tan\beta
\sim 50$ values typically favored by SO(10) Susy GUTs) of
supersymmetric particle  threshold corrections on the down type
quark masses. This is done in order to lower the Yukawa couplings
of the down and strange quarks to values that can be accommodated
by the NMSGUT specific fitting formulae.   These fits are
manifestly distinct from the accurate generic fits found in
\cite{grimus2,grimus3}, which besides being un-utilizable in the
NMSGUT\cite{pinmsgut} also give a distinct hierarchy  of right-handed
neutrino masses. In our fits we find that right-handed
neutrinos are much lighter than the GUT scale and strongly
hierarchical while neither statement applies to the fits of
\cite{grimus2,grimus3}. Moreover \cite{pinmsgut} the translation
between generic and NMSGUT parameters implies the NMSGUT
parameters are subject to insolubly non-linear constraints
rendering the usefulness of the generic fits problematic.
 We showed that in order to obtain fits that were
compatible with the   NMSGUT it would be necessary to search for
fits using the detailed NMSGUT specific formulae developed by us
in  \cite{nmsgut1} since the the generic parameters used in
\cite{grimus1,grimus2} were in fact subject to highly non-trivial
constraints (due to their origin in the NMSGUT) that were in
practice unverifiable since they could not be inverted.

Thus the GUT based on the ${\bf{210\oplus 10\oplus 120\oplus
126\oplus {\overline {126}} }}$ Higgs system   emerges as a New
Minimal Supersymmetric GUT (NMSGUT) capable of fitting all the
known fermion mass and mixing data. The New MSGUT calls for and
deserves the same detailed analysis of its superheavy
Renormalization Group(RG) flow threshold effects, fermion mass fit
compatibility and exotic effect effective superpotential that we
earlier provided\cite{aulmoh,ag1,ag2,gmblm,blmdm} for the theory
without the $\mathbf{120}$. We previously called that the
MSGUT\cite{abmsv} but it's claim on that name is now tenuous and
faded. In this paper we present  the required spontaneous symmetry
breaking, spectra, Higgs doublet fine-tuning and
``Higgs-fraction'' determination leading to matter fermion  Yukawa
expressions in terms of GUT parameters (as well as Weinberg
operator coefficients leading to seesaw neutrino masses) and
superheavy  threshold  effect formulae in the gauge RG flow from
$M_Z$ up to $M_X$. With these formulae in hand we can face the
burning question as to  whether the   renormalizable
supersymmetric SO(10) GUT based on the Higgs set ${\bf{210\oplus
10\oplus 120\oplus 126\oplus {\overline {126}} }}$  i.e. the so-
called Next or New Minimal Supersymmetric GUT(NMSGUT)
\cite{nmsgut1} is capable of encompassing the available fermion
mass-mixing data.

 As mentioned    the
  fit to charged fermion data using only $\mathbf{10}$ and
 $\mathbf{120}$ representations   fails badly to account for the
   down  and strange quark Yukawa couplings  at $M_X$. These Yukawas
    emerge much  too  small if the 3rd generation is fitted correctly. In retrospect
 this failure was  to be expected since the $\mathbf{\oot}$
 contains the Higgs irreps required to implement the
 Georgi-Jarlskog\cite{gjrl} mechanism that can account for
 $y_{\mu}/y_s(M_X)\sim 3$  in terms of the vev of the (15,1,1)
 sub-rep of the  $\mathbf{\oot}$ and third generation Yukawas
  dominated by the \textbf{10}-plet. This is not possible when the
  $\mathbf{\oot}$ couplings are highly suppressed.

Till  recently these  discussions \textbf{--}in the `MSGUT
community'   \textbf{--}proceeded under the somewhat sanguine
assumption that the  appropriate uncertainties  for the fitting
exercise could be estimated by ignoring all   threshold
corrections  right up to the GUT scale and hence were essentially
the  extrapolated errors at the low scale : thus for example the
 lepton  masses were sought to be fitted to one part in a million or
 better\cite{allferm}.
 In fact however it has been known\cite{hallratt,carena} from
 the early days of (large $\tan\beta$)  $  b-\tau-t$ unification
 that the $T_{3L}=-1/2$ charged fermions suffer large threshold
 corrections. The relevance of these corrections for estimating
  the uncertainties in the  GUT scale fermion masses
  has recently been re-emphasized\cite{rosserna,antusch}.
    The NMSGUT   like   other  $SO(10)$ GUTs   prefers
 large $ \tan\beta$ to naturally explain third generation charged
 fermion  Yukawas $(y_t, y_b, y_\tau)$ as nearly equal due to their  domination
  by the $33$ component of the $\mathbf{10}$-plet Yukawa coupling.
  In this paper we shall see that the NMSGUT also requires large
  $\tan\beta\sim 50$ to drive  threshold corrections that lower
  the values of the down and strange quark Yukawa couplings
  in the MSSM by a factor of about 5 while
  preserving, or  rather slightly  raising, the bottom
  quark Yukawa coupling in the MSSM.  This allows us to
  match the rather low values (at $M_X^0=10^{16.33}$ GeV) found by our ``downhill
  simplex''   search and fitting program. We   run  the
   couplings between $M_Z$ and $M_X^0$ using the   two-loop
    Renormalization Group(RG) equations  for the MSSM\cite{ramondSMMSSM,martinRG};
      we also include running of the $d=5$  Weinberg
  operator\cite{babpanta,antudrees}  but not  any right-handed neutrino thresholds.
This fitting  determines characteristic patterns(but not the
overall scale) of Susy breaking parameters and thus a distinctive
phenomenology for the NMSGUT which is  potentially falsifiable or
verifiable   at the LHC.

A significant improvement relative to the initial version of this
paper\cite{nmsgut2} is that we   carried out the fitting in terms
of soft Susy breaking parameters at $M_X$ that manifestly  respect
the SO(10) GUT structure. We generated the soft Susy breaking
terms from the values of
  $\{m_0,m_{1/2},A_0,B,M^2_{H,\bar H}\}(M_X^0)$ together with $\mu(M_X^0)$
by imposing electroweak  symmetry breaking (at one loop) at
$M_Z$.  These Susy breaking parameters respect the GUT symmetry.
 The light Higgs masses derive from the soft masses of several
NMSGUT Higgs multiplets and may thus justifiedly taken to be
different and independent of the sfermion mass parameter. The two
scale fitting is achieved by the usual iterative process that
loops from low to high scales and back\cite{ramondSMMSSM}.  The
constraints determined by the peculiar route taken by   NMSGUT
fits may become acute if LHC discovers at least one of the Susy
particles and anchors the floating scale of the Susy breaking soft
mass patterns required in our model. Current\cite{mumbai} LHC data
indicates that if a light Higgs exists it must be in the Susy
preferred range of $115-145 \,GeV$. In addition colored sparticles
should be heavier than at least 500 GeV and singly charged exotics
than 200 GeV. Our analysis predicts a scenario with light Bino
LSP($\chi^0\sim {\tilde B}$), light Higgs($h^0$) and $\tilde
W^\pm$, heavy third generation sfermions, very heavy exotic
Higgs(inos)($A, H^0,H^{\pm},\chi^0_{3,4}$) and with 1-2
generation sfermions spanning the range between the gaugino masses
and the third sgeneration. In our model this novel pattern is made
possible by values of $\mu,A_0,\sqrt{B} \sim 100\, TeV$. Since
accurate unification is the main motivation for supersymmetry
itself, the need at this juncture is for GUT driven insight into
the possible mass patterns of Susy partners that might be unveiled
by LHC. We offer the preferred soft parameter patterns of the
NMSGUT as our `bet' in the LHC-Susy `sweepstakes'.

   Even after it fits  the fermion data the
NMSGUT   must still  face the challenge posed by  the non-observation of
proton decay. Minimal  Susy GUTs generically imply
  proton  decay rates via $d=5$ operators which are higher than the
current experimental upper bounds\cite{murpierce} (but see also
\cite{bperezsenj}).  Soft susy
 breaking parameters and in particular the sfermion masses and
 mixing matrices have a crucial influence on the rates of the
 $d=5$ mediated proton decay.   In this paper
     we derive also  the $\Delta  B\neq 0$ effective superpotential
      by integrating out heavy fields
  and obtain the (operator dimension 4) B violation superpotential.
  We use these to calculate the nucleon decay rates implied by the
  fermion fit parameters found. The rapidity of the baryon decay rates  presents a
  difficulty whose resolution  requires  further development
  of the NMSGUT to account for threshold effects also on the GUT
  scale effective Yukawa couplings which is accomplished in a subsequent
  paper\cite{gutupend}.

In Section \textbf{2} We discuss the modifications of the
superpotential  introduced by the \textbf{120}-plet. For
completeness  some details regarding the   parameters of the MSGUT
superpotential are summarized  in Appendix \textbf{A}. In Section
\textbf{3} we discuss
   the  modified mass spectra of the NMSGUT and give
   the detailed analytic expressions in  Appendix \textbf{B}.
   In Section \textbf{4}    we  give the renormalization group  formulae
used to check the effect of superheavy thresholds on the
unification parameters. In section \textbf{5 } we give formulae
relating the effective MSSM parameters in terms of SO(10)
parameters. In sub-section \textbf{5.1} we give the formulae for
the fermion Yukawas  of the effective MSSM in terms of the GUT
couplings and the so-called ``Higgs fractions ''. In sub-section
\textbf{5.2} we integrate
  out the heavy triplets  that mediate baryon decay via $d=5$ operators
  and give the resultant effective  superpotential in terms of the matter superfields
 of the effective MSSM. In Section \textbf{6} we describe our numerical
 procedure  but  give details regarding Susy threshold corrections
 and electroweak breaking in Appendix \textbf{C}. In Section
 \textbf{7} we discuss(\textbf{Section 7.1}) the features of two example fits (the numerical parameter values of which are
 given in tabular form in Appendix \textbf{D}) and  dissect(\textbf{Section 7.2}) their structure.
     In Section \textbf{7.3} we use  the  explicit sample sets of NMSGUT parameter
    values and Susy soft breaking data   to estimate
     typical  predictions  of a \emph{fully specified }   NMSGUT for
  nucleon decay rates using the formalism developed in  \cite{lucasraby,gotonihei}
  (and  by earlier workers cited  therein).
 We conclude, in Section \textbf{8}, with a summary,  a  discussion
    of issues  and the way forward.
     \cite{gutupend}.

 \section{The New  Minimal Susy GUT }

 The  original  MSGUT \cite{aulmoh,ckn,babmoh,abmsv} was the
  renormalizable   supersymmetric $SO(10)$ GUT
 whose Higgs chiral supermultiplets  consist of AM(Adjoint Multiplet) type   totally
 antisymmetric tensors : $
{\bf{210}}(\Phi_{ijkl})$,   $
{\bf{\overline{126}}}({\bf{\Sigb}}_{ijklm}),$
 ${\bf{126}} ({\bf\Sigma}_{ijklm})(i,j=1...10)$ which   break the GUT symmetry
 to the MSSM, together with Fermion mass (FM) Higgs {\bf{10}}-plet(${\bf{H}}_i$).
  The  ${\bf{\overline{126}}}$ plays a dual or AM-FM
role since  it also enables the generation of realistic charged
fermion   and    neutrino masses and mixings (via the Type I
and/or Type II seesaw mechanisms);  three  {\bf{16}}-plets
${\bf{\Psi}_A}(A=1,2,3)$  contain the matter  including the three
conjugate neutrinos (${\bar\nu_L^A}$). Some further details are
provided in Appendix \textbf{A} and a full treatment is available
in\cite{abmsv,ag1,bmsv,ag2}.

The introduction of the three    anti-symmetric index
${\bf{120}}$-plet Higgs representation $\mathbf{\Theta}_{ijk}$
leads to new couplings in the superpotential .
  The additional terms are
 \begin{eqnarray*}
W_{NMSGUT}&=& \frac{m_{\Theta}}{2(3!)}\Theta_{ijk}\Theta_{ijk} +
\frac{k}{3!}\Theta_{ijk}H_m \Phi_{mijk}
  +
 \frac{\rho}{4!}\Theta_{ijk}\Theta_{mnk}\Phi_{ijmn} \\
  &+&  \frac{1}{2(3!)}
 \Theta_{ijk}\Phi_{klmn}(\zeta \Sigma_{lmnij}
  +  \bar\zeta \bar\Sigma_{lmnij})
 +  \frac{1}{3!}g_{AB} \Psi_A^T
 C_2^{(5)}\gamma_{i }\gamma_{j}\gamma_{k}\Psi_B \Theta_{ijk}
 \end{eqnarray*}

 The Yukawa coupling $g_{AB}$ is a complex
antisymmetric $3\times 3 $ matrix. We will frequently need to
decompose SO(10) fields and invariants according to the Pati-Salam
and SM subgroups of SO(10). This involves translating the SO(10)
vector and spinor labels to those of these subgroups. A complete
method and handbook for this translation is available in
\cite{ag1}. Some relevant details about our index conventions are
as follows : The real indices of the vector representation of
SO(10)  are denoted by $i,j =1..10 $.  The indices of the doublet
of SU(2)$_L(\rm{SU(2)}_R$) are denoted
$\alpha,\beta=1,2$($\dot\alpha,\dot\beta=\dot{1},\dot{2})$.
Finally the index of the fundamental {\bf{4}}-plet of SU(4) is
denoted by a (lower) $\mu,\nu = 1,2,3,4$ and its upper-left block
SU(3) subgroup indices are $\bar\mu,\bar\nu = 1,2,3$. The
corresponding indices on the ${\bar{\bf 4}}$ are carried as
superscripts. These doublets and quartets correspond to the chiral
spinor representations of the $SO(4)$ and $SO(6)$ subgroups of
$SO(10)$. Details of the spinor decomposition techniques may be
found in \cite{ag1}.
  The $SU(4)\times SU(2)_L\times SU(2)_R\quad $ ( $(n,m,l)$ denotes a
Pati-Salam representation of SU(4) dimension n , $SU(2)_L$
dimension m and $SU(2)_R$  dimension $l$) decomposition of the
${\bf{120}}$-plet is as follows : \bea \Theta_{ijk}(120)
&=&\Theta^{(s)}_{\mu\nu}({10,1,1})+
{\Theta}^{\mu\nu}_{(s)}(\overline{10},1,1)
+{\Theta_{\nu\alpha\dot\alpha}}^{\mu}(15,2,2)\nonumber\\
&+& \Theta^{(a)}_{\mu\nu\dot\alpha\dot\beta}(6,1,3)+
{\Theta^{(a)}_{\mu\nu}}_{\alpha\beta}(6,3,1)+
\Theta_{\alpha\dot\alpha}(1,2,2) \eea The sub/superscripts ``(s),
(a)'' denote symmetry and antisymmetry in SU(4) indices $\mu,\nu$.
Note that the \textbf{120}-plet contains no SM singlet so that the
MSGUT high scale spontaneous symmetry breaking analysis remains
the same. The arbitrary phase of the $\bf{120}$ reduces the
effective number of the extra couplings
($m_{\Theta},\rho,\zeta,\bar\zeta,k$ ($5_c-1_r=9_r$) and
$g_{AB}(3_c=6_r)$) so they  amount to 15 additional parameters.
Thus the relative advantage\cite{abmsv,melsen} with respect to
$SU(5)$ theories using additional fields or higher dimensional
operators to correct the fermion mass relations of the simplest
$SU(5)$ model  seems weakened but is not lost.

In fact the old MSGUT fails to fit the fermion mass data due to
difficulties with the overall neutrino  mass
scale\cite{gmblm,blmdm}. An alternative scenario within the NMSGUT
  to remove  the problems of the old MSGUT was proposed and
elaborated   in \cite{blmdm,nums,msgreb}.   In this scenario the
Yukawa  couplings ($f_{AB}$) of the $\mathbf{\oot}$ are much
smaller   than those of the $\mathbf{10,120}$. This boosts the
value of the Type I seesaw masses (which were in any case dominant
over the Type II seesaw masses but still too small) so that they
are generically viable.

  The NMSGUT even with only real Yukawas(except the fine tuned complex parameter
  $M_H$) i.e. with a total of 23($=12~ \rm{fermion ~ Yukawas} +  11 ~ \rm{AM ~
Yukawas}$) real parameters (further reduced to 22 by the  fine
tuning condition ) may be first tried to fit  the fermion data (
12 masses + CKM phase + 3 CKM angles + 3 PMNS angles + 3 PMNS
phases = 22 parameters, of which  the last three and one neutrino
mass are unknown leaving 18 targets for the fitting). Such a
theory will have less parameters than even the (unsuccessful) old
MSGUT.  However   restriction to real values is somewhat arbitrary
 since it cannot be justified by a CP symmetry in view of the
 complexity of the fine tuned value of $M_H$.
  A preliminary search with real parameters is useful for locating
   promising approximate solutions to use as starting points
    in the full parameter space.Thus we shall allow all
 parameters to be complex. In that case number of free NMSGUT superpotential  parameters
 mounts to 37.

 The FM Higgs ${\bf{120}}$ does not contain any
 SM   singlets and hence the analysis of the GUT scale
 symmetry breaking in MSGUT carries over unchanged to the NMSGUT.
 In particular there is still only one complex parameter($x$) whose
 variation directly affects the vevs and thus the masses in the theory.
The additional kinetic terms  are given by covariantizing in the
standard way the global SO(10) invariant D-terms $[{\frac{1}
{3!}}\mathbf{\Theta}_{ijk}^*\mathbf{\Theta}_{ijk}]_D$.

\section{  AM Chiral masses via PS}

 As in the case of the MSGUT \cite{ag1,ag2} we open up the maze of
 NMSGUT interactions by  decomposing  $SO(10)$
invariants in the superpotential first into Pati-Salam invariants
and then, after substituting  the GUT scale vevs, we obtain the
superpotential in the MSSM vacuum in terms of MSSM invariants. The
results for the old  MSGUT case were already given in
\cite{ag1,ag2} thus we list only the effect of the additional
terms in the superpotential.

An example of the   PS form of    $W_{NMSGUT}$ is (we have
inserted line numbers for easy reference):
\bea \frac{k}{3!}H_i {\Theta}_{jkl} \Phi_{ijkl} &=&
k[\frac{1}{\sqrt{2}}i(\tilde{H}^{\mu\nu}_{(a)}{\Theta}_{\mu\lambda}^{(s)}{\Phi}_{\nu}^{~\lambda}+
H_{\mu\nu}^{(a)}
{{\Theta}}^{\mu\lambda}_{(s)}{\Phi}_{\lambda}^{~\nu})\\
& +&
\frac{1}{\sqrt{2}}(\tilde{H}^{\mu\nu}_{(a)}{{\Theta}}_{\nu}^{~\lambda\alpha\dot{\alpha}}
     \Phi_{\mu\lambda\alpha\dot{\alpha}}^{(s)}-H_{\mu\nu}^{(a)}{{\Theta}}_{\lambda}^{~\nu\alpha\dot{\alpha}}
     {\Phi}_{\alpha\dot{\alpha}}^{\mu\lambda(s)})\\
     & &
     +H_{\mu\nu}^{(a)}(\vec{{\Theta}}_{(a)(R)}^{\nu\lambda}\cdot\vec{{\Phi}}_{\lambda(R)}^{~\mu}+
      \vec{{\Theta}}_{(a)(L)}^{\nu\lambda}\cdot\vec{{\Phi}}_{\lambda(L)}^{~\mu})\\
     & & -\frac{1}{2}
     \tilde{H}^{\mu\nu}_{(a)}{\Theta}^{\alpha\dot{\alpha}}\Phi_{\mu\nu\alpha\dot{\alpha}}^{(a)}\\
     & & +\frac{1}{2}H^{\alpha\dot{\alpha}}({\Theta}_{\mu\nu}^{(s)}{\Phi}_{~~\alpha\dot{\alpha}}^
     {\mu\nu(s)}+
     {{\Theta}}^{\mu\nu}_{(s)}\Phi_{\mu\nu\alpha\dot{\alpha}}^{(s)})\\
     & & -\frac{1}{\sqrt{2}}H^{\alpha\dot{\alpha}}({{\Theta}}_{\nu\alpha}^{~\mu\dot{\beta}}
     {\Phi}_{\mu\dot{\alpha}\dot{\beta}}^{~\nu}+
     {{\Theta}}_{\nu\dot{\alpha}}^{~\mu\beta}{\Phi}_{\mu\alpha\beta}^{~\nu})\\
     & & - \frac{1}{2\sqrt{2}}
     H^{\alpha\dot{\alpha}}(\tilde{{\Theta}}^{\mu\nu}_{(a)\dot{\alpha}\dot{\beta}}
     \Phi_{\mu\nu\alpha}^{(a)\dot{\beta}}-\tilde{{\Theta}}^{\mu\nu}_{(a)\alpha\beta}
     \Phi_{\mu\nu\dot{\alpha}}^{(a)\beta})\\
     & & + H^{\alpha\dot{\alpha}}{\Theta}_{\alpha\dot{\alpha}}\Phi ]
\eea
These terms are now easy to decompose into MSSM invariants using
the decompositions $  SU(4)\rightarrow SU(3)\times U(1)_{B-L} $
and the weak hypercharge identification $Y=2 T_{3R} + B-L $ in the
left-right symmetric gauge group. The decomposition of the rest of
the superpotential terms due to the \textbf{120}-plet is given in
the second part of Appendix \textbf{A}.

The purely chiral superheavy supermultiplet masses can be
determined from these expressions simply by substituting in the AM
Higgs vevs and breaking up the contributions
 according to MSSM labels.

It is again easiest to keep track of Chiral fermion masses since
all others follow using supersymmtery and the organization
provided by the gauge super-Higgs effect.

There are three types of   mass terms involving fermions from
chiral supermultiplets in such models :

\begin{itemize}
\item   Unmixed Chiral \item  Mixed pure chiral \item  Mixed
chiral-gauge. We briefly discuss the notable features of the mass
spectrum calculation and give the actual mass formulae in the
Appendix \textbf{B}.
\end{itemize}
\subsection{Unmixed Chiral}
There are 26 different SM multiplet types in the MSGUT and NMSGUT.
Thus the  alphabetical naming convention specified in Appendix
\textbf{B} is very natural.  A pair of chiral fermions
transforming as \GSM conjugates pairs up to form a massive Dirac
fermion.   In the  MSGUT case there are 17 MSSM types of unmixed
chiral multiplets which form  Dirac supermultiplets pairwise and
two Majorana singletons. In the NMSGUT 6 of these
types(C,D,E,K,L,P) become further mixed with  others of the same
type leaving 11 Dirac supermultiplets (A,B,I,M,N,O,U,V,W,Y,Z) and
2 Majorana supermultiplets (S,Q)  unmixed.
 If the representation is real rather than complex
one obtains an extra factor of 2 in the   masses. The
representations,  field  components and masses  of all fields
are given in   in Appendix \textbf{B}.

\subsection{Mixed Pure Chiral}

For  such multiplets there is no mixing with the massive coset
gauginos   but there is a mixing among several multiplets with the
same SM quantum numbers. There were only three such multiplet
types in the MSGUT (i.e. $R[8,1,0], h[1,2,\pm 1],
t[3,1,\pm{\frac{2}{3}}]$) but in the NMSGUT, there are an
additional 5 mixed pure chiral types namely the $C[8,2,\pm 1],
D[3,2,\pm {\frac{7}{3}}], K[3,1,\pm {\frac{8}{3}}],\hfil\break
L[6,1,\pm {\frac{2}{3}}], P[3,3,\pm {\frac{2}{3}}]$. As for the
multiplet types  which had  mixed pure chiral mass terms in the
MSGUT,  the type   $ R[8,1,0] $
  acquires no new partners  and has an unchanged mass
matrix since the ${\bf{120}}$ has no such submultiplets. However
the other two mixed pure chiral multiplet types of the MSGUT do
acquire new contributions  :
\begin{itemize}

\item
$[1,2,-1](\bar{h}_1,\bar{h}_2,\bar{h}_3,\bar{h}_4,\bar{h}_5,\bar{h}_6)\bigoplus
[1,2,1](h_1,h_2,h_3,h_4,h_5,h_6)\equiv(H^{\alpha }_{\dot{2}},
\bar\Sigma_{\dot{2}}^{(15)\alpha},\\\Sigma_{\dot{2}}^{(15)\alpha},
\frac{\Phi_{44}^{\dot{2}\alpha}}{\sqrt{2}},\Theta^{\alpha}_{\dot{2}},
\Theta_{\dot{2}}^{(15)\alpha}\hspace{2mm}) \bigoplus (H_{\alpha
\dot{1}},\bar\Sigma_{\alpha \dot{1}}^{(15)},\Sigma_{\alpha
\dot{1}}^{(15)},\frac{\Phi_{\alpha}^{44\dot{1}}}{\sqrt{2}},
\Theta_{\alpha\dot{1}},\Theta_{\alpha\dot{1}}^{(15)})$\\
Here one gets an additional  2 rows and 2 columns relative to the
MSGUT since the \textbf{120}-plet contains two pairs of doublets
with MSSM type Higgs doublet quantum numbers so that the mass
matrix ${\cal H}$ is  $6\times 6 $. To  keep   one pair  of light
doublets in the low energy effective theory, it is necessary to
fine tune one of the parameters of the superpotential (e.g. $M_H$)
so that $Det{\cal H} =0$. By extracting the null eigenvectors of
${\cal H}^{\dagger}{\cal H}$ and ${\cal H}{\cal H}^{\dagger}$ one
can compute the composition of  the light doublet pair in terms of
the doublet fields in the full SO(10) GUT, and, in particular, we
can find the proportions of the doublets coming from the ${\bf 10,
{\overline{126}},120} $ multiplets which couple to the matter
sector\cite{nmsgut1}.

 \item $[\bar{3},1,\frac{2}{3}](\bar{t}_1,\bar{t}_2,\bar{t}_3,\bar{t}_4,
 \bar{t}_5,\bar{t}_6,\bar{t}_7)\bigoplus
[3,1,-\frac{2}{3}](t_1,t_2,t_3,t_4,t_5,t_6,t_7)\equiv(H^{\bar\mu 4
},\bar\Sigma_{(a)}^{\bar\mu4},\Sigma_{(a)}^{\bar\mu4},\Sigma_{(R0)}^{\bar\mu4},\hfil\break
\Phi_{4(R+)}^{\bar\mu},\Theta^{\bar\mu
4(s)},\Theta^{\bar\mu4}_{(R0)}) \bigoplus
(H_{\bar\mu4},\bar\Sigma_{\bar\mu 4(a)},\Sigma_{\bar\mu
4(a)},\bar\Sigma_{\bar\mu
4(R0)},\Phi_{\bar\mu(R-)}^{4},\Theta_{\bar\mu 4
(s)},\Theta_{\bar\mu 4(R0)})$

With the contribution of the  {\bf{120}}-plet one gets two
additional rows and columns and  the dimension of $[
 {3} ,1,\pm\frac{2}{3}]$ mass matrix ${\cal T}$  becomes 7
$\times$ 7. These triplets and antitriplets participate in baryon
violating processes  since the exchange of \hfil\break
$(t_1,t_2,t_4,t_6,t_7)\oplus (\bt_1,\bt_2,\bt_6,\bt_7) $ Higgsinos
generates $d=5$ operators of type QQQL and ${\bar l \bar u \bar
u\bar d}$. The strength of the operator is controlled by the
inverse of the $\bt-t$ mass matrix ${\cal T}$.

\end{itemize}

\subsection{ Mixed Chiral-Gauge}

Finally we come to the mixing matrices for the chiral modes that
mix with the gauge particles as well as among themselves. There is
no direct mixing between MSSM fields contained in
\textbf{120}-plet with gauge particles. However mixing \emph{is}
present  via other MSSM submultiplets present in NMSGUT Higgs
fields which further mix with gauge fields. This occurs for   all
such    multiplet types except $G[1,1,0]$ and
$X[3,2,\pm{\frac{4}{3}}]$ which are unchanged, while   for
$E[3,2,\pm{\frac{1}{3}}],F[1,1,\pm 2],J[3,1,\pm{\frac{4}{3}}]$
mass matrices  acquire  additional rows and columns. Thus

\begin{itemize}

\item{ } $[\bar 3,2,-{\frac{1}{ 3}}](\bar E_1, \bar E_2,\bar
E_3,\bar E_4,\bar E_5,\bar E_6) \oplus [3,2,{\frac{1}{
3}}](E_1,E_2,E_3,E_4,E_5,E_6)$\hfil\break $\qquad\qquad \equiv
(\Sigma_{4 \dot 1}^{\bar\mu\alpha}, \Sigb_{4\dot 1}^{\bm \alpha},
\phi^{\bm 4\alpha}_{(s)\dot 2} , \phi^{(a) \bm 4\alpha}_{\dot
2},\lambda^{\bm 4\alpha}_{\dot 2},\Theta_{4
\dot{1}}^{\bar\sigma\alpha}) \oplus  (\bar\Sigma_{\bar\mu \alpha
\dot{2}}^{4}\s_{\bm\alpha\dot 2}^4,\phi_{\bm 4\alpha\dot 1}^{(s)},
\phi_{\bm 4\alpha\dot 1}^{(a)},\lambda_{\bm\alpha\dot
1},\Theta_{\bar\sigma\alpha}^{4 \dot{1}}) $

The $6\times 6$ mass matrix $\cal E$
  has the usual super-Higgs structure : complex conjugates of
 the 5th row and column (omitting the diagonal entry)
 furnish  left and right null eigenvectors
of the  chiral $5 \times 5 $ submatrix ${\bf E}$ obtained by
omitting the fifth row and column.
 ${\cal E}$ has non-zero determinant
although the determinant of ${\bf{E}}$ vanishes.

\item{ }
$[1,1,-2](\hspace{2mm}\bar{F}_1,\bar{F}_2,\bar{F}_3,\bar{F}_4\hspace{2mm})\bigoplus
[1,1,2](\hspace{2mm}F_1,F_2,F_3,F_4\hspace{2mm})\equiv
(\bar\Sigma_{44 (R0)}/\sqrt{2}, \Phi_{(R-) }^{(15)},\lambda_{(R-)
},\\\Theta_{44}/\sqrt{2}\hspace{2mm}) \bigoplus (\Sigma_{(R0)
}/\sqrt{2},\Phi_{(R+)}^{(15)},\lambda_{(R+)},\Theta^{44}/\sqrt{2})$

The same comments as for $E[3,2,\pm{\frac{1}{ 3}}]$ apply
\emph{mutatis mutandis} i.e. $4\times 4$ mass matrix $\cal F$ with
$3 \times 3 $ submatrix ${\bf F}$ and so forth.

\item  $[\bar 3,1,-{\frac{4}{ 3}}](\bar J_1,\bar J_2,\bar J_3,\bar
J_4,\bar J_5) \oplus [3,1,{\frac{4}{ 3}}](J_1,J_2,J_3,J_4,J_5)$
\hfil\break $\qquad\qquad\equiv (\s^{\bm4}_{(R-)},\phi_4^{\bm},
\phi_4^{~\bm(R0)},\lambda_4^{~\bm},\Theta^{\bar{\mu}4}_{(R-)})
\oplus (\Sigb_{\bm4(R+)},\phi_{~\bm}^4,
\phi_{\bm(R0)}^{~4},\lambda_{\bm}^4,\Theta_{\bar\mu 4}^{(R+)})$

Here we have $5\times 5$ mass matrix $\cal J$ with $4 \times 4 $
submatrix ${\bf J}$ and so forth.
\end{itemize}
This concludes our description of the superheavy mass spectrum of
the  NMSGUT.

\section{ RG Analysis}

The analysis of the flow of gauge couplings is the most familiar
element in   papers on grand unification,  with the iconic diagram
of merging gauge couplings at the GUT scale as   an easily
recognized referent. Even the failure to unify is often quantified
in terms of a failure of the three running couplings to intersect
at a point. However in  the conceptually clearer  approach of
Weinberg\cite{weinberg}, explicated at length by Hall\cite{hall},
  based upon matching the low and high energy effective
gauge field theories in an overlap region, the flow of the three
gauge couplings of the SM intersects the flow of the gauge
coupling of the unifying group at three distinct scales.
Nevertheless the three equations relating the MSSM gauge couplings
at $M_Z$ to the GUT coupling at any  convenient superheavy scale
 may be solved for the unification predictions.
Perturbative consistency demands that the three scales are closely
spaced so that the calculations are consistent. To counter the
fixed idea that the aim of unification is a  three fold
intersection of gauge couplings we explain the formalism and our
procedure at some length.
   With the superheavy spectrum formulae in hand the threshold
 corrections to the matching relations between the couplings in the low and high energy effective
 theories   can be evaluated. The relation between the MSSM couplings at the  low scale  where
they are specified from measurement (which we always take to be
the Z boson mass  $M_Z= 91.1876$ GeV to allow easy use of the
standard calculation of 1-loop effects in the MSSM
\cite{piercebagger}) and the GUT coupling at the scale $M_X$ is
given by \cite{weinberg, hall} : \bea {\frac{1}
{\hat\alpha_i(M_Z)}}
 ={\frac{1} {\alpha_G(M_X)}} +
 8 \pi b_i \ln{{M_X}\over{M_Z} }
  + 4 \pi \sum_j {{{b_{ij}} \over {b_j}}} \ln X_j -
4\pi\lambda_i(M_X) +\ldots \eea Here $\hat\alpha_i=g_i^2/4\pi,
i=1,2,3$, where $g_i$ are the MSSM gauge couplings in (SU(5) ) GUT
normalization, should not be confused with the Higgs fractions
$\alpha_i, i=1...6$.  The second, third, and fourth terms
correspond to the one-loop gauge running, two loop gauge running
and superheavy threshold corrections respectively. The ellipsis
refers to the (small) contributions from matter Yukawa couplings
at two loops. Here $ X_j= 1 + 8 \pi b_j \alpha_G(M_X ) \ln{{M_X
}\over{M_Z }}$ and $\lambda_i(M_X)$ may be evaluated at \emph{
any} superheavy scale (concretely : any scale whose logarithm
differs from that of the 1-loop crossing point($M_X^0$) of the
$SU(2)$ and $U(1) $ couplings by terms of order $\alpha$
\cite{hall}). We shall  use
    $M_X^0$    as the leading approximation to the  superheavy   matching scale between
MSSM and GUT    in all our calculations, while the unification
scale itself  is taken as  the mass $M_X$ of the $X[3,2,\pm 5/3]$
type gauge supermultiplet.
   Similarly for ease of interpretation we shall use the mass  of the
   Z boson $M_Z=91.1876$ GeV as the low scale renormalization and matching
    point  between the SM and MSSM.
    The coefficients  $b_i\{b_1,b_2,b_3\}=(1/16\pi^2) \{{{33}\over 5},1,-3\}$ are the
one-loop gauge beta function coefficients in the MSSM and
     similarly $b_{ij}$ are the   two-loop coefficients\cite{martinRG} .
   The term containing $\lambda_i$  represents the leading contribution of the
superheavy thresholds calculated within the Weinberg-Hall
formalism\cite{weinberg,hall}. In the ${\overline{MS}}$ scheme one
has : \be \lambda_i (\mu)=-{2\over {21}} (b_{iV} + b_{iGB})
 + 2(b_{iV} + b_{iGB})\ln{{M_V}\over{\mu }} +2 b_{iS}\ln{{M_V}\over{\mu }}+2
b_{iF}\ln{{M_F}\over{\mu }} \ee where V,GB,S,F refer to vectors,
Goldstone bosons,
 scalars and fermions respectively and $b_i$ are one-loop beta function coefficients,
   and a sum over heavy  mass eigenstates is implicit. The
   coefficient of the $\mu$ independent term arises\cite{hall} in an ${\overline{MS}}$ scheme
   where metric traces are carried out in $D-4$ but the  gamma matrix
   traces for fermions are in 4 dimensions.
However for supersymmetric theories the ${\overline{DR}}$
scheme\cite{DRbarscheme}  should be  used so that the equality of
 bosonic and fermionic degrees of freedom is preserved.   In this
scheme only the momentum integrations, but not the metric or gamma
matrix traces, are continued  away from 4 dimensions. Hence the
$\mu$ independent term is absent from the threshold corrections in
the ${\overline{DR}}$ scheme. Thus for consistency we should
convert the input parameters $\alpha_{1,2}(M_Z)$ calculated from
the precisely known\cite{pdb} electroweak data  in the
${\overline{MS}}$ scheme ( $1/\alpha(M_Z)  = 127.916 \pm 0.015,
\sin^2 \theta_W=0.23116 \pm 0.00013$ ) into the corresponding
${\overline{DR}}$ parameters.

Next one uses the three equations to determine
  ${\hat\alpha_3}(M_Z), M_X,\alpha_G(M_X)$  as a perturbation expansion around the one-loop values.
 In view of the above remarks the unified coupling $\alpha_G$ and the prediction for
${\hat\alpha_3}(M_Z) $ will also emerge in the ${\overline{DR}}$
scheme.
  The required formulae were conveniently summarized in \cite{langpolo1}
   and   analysis of the difficulty concerning a too high
  value of ${\hat\alpha_3}(M_Z)$ can be found in \cite{langpolo2}.

Inserting the ${\overline{DR}}$   converted values we find the one-loop predictions \bea M_X^0=10^{16.33}\, GeV \quad
(\alpha_G^0)^{-1} =24.2496 \quad {\hat\alpha_3}^0(M_Z)=0.1183\eea
Including the two-loop corrections and iterating once gives the
two-loop values
 \bea
M_X =10^{16.553}\, GeV \quad \alpha_G^{-1} =22.9803 \quad
{\hat\alpha_3}(M_Z) =0.1307\eea This should be compared with the
${\overline{DR}}$ experimental value
$({\hat\alpha_3}(M_Z))^{expt}=0.1194 $.   Two-loop corrections
$H_s(y_t,y_b,y_\tau, ...,\tan\beta)$ due to  third generation
Yukawas are usually negative but typically\cite{langpolo2} $ H_s >
-.003 $ even for the large values of  $\tan\beta\sim 50$ relevant
for SO(10) Susy GUTs. Thus it is clear that there is, \emph{prima
facie},  a discrepancy of  6 to 8 \% between the larger value of
the prediction and the current best estimate for
${\hat\alpha_3}(M_Z)^{\overline {DR}}= 0.119$. However the effects
of threshold corrections due to the supersymmetric thresholds
between $M_Z$ and (up to, say,)  $100 \, TeV$   \emph{as well as}
the thresholds associated with the very large number ($\sim 500$ )
superheavy fields present in the model (with  masses ranging in
the band $10^{15\pm 3}$ GeV) could well explain this discrepancy.
It is the effect of superheavy thresholds that is the focus of the
gauge RG investigations of the NMSGUT  in this paper. With the
above motivation we shall search for   GUT  parameter sets for
which the corrections to the unification parameters due to
superheavy thresholds stay  within the following prima facie
reasonable but open-minded  limits : \bea
-20.0\leq \Delta_G &\equiv&  \Delta  (\alpha_G^{-1}(M_X))  \leq 25 \nonumber \\
3.0 \geq  \Delta_X &\equiv &\Delta (Log_{10}{M_X}) \geq - 0.3\nonumber \\
-0.017< \Delta_{3} &\equiv & {\hat\alpha_3}(M_Z)  <
-0.004\label{criteria} \eea  These limits are imposed to respect
perturbativity, $M_X <M_{Planck}$ and yield a negative correction
to ${\hat\alpha_3}(M_Z)$ in the relevant range. Here
$\Delta_{G,X}$ include the two-loop corrections as well as the
threshold corrections but $\Delta_3$ is only due to superheavy
thresholds. Eventually- when likely supersymmetric thresholds are
also established - we can combine all contributions  in order to
fit ${\hat\alpha_3}(M_Z)$ to within its error bars (currently
given, somewhat optimistically at  around 1.5 \% but with a rather
larger variance across measurement methods !).

The threshold correction \cite{ag2,gmblm} formulae are \bea
\Delta^{(th)}(\ln{M_X}) &=&{{\lambda_1(M_X) - \lambda_2(M_X) }
\over{2(b_1 - b_2)}} \nonumber \\
 \Delta_X &\equiv& \Delta^{(th)}(Log_{10}{\frac{M_X}{1 GeV}}) + \Delta^{(2-loop)}(Log_{10}{\frac{M_X}{1 GeV}}) \nonumber\\
   &=&  0.222 + {\frac{5({{\bar b}'}_1 -{{\bar b}'}_2
 )}{28}}
 Log_{10}{{M'}\over  {M_X}} \label{Deltasw} \nonumber \\
 \Delta_3 &\equiv &\Delta^{(th)} ({\hat\alpha_3} (M_Z))  \nonumber\\&=&
 {{100 \pi (b_1-b_2)\alpha(M_Z)^2}\over{[(5b_1+3b_2-8b_3)sin^2\theta_w(M_Z)-3(b_2-b_3)]^2}}
 \sum_{ijk}\epsilon_{ijk}(b_i-b_j)\lambda_k(M_X)\nonumber\\
&=&   .000311667 \sum_{M'} (5 {{\bar b}'}_1
 -12{{\bar b}'}_2 +7{{\bar b}'}_3) \ln{{M'}\over  {M_X
 }} \nonumber \\
 \Delta_G &\equiv &\Delta^{(th)}(\alpha_G^{-1}(M_X)) +\Delta^{(2-loop)}(\alpha_G^{-1}(M_X))  =  \frac{4 \pi(b_1
\lambda_2(M_X)-b_2 \lambda_1(M_X))}{b_1-b_2}\nonumber \\
&=& -1.27 +  {\frac{1}{56 \pi}} \sum_{M'}( 33 {{\bar b}'}_2 - 5
{{\bar b}'}_1) \ln{{M'}\over {M_X }}
  \label{Deltath} \eea
Where ${\bar b'}_i = 16\pi^2 b_i'  $ are   1-loop $\beta$ function
coefficients ($ \beta_i=b_i g_i^3 $)
 for multiplets with  mass $M'$ and $\lambda_i$ are
 the leading contributions of the superheavy thresholds\cite{hall,ag2}.

We   identify the scale $M_X$ with the physical  mass of the gauge
supermultiplet transforming as $[3,2,\pm {\frac{5}{3}}]$
responsible for $d=6$ baryon violation.  A crucial
point\cite{gmblm} is that the threshold corrections depend only on
ratios of masses and are independent of the overall scale
parameter which we choose to be the mass parameter of the
$\mathbf{210}$-plet $m$. Since on the one hand  $M_X= M_X^0
10^{\Delta_X} $ and on the  other  $M_X=m_{\lambda_X} =
|m/\lambda|{g\sqrt{ 4 |\tilde{a} + \tilde{w}|^2 +
   2 |\tilde{p}+ \tilde{\omega}|^2 }}   $  it
follows that the parameter $m$ is determined by the RG analysis to
be :\bea \Delta_X &=& \Delta (Log_{10}{{M_X} \over {1 GeV}})\nonumber \\
  | m| &=& M_X^0 10^{  + \Delta_X }
   {{|\lambda|}\over
   {g\sqrt{ 4 |\tilde{a} + \tilde{w}|^2 +
   2 |\tilde{p}+ \tilde{\omega}|^2 }}} GeV \label{mvalue}\eea
In the (N)MSGUT   $m$ sets the scale of all the superheavy vevs so
every superheavy mass   must   rise or fall in tandem with $M_X$
i.e with $\Delta_X$. The SO(10) gauge coupling in this formula may
be improved by using its threshold corrected value.

These corrections   modify  the one-loop values corresponding to
the successful gauge unification of the MSSM. In   spite of the
large number of superheavy fields they  can still give viable
unification over extended regions of the GUT parameter space thus
belying\cite{ag2,gmblm,blmdm}) early expectations that the
unification exercise is futile in SO(10) Susy
GUTs\cite{dixitsher}. The parameter $\xi= \lambda M/ \eta m$ (see
Appendix \textbf{A} of this paper for the parameter naming) is the
only numerical parameter that enters into the cubic
eqn.(\ref{cubic}) that determines the parameter $x$
   in terms of which all the  superheavy vevs are given.
    It is thus  the most crucial  determinant of
     the mass spectrum.      The rest of the coupling parameters divide into
     ``diagonal''($\lambda,\eta,\rho$) and ``non-diagonal''
     ($\gamma,\bar{\gamma},\zeta,\bar{\zeta }, k$)  couplings
     with the latter exerting a  weaker influence
     on the unification parameters\cite{ag2,nmsgut1}.
      The dependence of the threshold corrections on the
``diagonal'' couplings is also comparatively mild  except when
coherent e.g when  many masses are lowered  together leading to
$\alpha_G $ explosion, $ Log M_X$ collapse or large changes  in
${\hat\alpha_3} (M_Z)$. The development of the NMSGUT presented
here was motivated by the need to reconcile the demands of
unification and constraints imposed by a fit of the fermion data
using the GUT specific fermion mass formulae. In order to proceed
toward the example fits that prove by construction that such
completely realistic fits exist, and for reasons of space, we do
not present a survey of RG corrections over the huge parameter
space. However certain plots that attempt to illustrate some
typical behaviors are given   in\cite{nmsgut1}.

\section{NMSGUT derived parameters of the  Effective Action}
\subsection{Effective fermion Yukawas and Weinberg Operator
coefficients from the NMSGUT} As in the case of the MSGUT one
imposes the fine tuning condition $Det {\cal H} =0$ to keep a pair
of Higgs doublets $H_{(1)}, {\bar H}_{(1)}$ (left and right  null
eigenstates of the mass matrix ${\cal H}$) light. The composition
of these null eigenstates in terms of the GUT scale doublets then
specifies how much the different doublets contribute to the low
energy EW scale symmetry breaking. In the   Dirac mass matrices we
can replace $<h_i>\rightarrow \alpha_i v_u,  <\bar h_i>\rightarrow
\bar\alpha_i v_d$  (see Appendix \textbf{A} and \cite{ag1}). The
fermion Dirac masses may be read off the
 decomposition of $\bf{16\cdot16\cdot (10 \oplus 120\oplus \oot)}$ given
 in \cite{ag1,ag2}  and this yields
(we have  made slight  changes in notation relative
to\cite{blmdm}). The  \bea { y}^u &=&  ( {\hat  h} + {\hat f} +
{\hat g} )\quad ;\quad {\hat {r}}_1= \frac{
\bar\alpha_1}{\alpha_1} \quad ;\quad {\hat {r}}_2= \frac{
\bar\alpha_2}{\alpha_2} \nnu
 { y}^{\nu}&=& ({\hat  h} -3 {\hat  f}  + ({\hat {r}}_5 -3) {\hat {g}})\quad
;\quad  {{\hat {r}}_5}= \frac{4 i \sqrt{3}{\alpha_5}}{\alpha_6+ i
   \sqrt{3}\alpha_5}
 \nnu
 { y}^d &=& {  ({\hat {r}}_1} {\hat  h} + { {\hat {r}}_2} {\hat  f}  +
{\hat {r}}_6 {\hat  g}); \quad {\hat {r}}_6 =
\frac{{{\bar{\alpha}}_6}+ i \sqrt{3}{{\bar{\alpha}}_5}}{\alpha_6+
i \sqrt{3}\alpha_5} \nnu
   { y}^l &=&{ ( {\hat {r}}_1} {\hat  h} - 3 {  {\hat {r}}_2} {\hat  f} +
   ( {\hat {\bar{r}}_5} -
   3{\hat {r}}_6){\hat  g});\quad {\hat {\bar{r}}_5}=
\frac{4 i \sqrt{3}{{\bar{\alpha}}_5}}{\alpha_6+ i
   \sqrt{3}\alpha_5}
\label{120mdir}\\
{\hat  g} &=&2i g {\sqrt{\frac{2}{3}}}(\alpha_6 + i\sqt \alpha_5)
\quad;\quad \hat  h = 2 {\sqrt{2}} h \alpha_1 \quad;\quad\hat  f =
-4 {\sqrt{\frac{2}{3}}} i f\alpha_2 \nonumber
 \eea
The Yukawa couplings of matter fields with \textbf{120} Higgs
field give no contribution to the  Majorana mass matrix
 of the superheavy neutrinos $\bar\nu_A$  so it remains
 $M^{\bar\nu}_{AB}=  8 {\sqrt{2}} f_{AB} {\bar\sigma}$. Thus the
 Type I contribution  is obtained by eliminating $\bar\nu_A$
\bea W &=& {\frac{1}{2}}  M^{\bar\nu}_{AB} \bar\nu_A\bar\nu_B  +
\bar\nu_A m^{\nu}_{AB} \nu_B  + .....\rightarrow {\frac{1}{2}}
M^{\nu (I)}_{AB} \nu_A\nu_B  + ....\nnu M^{\nu(I)}_{AB} &=&
-((m^{\nu})^T (M^{\bar\nu})^{-1} m^{\nu})_{AB} \eea As  shown
in\cite{gmblm,blmdm} it is likely that the Type II seesaw
contribution is subdominant to the Type I seesaw. However the
consistency of the assumption that it is negligible must be
checked and quantified so we also evaluate the tadpole that gives
rise to the Type II seesaw since the ${\bf{120}}-$ plet does
contribute new terms.

  For computing the vev  $ < {\overline O}\{1,3,1,2\}_{\oot}>$
   which    gives rise to the Type II seesaw mass,   the relevant
terms in the superpotential are \bea W_{FM}^{\Sigb} &=&
M_O{\vec{\bar O}}\cdot  {\vec O} -{{\bar\ga}\over {\sq}}H^{\al\da}
\Phi_{44\da}^{\bet} {\bar O}_{\al\bet} -{{\ga}\over
{\sq}}H^{\al\da}
\Phi^{44\bet}_{\da} { O}_{\al\bet}\nonumber \\
&-&2\sq i \e ( \s_4^{~4\al\da} \Phi_{44\da}^{\bet}
 {\bar O}_{\al\bet}
+\Sigb_4^{~4\al\da} \Phi^{44\bet}_{\da}
 {O}_{\al\bet} )\nonumber\\
 &+& {\sqrt{2}}\zeta[\frac{1}{2}\Theta_{\dot{\alpha}}^{\beta}
 \bar{\Phi}^{44\alpha\dot{\alpha}}+{\Theta}_{4}^{~4\alpha\dot{\alpha}}
 \overline{\Phi}_{\dot{\alpha}}^{44\beta}]O_{\alpha\beta}\nonumber\\
  &+&
  {\bar\zeta}{\sqrt{2}}[\frac{1}{2}\Theta_{\dot{\alpha}}^{\beta}
 {\Phi}^{\alpha\dot{\alpha}}_{44}- {\Theta}_{4}^{~4\alpha\dot{\alpha}}
 {\Phi}_{44\dot{\alpha}}^{\beta}] {\bar{O}}_{\alpha\beta}    \nonumber\\
 &=&  M_O{\bar O}_{-} O_+  + {\bar O}_{-}
({{ {{i\bar\ga}}{ } }} \bar\al_1 +  2i{\sqrt 3}\e \bar\al_3+
{\sqrt{3}}\bar\zeta \bar\al_6 + i \bar\zeta \bar\al_5) \bar\al_4
v_d^2 \sqrt{2}\nonumber\\ & & - { O}_{+} ({ {{ i\ga}}{ } } \al_1 +
{ 2i{\sqrt  3}}\e \al_2- {\sqrt{3}}\zeta \al_6 + i \zeta \al_5)
\al_4 v_u^2 \sqrt{2} \eea Here the fields named $O,{\overline{O}}$
transform as $\{1,3,1,\mp 2\}$ with respect to the   LR symmetric
gauge group $G_{3221}$
  (with $U(1)_{B-L}$  as the Abelian  factor)  and are the left-handed
triplets contained in $(10,3,1),   ({\overline {10}},3,1)$ PS
sub-representations of
$\mathbf{\Sigma(126),{\overline{\Sigma}}(\oot)}$ respectively. Our
phase conventions\cite{ag1} have for  SU(4) 15 plet components
$\phi^A,A=1...15 : \, \phi_\mu^{~\nu}=\sum_{A=1}^{15}
i\phi^A{\lambda^A}_\mu^{~\nu}/{\sqrt{2}}$.
 So the electrically neutral component $ {\overline O}_-$
  of ${\overline{O}}\{1,3,1, 2\}$
has vev
 \be <{\bar O}_{-}>  = ( i\ga {\sqrt  2} \al_1 +
{ 2 i{\sqrt  6}}\e \al_2- {\sqrt{6}}\zeta \al_6 + i
{\sqrt{2}}\zeta \al_5) \al_4
   {\frac{v_u^2}{M_O}}     \qquad \ee
 and $M_O$
can be read off from Table in Appendix \textbf{B}   to be $M_O= 2
(M + \eta (3a-p))$. The Type II neutrino mass is then simply
$M^{\nu}_{AB} =16 i f_{AB} <{\bar{O}}_- >$. However, as found in
\cite{gmblm,blmdm} when computed in the  MSGUT these masses are
always negligible.

The NMSGUT derived formulae for the matter fermion Yukawas  given
in this section combined with the explicit formulae\cite{nmsgut1}
for the Higgs fractions $\alpha_i,\bar\alpha_i$, serve as the
basis for our investigation of the ability of the NMSGUT to fit
all the fermion mass data now available.

   As discussed in \cite{fukrebut,babel}
 the influence of conventions for phase choices is crucial and and
 requires  painful tracing of equivalences before numerical
 results for mass spectra  can be compared. Comparison
 with other works \cite{grimus1} that attempt  fits with the
 same Higgs representations but based on \emph{generic} fermion Yukawa formulae
 (rather than those specified by the explicit  GUT scale symmetry
 breaking which are used here) has become less interesting  since our
 demonstration that the generic fits are
 probably unrealizable in the NMSGUT due to the  extreme
 non-linearity of the translation and the hidden and insoluble
  constraints on fundamental  GUT parameters that the generic fits imply \cite{pinmsgut}.
 Hence we do not discuss this relation further here but refer the interested
 reader  to the previous arXiv versions of this paper if she wishes to find the
 relations between the parameters. See however our remarks on the
 magnitudes of parameters used in generic and NMSGUT specific fits in  subsection \textbf{7.1}.

\subsection{  Dimension $5 $ Operators for B,L violation}

In\cite{ag2} we worked out the effective $d=4$ superpotential for
$B+L$ violating processes  due to   exchange of color triplet
superheavy chiral supermultiplets contained in the
$\mathbf{10,\oot}$ Higgs multiplets. These included a novel
channel due to decays mediated by exchange of triplets
$t_{(4)}[3,1,\pm {\frac{2}{3}}]$ contained in the $\mathbf{\oot}$
Higgs irrep. Evidently the inclusion of   $\mathbf{120}$ plet
Higgs will lead to additional channels    for baryon violation.
These can be easily derived using the Pati-Salam decomposition of
the $\mathbf{16.16.120}$ SO(10) invariants\cite{ag1} : \bea
 { 1 \over
(3!)}\psi C^{(5)}_{2}\gamma_{i}\gamma_{j}\gamma_{k}\chi
{\Theta}_{ijk} & =&
-2{(\bar{{\Theta}}^{\mu\nu}_{(s)}\psi_{\mu}^{\alpha}\chi_{\nu\alpha}
+{\Theta}_{\mu\nu}^{(s)}\widehat\psi^{\mu\dot\alpha}{\widehat\chi_{\dot\alpha}^{\nu}}
)} -2\sqrt{2}{{\Theta}^{~\mu\alpha\dot\alpha}_{\nu}}
{{(\widehat\psi_{\dot\alpha}^{\nu}\chi_{\mu\alpha}-
{\psi_{\mu\alpha}\widehat\chi_{\dot\alpha}^\nu})}} \nonumber\\
&-&2({{\Theta}_{\mu\nu}^{(a)}}^{\dot\alpha\dot\beta}
\widehat\psi^{\mu}_{\dot\alpha}\widehat\chi^{\nu}_{\dot\beta}+\widetilde{{\Theta}}^
{\mu\nu\alpha\beta}_{(a)}\psi_{\mu\alpha}\chi_{\nu\beta})
+\sqrt{2}{\Theta}^{\alpha
\dot\alpha}(\psi_{\dot\alpha}^{\mu}\chi_{\mu\alpha}-\psi_{\mu\alpha}
\widehat\chi_{\dot\alpha}^{\mu}) \nonumber\eea \bea
  W_{FM}^{\Theta}&=&
  2\sqrt{2}g_{AB}[\bar{h}_5 ({\bar{ d}}_A { Q}_{B } +
\bar{ e}_A { L}_{B }) - h_5(\bar{{ u}}_A { Q}_{ B } + \bar{ \nu}_A
{ L}_{B})]
\nonumber\\
& &  -2\sqrt{2} g_{AB}[{\sqrt{2}} \bar{{ L}}_2 {Q}_A { Q}_{B} +
F_4 { L}_A { L}_{ B} + \sqrt{2}
\bar{t}_6 { Q}_A { L}_{ B} \nonumber\\
& &  + 2{\sqrt{2}}{ L}_2 \bar{{ u}}_A \bar{ d}_B +
\sqrt{2}t_6(\bar{{ u}}_A \bar{ e}_B - \bar{ d}_A \bar{ \nu}_B )+
2\bar{F}_4 \bar{ \nu}_A
\bar{ e}_B]\nonumber\\
 & & -2\sqrt{2}g_{AB}[2\bar{C}_3 \bar{ d}_A { Q}_B - 2 C_3 \bar{ u}_A { Q}_B +
\frac{i}{\sqrt{3}}\bar{h}_6( \bar{ d}_A { Q}_B- 3\bar{ e}_A { L}_B )\\
& & - \frac{i}{\sqrt{3}}h_6 (\bar{ u}_A { Q}_B -3 \bar{ \nu}_A {
L}_B)+2
\bar{D}_3 \bar{ e}_A { Q}_B - 2\bar{E}_6 \bar{ \nu}_A { Q}_B\nonumber\\
& & +2 E_6 \bar{ d}_A { L}_B - 2D_3 \bar{ u}_A { L}_B ]
-2i{\sqrt{2}} g_{AB}[\epsilon \bar{J}_5 \bar{ d}_A \bar{ d}_B
\nonumber\\ && + 2 K_2 \bar{ d}_A \bar{ e}_B - \epsilon \bar{K}_2
\bar{ u}_A \bar{ u}_B - 2
J_5 \bar{ u}_A \bar{ \nu}_B \nonumber\\
& & -{\sqrt{2}} \epsilon  \bar{t}_7 \bar{ d}_A \bar{ u}_B
-{\sqrt{2}} t_7(\bar{ d}_A \bar{ \nu}_B - \bar{ e}_A \bar{ u}_B)]
- 2 g_{AB}[\epsilon P_2 { Q}_A { Q}_B+ 2 \bar{P}_2 { Q}_A { L}_B
]\nonumber
 \eea
We have suppressed SM  indices and used a sub-multiplet naming
convention specified in Appendix \textbf{B}. Conversion to fields
of unit norm in the terms containing color sextets ($L_2,\bar
L_2$) is explained in the caption to the table   of unmixed masses
in Appendix \textbf{B}.

In order that   exchange of a Higgs  that couples to matter with a
given $B+L$ lead to a $B+L$ violating $d=5$ operator in the
effective theory at sub GUT  energies it is necessary that it have
a non-zero contraction with  a conjugate (MSSM) representation
Higgs  that couples to a matter chiral bilinear with a  $B+L$
different from the conjugate of the first $B+L$ value. On
inspection   one finds that not only the  familiar triplet types
 $[\bar 3,1,\pm {\frac{2}{3}}]\subset {\bf{120}}$  i.e  $\{ \bar
t_{(6)},\bar t_{(7)}\}[\bar 3,1,{\frac{2}{3}}]$ and
$\{t_{(6)},t_{(7)}\} $ but
 also   novel exchange modes from the
$P[3,3,\pm{\frac{2}{3}}]$ and $K[3,1,\pm{\frac{8}{3}}]$ multiplet
types can contribute to baryon violation. In the case of the
${\bf{\oot}}$  the $\bar P_1, K_1 \subset {\bf{\oot}}$ multiplets
did couple   to the fermions but $P_1,{\bar K}_1 \subset
{\bf{126}}$ did not. The ${\bf{120}}$ however contains both
$P_2,\bar P_2$ and $K_2,\bar K_2$. Since these mix with $P_1,\bar
P_1$ and $K_1,\bar K_1$,  a number of fresh contributions appear.

   Note in particular that these novel exchanges always lead to
contributions in which at least
 one and possibly both pairs of final state family indices
are antisymmetrized. On integrating out the heavy triplet Higgs
supermultiplets one obtains
 the following  effective
$d=4$ superpotential for baryon Number violating processes  in the
NMSGUT  to leading order in $m_W/M_X$ : \bea  W_{eff}^{\Delta
B\neq  0} = -{ L}_{ABCD} ({1\over 2}\epsilon { Q}_A { Q}_B { Q}_C
{ L}_D) -{ R}_{ABCD} (\epsilon {\bar{ e}}_A {\bar{ u}}_B { \bar{
u}}_C {\bar{ d}}_D) \eea where the coefficients are

\bea  L_{ABCD} &=& {\cal S}_1^{~1} {\tilde h}_{AB} {\tilde h}_{CD}
+ {\cal S}_1^{~2} {\tilde h}_{AB} {\tilde f}_{CD} +
 {\cal S}_2^{~1}  {\tilde f}_{AB} {\tilde h}_{CD} + {\cal S}_2^{~2}  {\tilde f}_{AB} {\tilde
 f}_{CD}\nnu
&-&  {\cal S}_1^{~6}  {\tilde h}_{AB} {\tilde g}_{CD} -
 {\cal S}_2^{~6}  {\tilde f}_{AB} {\tilde g}_{CD}
 +  \sqrt{2}({\cal P}^{-1})_2^{~1} {\tilde g}_{AC}{\tilde f}_{BD}\nonumber\\
 &-&   ({\cal P}^{-1})_2^{~2} {\tilde g}_{AC}{\tilde g}_{BD}
 \eea
and \bea  R_{ABCD} &=&{\cal S}_1^{~1} {\tilde h}_{AB} {\tilde
h}_{CD}
 - {\cal S}_1^{~2}  {\tilde h}_{AB} {\tilde f}_{CD} -
 {\cal S}_2^{~1}  {\tilde f}_{AB} {\tilde h}_{CD} + {\cal S}_2^{~2}  {\tilde f}_{AB} {\tilde f}_{CD} \nonumber \\
 &-& i{\sqrt 2} {\cal S}_4 ^{~1} {\tilde f}_{AB} {\tilde h}_{CD}
+i {\sqrt 2} {\cal S}_4 ^{~2} {\tilde f}_{AB} {\tilde f}_{CD}
\nnu& +& {\cal S}_6 ^{~1} {\tilde g}_{AB} {\tilde h}_{CD} - i
{\cal S}_7 ^{~1} {\tilde g}_{AB} {\tilde h}_{CD} -  {\cal S}_6
^{~2} {\tilde g}_{AB} {  {\tilde f}}_{CD}+ i   {\cal S}_7^{~2}
{\tilde g}_{AB} { {\tilde f}}_{CD}
\nonumber \\
&+&   i{\cal S}_1 ^{~7} {\tilde h}_{AB} {\tilde g}_{CD} -i  {\cal
S}_2 ^{~7} {{\tilde  f}}_{AB} {\tilde g}_{CD}+ \sqrt{2} {\cal S}_4
^{~7} {{\tilde  f}}_{AB} {\tilde g}_{CD}\nonumber
\\&+&  i {\cal S}_6 ^{~7} {\tilde g}_{AB} {\tilde g}_{CD}  +{\cal S}_7
^{~7} {\tilde g}_{AB} {\tilde g}_{CD}- \sqrt{2} ({\cal
K}^{-1})_1^{~2}
{{\tilde  f}}_{AD}{\tilde g}_{BC}\nonumber\\
&-&  ({\cal K}^{-1})_2^{~2} {\tilde g}_{AD}{\tilde g}_{BC}
 \eea

here ${\cal S}= {\cal T}^{-1} $ and ${\cal T} $ is the mass matrix
for $[3,1,\pm 2/3]$-sector  triplets :
 $W={\bar t}^i {\cal T}_i^j t_j
+...$, while
 \bea {\tilde h}_{AB} = 2 {\sqrt 2} h_{AB}  \qquad {\tilde f}_{AB} = 4
{\sqrt 2} f_{AB} \qquad  {\tilde g_{AB}} = 4 g_{AB} \eea These
operators are dressed by sparticles
 to yield the $d=6$ effective 4-fermi operators for baryon decay.
 This dressing requires knowledge of the sparticle spectra and mixing
 angles. This information is   determined  via  the invocation of  threshold
 corrections used to fit the down  and strange quark masses which
 assume  adequate(diagonal)sfermion spectra for the purpose.
 However the scalar mixing which is so crucial to the baryon decay
 rate is assumed minimal i.e to be determined simply by evolution of the
 GUT scale (super)CKM mixing.  The rates for B violation via the
 dominant $d=5$ operators are evaluated  using the above formulae
 and the usual dressing by Gaugino/Higgsino exchange.

\section{Method for Numerical searches }
 We now describe our numerical procedures and application of Susy threshold corrections.
  We use the 2-loop RG equations for the
  MSSM \cite{RGeqnsMSSM1,martinRG} to extrapolate the central values of fermion  Yukawa  data at
  $M_Z$ (including neutrino mass splitting and mixing data,
 see  \cite{xzz,strumviss,theta13ref} for recent values), up
to the MSSM one-loop gauge   unification scale
$M_X^{0}=10^{16.33}$ GeV  while ignoring, for the time being,
complications such as right-handed neutrino and supersymmetric
thresholds.  In the case of the neutrino masses(Type I seesaw) we
extrapolate the coefficient\cite{babpanta} $\kappa_{AB}=M_{\nu}/(2
v_u^2)$ of the $SU(2)\times U(1)$ invariant dimension 5
operator\cite{weinberg,akhsenj} that gives rise to neutrino mass
when the electroweak vev ($v_u=174 \sin\beta~ GeV$; for the SM
$\sin\beta $ is 1) is substituted for the two electroweak doublets
in it. We assume that the neutrino mass hierarchy is normal and
that the lightest neutrino species has a very small mass $< 1\,
meV$.

   After two-loop RG evolution\cite{martinRG} up-to $M_X^0$ we re-extract canonical input parameters i.e the
9 charged fermion Yukawas, the 4 CKM parameters, and the 5
neutrino mass data. We use  percentage uncertainties in fermion
 Yukawas   at $M_X^0=10^{16.33} GeV$
 recently re-estimated in\cite{antusch},   which
include the large uncertainties ($ \delta O_i$) due to large
threshold effects on down type quarks and charged leptons at large
$\tan\beta$.  We form $\chi^2=\sum_i((O_i-{\overline{O}}_i)/\delta
O_i)^2$ which compares  the central values ($
{\overline{O}}_i$)with the values of the same parameters($O_i$)
calculated using the formulae described in Section \textbf{5} and
normalized by the errors ($\delta O_i$).   The `down hill simplex'
or ``amoeba'' algorithm of Nelder and Mead\cite{numrecipes} is
then used  to randomly troll the 38 dimensional (real)  hyperspace
of  NMSGUT parameters.

 After diagonalization of the  \textbf{10}-plet Yukawas
using the $U(3)$ generation basis freedom with respect to the
matter \textbf{16}-plets
 and after  using the rephasing  freedom for Higgs fields, the
 parameters consist of (numbers in brackets denote the number of real parameters)
 $\{h_{ii}(3),f_{ij}(12),g_{ij}(6),\hfil\break
 \eta(2),\rho(2),k(2), \gamma(1), \bar\gamma(1), \zeta (2),
  \bar\zeta (2)\}$ which are 33 in number
 and an additional 5  $\{m(1),{\tilde{m_{\Theta}}}(1),\hfil\break
 |\lambda|(1),\xi \Leftarrow x(2)\}$.
    $M_H$ has been fixed by fine-tuning the doublet
masses. The randomly chosen complex parameter $x$,  which is
central to the explicit solution of the GUT symmetry breaking
problem, is used to generate $\xi ={{ \lambda M}\over {\eta m}}$
(using(\ref{cubic})). $\xi$ stands in for the real parameter M and
hence the phase of $\lambda$ is determined to be
$Arg(\xi)+Arg(\eta)$ for consistency. Note that with the rescaled
form ${\tilde m}_\Theta=m_{\Theta} \lambda/m$ chosen to be real
the value of $m_\Theta $ will have the phase of $\lambda^*$ : as
seen in the explicit fits given in Appendix \textbf{D}.
    While searching we imposed penalties to keep all the Yukawas and GUT
superpotential parameters  within the perturbative range and the
threshold corrected unification parameters\cite{ag1,blmdm}
$\Delta_X=\Delta Log_{10}M_X,\Delta_G=\Delta\alpha_G^{-1}(M_X),
\Delta_3=\Delta {\hat\alpha_3}(M_Z)$  within reasonable limits
given in
 eqn(\ref{criteria}). It is worth emphasizing that the single
superheavy mass scale $m$ (the \textbf{210} mass parameter in the
superpotential)  of the NMSGUT, which controls both  baryon
violation and neutrino masses, is determined by the RG analysis
that fixes $\Delta_X$ \cite{blmdm}. Even though the vev  of the
${\bf{\oot}}$  is $O(M_X)$ it's small couplings confer a freedom
to lower right-handed neutrino masses and thus raise the light
neutrino masses to acceptable values.

The programs were written in FORTRAN95 and run on PC's with a
random starting point. Currently the code runs to a few  tens
thousands of lines including borrowed subroutines, mainly from
SPHENO\cite{porod}.  The solutions found on the first run (before
applying the Susy threshold corrections to lower $y_{d,s}^{MSSM}$)
present the following salient features :

\begin{itemize}
  \item  Neutrino masses and large mixing angles were accurately
  fitted.
  \item  The values of the $\mathbf{\oot}$ couplings were so tiny that
  the three  right-handed neutrino masses were much lighter than $M_X$ and
   spanned the range $\sim 10^8 -  10^{13}\, GeV$, just as conjectured by us earlier
  \cite{blmdm,nums,msgreb}. Note that this is   the range
  required by most (Type I) leptogenesis models.

\item  The bulk  of the residual $\chi^2$ typically consisted of
errors in the d,s quark Yukawa couplings.

 \item The right-handed  neutrino
 mass scale was not independently chosen and then checked for
 approximate consistency(as in an early  version of this work\cite{nmsgut2})  but taken to be the exact expression
 dictated by the NMSGUT vevs.

\end{itemize}

Having found solutions acceptable except for the small down and
strange quark Yukawas, we extrapolate the fermion Yukawas
(neglecting generation mixing for simplicity)  and randomly chosen
Supergravity(SUGRY) type supersymmetry breaking parameters with
non-universal Higgs masses(NUHM) at $M_X^{0}$
($M_{gaugino}=m_{1/2},m_{\tilde f}= m_0 ,
A^0,m^2_{H_u},m^2_{H_d}$) back down to $M_Z$ and use the diagonal
Yukawa couplings as targets of a downhill simplex fitting
procedure that incorporates the large $\tan\beta$ Susy threshold
corrections. We use the formulae given in \cite{piercebagger}
 to estimate the leading (gluino and chargino diagram) corrections to the charged
fermion masses but slightly modify them using
\cite{freitas,antusch} to also include Bino corrections. We also
crosschecked that the corrections to the $T_{3L}=-1/2$ fermion
masses obtained therefrom are equal within a few percent to the
large $\tan \beta$ corrections calculated using the formulae
of\cite{freitas} which are in the limit of unbroken electroweak
symmetry. Although the threshold corrections to the up and charm
quark Yukawas(calculated using the formulae of
\cite{piercebagger}) are very small the top quark mass suffers
threshold corrections $\sim 10- 15 \%$  which must be included.
   Once a set of SUGRY-NUHM   parameters (at $M_X^{0}$) which gives corrected
MSSM values reasonably close to the target values is found we
extrapolate the corrected values back up to $M_X^{0}$ and start a
fresh search(starting from the last achieved values of the GUT
scale parameters) for a new fit to the complete set of fermion
Yukawa parameters.  Since the values of the threshold corrected
down and strange quark Yukawas are now much smaller due to the
Susy threshold corrections invoked to fit the small values
achieved in the earlier cycle, much lower values of $\chi^2$ can
now be achieved. However it is still much easier to find fits in
which $\sin^2\theta_{12}$  or $\sin^2\theta_{23} $ is very small
than to find fits with both leptonic mixing angles are small. In
fact the false minima in $\chi_X^2$ due to this type of solution
are a trap the search program  falls into repeatedly. It is
amusing to note that some years ago this could have been heard as
a \emph{vox ex machina} predicting the small mixing angle solution
of the solar neutrino problem ! Iterations of this procedure  can
give fairly accurate fits in which the 18 data values at the high
scale($M_X^{0}$) are fit with a \emph{total} $\chi_X < 10^{-1}$.
Also  at the scale $M_Z$ one may achieve ($\chi_Z< 10^{-1}$).  A
tension between the threshold corrected values of $y_b(M_Z)$ which
are lower than the values required by the fit using NMSGUT
formulae is often noted (see e.g. \cite{king}) when $A_0, \mu$ are
small and   there are thus only small  threshold corrections. This
is even used to argue for \emph{negative} values of $\mu$. However
since we have allowed quite a large little hierarchy and large
$A_0$ values  we can in fact operate with close to central values
($m_b(M_Z)^{SM}=2.9\pm 0.1$ GeV) as also for all the other SM and
neutrino oscillation parameters. In this sense the region of Susy
breaking soft parameters we focus on seems quite distinct and
novel compared to most earlier studies.

\section{Examples of Soft parameters sets and Susy Spectra found}
\subsection{Assumptions and generic features}

Recently preliminary results from the 1 fb${}^{-1}$ data
accumulation at LHC has been announced\cite{mumbai} and allowed a
($\tan\beta$ independent) inference that colored sparticles
masses are higher than about  $ 500 $ GeV and charged exotics like
$\tilde l, \tilde W^\pm$ more massive than about 200 GeV. When
searching for the soft parameters which would cure the too low
d-s-b quark Yukawa limitation\cite{grimus1} of the NMSGUT Yukawa
formulae we imposed the following constraints on the soft
supersymmetry breaking parameters \cite{sparticles} : \bea && |
\tilde{m}_{\bar l,L} /M_1| \geq 0.9
\qquad ;\qquad |\tilde{m}_{\bar q,Q} /M_3| \geq 0.75   \\
&&\mu,|A_0| < 150\,\, TeV \quad;\quad m_{\tilde l,H^{\pm}} > 200\,
GeV \nonumber
\\&& \tilde W : 200 \,\, GeV< M_2(M_Z)< 1000.0 \,\,GeV \nonumber
\\ && \tilde g : 500 \,\, GeV< M_3(M_Z)< 1000.0 \,\,GeV\quad ;\quad m_{\tilde q} >500 \, GeV\nonumber\eea We
also required  the SM Higgs to be heavier than $114 $ GeV and the
Bino lighter than the lightest sfermion.

In   Tables 3-14 in Appendix \textbf{D}  we present the
NMSGUT-SUGRY-NUHM parameters at $M_X^0=10^{16.33}$ GeV of two
fairly precise  fits to the complete fermion data. In Table 3 (
Table 9) we give NMSGUT superpotential parameters, superheavy mass
spectra, threshold corrected gauge-unification parameters, soft
susy breaking seed parameters and the strength of B-violation  at
$M_X^0$ for the first (second) fit. In Table  4 (Table 10) we give
the Target values, uncertainties and achieved values of the yukawa
couplings at $M_X^0$ along with the Pulls  for the first (second)
fit. In Table 5 (Table 11) we give NMSGUT derived fermion masses
at $M_Z$ before and after Susy threshold corrections for the first
(second) fit. In Table 6 (Table 12) we give the values of the soft
supersymmetry breaking parameters at $M_Z$ that arise after two
loop RG evolution(with no mixing) from $M_Z$ for the first
(second) fit. In Tables 7,8 (Tables 13,14) we give the sparticle
spectra calculated with and without the use of generation mixing
in the RG flow  for the first (second) fit.
 We see that the 18 observables are fitted
 rather  well with minimal  tension   visible even in the troublesome $ T_3=-\frac{1}{2}$ Yukawa couplings.
  {\emph{ However this is done only as a proof of principle and
    feasibility and as an example of the typical form of non pathological exemplars
 of our fitting program. It would be highly premature to subscribe
 to any one parameter set as   specifically demanded by NMSGUT
 fits.} }

    In the type of mass spectra we encounter here, the
LSP (stable due to the primary virtue of R-parity preservation
down to low energies \cite{rpar2,rpar3}) is  pure Bino.
Surprisingly it (often) happens that the  smuon emerges lighter
than the Bino and this needs to be avoided by a penalty in the
search to avoid a charged stable LSP. It is interesting that a
tendency to such a pairing of a Bino LSP and light smuon may make
the Bino  an   effective dark matter candidate by providing an
effective co-annihilation channel for Bino Cold Dark Matter. This
tendency is visible in both our example fits(see Tables 7 and 13)
where the (right) smuon is lighter than the selectron and thus
closest to the light Bino LSP.  We shall return to the cosmology
of these models elsewhere. On the other hand, since we are still
to implement loop corrections to scalar masses (which
 may be very significant for light sparticles) it would be wise to hold
enthusiasm for a particular low  tree level slepton mass in check
till it survives loop corrections.

    The conventional wisdom\cite{sparticles} that the sfermions  of the third
 generation will be much lighter than the first two generations
 of squarks due to strong renormalization of the common sfermion mass at
$M_X^0$ by top Yukawa driven effects
 is directly challenged by our spectra which     yield third
 generation sfermions much heavier than the first two generations.
This feature marks out  such sfermion spectra as linked to the
inner logic governing the  fermion mass pattern in the NMSGUT and
we proffer it as a  prediction regarding supersymmetric physics
that may be tested by LHC or its successors. The large value of
the third generation scalar trilinear may also become relevant
phenomenologically since it leads to large L-R mixing in the
sfermion sector for the third generation and thus to enhancement
of various processes  suppressed by the low value of $M^2_{LR}$ in
many scenarios.

The structure of the Susy threshold corrections  and the
electroweak symmetry breaking conditions which determine
$|\mu|,B$ in terms of $m^2_{H,\bar H}, \tan\beta$  are reviewed in
detail in Appendix C. From the discussion there it emerges \emph{
that the NMSGUT with viable Type I seesaw masses is inescapably
linked to third generation squarks more massive than the first two
generation squarks, large negative values of the trilinear scalar
parameters and also a large $ \mu$ parameter.} Although such
distinctive requirements on the soft sector are welcome we should
recognize that they do imply a considerable degree of fine tuning
to assure viable electroweak symmetry breaking :  as will be
visible in the small value of the ratio $M_W/\mu\sim 10^{-3} $.
However   such a fine tuning is generally needed by large
$\tan\beta$ models.   In contrast with the first version of this
paper(where we searched the soft Susy parameter space at $M_S$) we
have now achieved fits to the required low values of $y_{d,s}$
using constrained SUGRY-NUHM Susy breaking where the  5 element
(i.e $ m_{1/2}=M_{gaugino},m_{\tilde f}=m_{16},A_0,M^2_{ H,\bar H
} $) parameter sets are specified at $M_X^0$. These   5 parameters
  are chosen directly at $M_X^0 $ by the random
search while the parameters $\mu(M_X^0),B(M_X^0)$ are fixed by first
imposing the   one-loop corrected EWRSB conditions at
$M_Z$(calculated using a subroutine taken from the program
SPHENO\cite{porod}) to fix $\mu(M_Z),B(M_Z)$ to ensure acceptable
electroweak symmetry breaking with the given  $\tan\beta$ and
vacuum expectation values and then running these back up to $M_X^0$.
Notice that the freedom to do this comes from  the non-dependence
of the RG flows of the rest of the    couplings and masses,   on
the $\mu ,B$ RG flows. This  is a crucial enabling feature of RG
flows in Susy GUT scenarios. The non-universal Higgs mass
parameters used are justifiable even in the Grand Unified context
since the light Higgs doublets arise as a mixture of 6 doublets
from 5 different GUT Higgs multiplets namely the
$\mathbf{10,120,126, \oot ,210}$. The GUT symmetry justifies
assuming distinct soft masses for each representation and the
light doublets are linear combinations of Higgs doublets from 5
sources in the GUT (with coefficients determined by the unitary
matrices that diagonalize the doublet mass matrices). Thus freedom
to choose two non-universal Higgs masses is certainly conceivable.
We also note that the
 large $\mu,A_0$ parameters help avoid problems with
  FCNC and CCB/UFB instability\cite{gutupend}.

\subsection{Dissecting  Specimens }

The basic difficulty in building a grand unified theory of fermion
masses is the achievement of sufficiently large neutrino masses
and  the reconciliation between small CKM mixing angles and large
PMNS ones. So it is interesting  to  dissect the solutions we have
found to see how exactly they circumvent the difficulties
typically encountered\cite{gmblm,blmdm,nums,msgreb}. The roles
played in the effective matter fermion Yukawa couplings
 $Y^{u,d,e}$ and the Weinberg $d=5$ neutrino mass operator coefficients
 $\kappa_{AB}$  by the Yukawa couplings
$h_{AB},g_{AB},f_{AB}$ of the three FM Higgs representations are,
so to speak, the  vital organic functions of the NMSGUT and their
dissection reveals the integral unity and `intelligent design'
within  the fermion mass pattern. By killing the Yukawa couplings
individually or in groups and noting the resultant effect one
 uncovers their roles.

 The  most influential coupling  is  $h_{33}$ which
directly  affects every mass and mixing angle and may be called
the lynch pin of the  Yukawa coupling pattern\cite{nums,msgreb}.
The third generation masses can be fit to within 5\% or better by
this coupling alone : this is the content of `$b-\tau-t$
unification' in this context.   The next most important( and
sometimes largest) coupling is $g_{23}$. It is crucial for the
$\theta_{23}^{(q)}$ mixing and thus also for second generation
masses: specially the muon   mass. The importance of the couplings
$f_{AB}$ arises from their smallness : since they control the mass
and couplings of the superheavy right-handed neutrinos which leave
behind a characteristic signature in the Type I seesaw masses
(where they enter in the denominator). The limit
$f_{AB}\rightarrow 0$ is \emph{singular } i.e. decreasing $f_{AB}$
changes the effective theory (MSSM with massive left-handed
neutrinos) drastically. The couplings
$h_{11},h_{22},g_{12},g_{13}$ are relevant for the masses and
mixing  of the first and second generations.

If we first set  $f_{AB}=0$ then  all neutrino masses and mixings
become  unphysical.  However one finds that the fractional
changes in the charged fermion masses are  typically
 less than one part in $10^4 $. Thus
the $\mathbf{\oot}$ coupling is irrelevant to the charged fermion
sector in our fits. This feature together with   much smaller
values of the parameter $r_D\sim 10^2$ (as compared to $r_D \sim
10^5$ ; see\cite{grimus1,grimus2} for the definition and values of
$r_D$ which controls the strength of the contribution of $g_{AB}$
to the neutrino Yukawa coupling) marks out the solutions we have
found as distinct from the class of generic fits explored in
\cite{grimus1,grimus2}. Moreover our $f_{AB}$ eigenvalues (and
thus the right-handed neutrino masses) are strongly
hierarchical($1 :10^2:10^4$) : so much so that they are effective
in compensating the large hierarchy in the neutrino Dirac Yukawas
($10^{-1}:10:10^{2}$) (which is amplified by the square of the
neutrino Dirac Yukawa in the Type I formula) to produce the rather
weak hierarchy of the light neutrino masses. This is again  very
different from the generic SO(10) solutions of
\cite{grimus1,grimus2} which have a hierarchy of
 at most $1:60:300$  for the right neutrino masses (which are also much
larger). There\cite{grimus1,grimus2} the dilution of the Dirac
mass hierarchy  relies on a strong boost of the coefficient $r_D$
of $g_{AB}$ in the  neutrino Yukawa matrix   to huge values $\sim
10^4-10^5$.   Our  parameters simply never venture towards such
large $r_D$ : they always produce  values of the coefficient $r_D$
below 500. In fact even if one removes constraints that the
parameters remain perturbative it proves extremely difficult to
force   the downhill simplex `amoeba'  to crawl towards the high
$r_D$ region. Of course we have already shown\cite{pinmsgut} that
generic numeric fits carry no direct implication for the NMSGUT
since their parameters are constrained, but still the generic
analysis propagated a scare\cite{grimus1} that the small
$\mathbf{\oot}$ scenario was  basically flawed and then compounded
it by the confusing conclusion that small(but not tiny) $f_{AB}$
solutions were in fact capable of finding accurate fits(by using
values of a Yukawa formula parameter $r_D\sim 10^5$ that are so
large as to make them unachievable in actual  NMSGUT Yukawa
formulae). In any case this history underlines once
again\cite{gmblm,blmdm} the emphatic \emph{non sequituur} with
which explicit UV theories must confront such generic no-go claims
unless they are founded on something more than numeric
projections.

In \cite{nums,msgreb} we argued that the 2-3 sector should be
regarded as the core of the fermion mass hierarchy and that the
complete hierarchy could and should  be understood
semi-analytically by perturbing around the couplings
$h_{33},g_{23}$ i.e by taking these couplings as $O(1)$ and the
others as order $O(\epsilon^n),\epsilon^2\sim \theta^q_{23}$. The
solutions we have found by random search bear this out.
   We could  now pursue  our earlier proposal\cite{nums,msgreb}
   to understand the fermion parameters by an expansion around the
    23 sector  in a  single hierarchy parameter with confidence that the solutions
   exist and have the required form. The phases and coefficient
   values we have found can be checked for their compatibility
   with such an expansion and thus a qualitative understanding of
   the hierarchy arrived at. Of course we must keep in mind that the values of the Yukawa
couplings   we find match the SM precision couplings extrapolated
to $M_X^0$ only after applying the large corrections to $y_{d,s}$ at
$M_S$  due to large $\tan\beta$  that   underpin  of our fits.
Thus the fermion masses explained above are themselves
significantly different from the ones that  we
earlier\cite{nums,msgreb} had only partial success in describing
via the $\epsilon-$expansion. We anticipate that the downwards
revision  of the target $y_d,y_s$ in the fit at $M_X^0$     by a
factor of $\epsilon$    will allow the determination of   fits
with two large angles in the lepton sector, which we could not
achieve earlier. As mentioned that
 the most  frequent ``false minimae'' encountered  have
  almost perfect charged fermion fits but one large  and one small leptonic mixing angle. Analytic
(perturbative) confirmation of this proposal  may be of some value
since it may  yield a more robust understanding of the innards of
the hierarchy.

It is worth mentioning that our perturbative analysis of both the
two  and three generation (real) tree level fermion Yukawa fitting
problem in SO(10) models based upon a \textbf{10-120}  FM Higgs
system led to the consistency constraint that $|{{y_b-y_\tau}
\over{y_s - y_\mu }}|= 1 + O(\epsilon^2) $. It is striking   that
this constraint is obeyed in   the fits without GUT scale
threshold corrections that we have found so far even though the
parameters are no longer real. Thus the NMSGUT seems constrained
to obey the stringent form of $b-\tau$ unification found in
\cite{nums,msgreb} so long as GUT scale threshold corrections are
unimportant. Conversely, as we show in \cite{gutupend}, the
inclusion of GUT threshold corrections to fermion Yukawas loosens
this constraint and permits one to find solutions that are also B
decay realistic. However if one-loop GUT scale threshold
corrections to Yukawas \cite{gutupend} are strong then the tree
level arguments of the previous paragraph become moot.

\subsection{Nucleon decay  rates}

  Viable RG flows in the NMSGUT can raise  $M_X$ by
  to near the Planck scale (as in the second example in Appendix \textbf{D}).
   Thus $d=6$(i.e gauge boson mediated)
   proton decay can  be suppressed by up to 8 orders of magnitude
   relative to the already long life times $\sim 10^{36} \, yrs$
   corresponding to the one-loop unification scale.
   The question of the rate of $d=5$ operator mediated baryon decay
 in the theory at hand is clearly the  critical one. Since the decay
 rates in  these channels will depend  on the soft scalar spectrum and
 mixing they could  provide  a welcome additional filtration of the
 parameter sets that pass the other criteria.

A complete calculation of B-violation  rates in all possible
channels is so  onerous that we would be justified in presenting
it  only after  we have optimized our fits and have otherwise
viable candidates. As mentioned above threshold corrections due to
{\bf{120}}-plet at $M_X$ are important in this
theory\cite{gutupend}. Moreover the searches for fits will
themselves need to be done keeping in mind the need to lower the B
violation rates below the $10^{-28} yr^{-1}$ that are generically
found (see below).   Since $d=5$ operators and among them the
$p\rightarrow M^+\nu$   channels are known to be the dominant
modes of B violation we  give  the rate of these reactions using
the (hard and soft) parameter values found by us. We emphasize
that sfermion mixing  at $ M_Z$ is assumed to be driven entirely
by  that present in the matter Yukawa couplings of the SO(10)
superpotential at $M_X^0$ (after renormalization down to $M_Z$).
Corrections due to renormalization between $M_X$ or $M_{Planck}$
and $M_X^0$ are ignored. The technology for  this calculation is
well known and we have used the calculations of
\cite{lucasraby,gotonihei} as starting points and cross checks for
developing a Mathematica program that calculates the decay rates
given the fundamental GUT and SUGRY-NUHM parameters (i.e. Tables
3-14).

The  (162 complex ) operator coefficients for B-decay ($
L_{ABCD},R_{ABCD}$) need to be renormalized down from the GUT
scale to $M_Z$ together with the other hard and soft parameters
and plugged into the formulae of \cite{gotonihei} after
appropriately removing the assumptions of an SU(5) GUT context and
negligible right-handed mixing. The formulae of\cite{lucasraby}
can be used as a cross check of the calculation. We find the
values given in Table \ref{BDEC}.

%\!\({8.061904848720777`*^28, {3.117060537064273`*^-30}, \
%{0.000025919606681482202`, 0.0885948321566031`,
 %     0.9113792482367155`}, {9.28695586804965`*^-30}, \
%{0.00011508232756201158`, 0.26662185449730974`,
%0.7332630631751282`}}\)
\begin{table}
 $$
 \begin{array}{|c|c|c|c|c|c|}
 \hline
 {\rm Fit }&\tau_p(M^+\nu)  &\Gamma(p\rightarrow \pi^+\nu) & BR( p\rightarrow
 \pi^+\nu_{e,\mu,\tau})&\Gamma(p\rightarrow K^+\nu) & BR( p\rightarrow K^+\nu_{e,\mu,\tau})\\ \hline
  Ex. 1
 &
                   8.1 \times 10^{  28}
 &
                   3.1 \times 10^{ -30}
 &
 \{
                 2.6  \times 10^{  -5}
 ,
                 0.09
 ,
                 0.91
 \}&
                   9.2 \times 10^{ -30}
 &\{
                 1.1   \times 10^{  -4}
 ,
                 0.27
 ,
                 0.73
 \} \\

%\!\({1.68864959872876`*^28, {7.22435556237068`*^-30}, \
%{0.00003037525900685337`, 0.01437132334478627`,
  %    0.9855983013962067`}, {5.199456118346069`*^-29}, \
%{0.000054483732504554036`, 0.010278381358522699`,
%0.9896671349089727`}}\)

 Ex. 2
 &
                   1.7  \times 10^{  28}
 &
                   7.2 \times 10^{ -30}
 &
 \{
                 3.04 \times 10^{  -5}
 ,
                 0.014
 ,
                 0.986
 \}&
                   5.2 \times 10^{ -29}
 &\{
                 5.45\times 10^{  -5}
 ,
                 0.01
 ,
                 0.99
 \} \\
 \hline\end{array}
 $$
\caption{\small{Table of $d=5$ operator mediated proton lifetimes
$\tau_p$(yrs), decay rates
 $ \Gamma ( yr^{-1} )$  and branching ratios in the dominant Meson${}^++\nu$ channels. }} \label{BDEC}\end{table}

  Clearly these proton  decay rates are too large given the
  current limits $\sim 10^{-33} yr^{-1}$\cite{superKBlimits}. Thus
the otherwise promising  fits imply an acute crisis  of  volatile
baryon number.  In the next paper of this series \cite{gutupend}
we   include also GUT scale threshold effects and conduct searches
that are optimized with respect to baryon stability. This permits
us to find much more realistic fits.

\section{Discussion and Outlook}

In this paper, motivated by successful fits of the fermion
data\cite{msgreb,grimus2} which evade the difficulties that forced
an abandonment\cite{gmblm,blmdm}  of the hope\cite{babmoh} that
the \bten,\boot FM Higgs system would be sufficient to describe
the entire fermion mass spectrum,  we specified the ingredients of
a New Minimal Supersymmteric GUT based on the gauge group $SO(10)$
and the  ${\bf{210\oplus 10\oplus 120\oplus 126\oplus {\overline
{126}} }}$ Higgs System. While inheriting the Higgs system
responsible for GUT scale symmetry breaking unchanged from the
MSGUT\cite{aulmoh,ckn,abmsv} but reassigning the roles of the FM
Higgs  the NMSGUT is able to describe   all the fermion data at
$M_X$ successfully provided recourse is had to relevant threshold
corrections at the Susy breaking scale. This alleviates a problem
with fitting down type Yukawa couplings using only the
$\bf{10,120}$ couplings to matter fields (since the \boot
couplings are lowered drastically to make the Type I seesaw
neutrino masses viable and are thus irrelevant to charged fermion
masses ).

Using the techniques we developed for the MSGUT\cite{ag1,ag2} we
computed the superheavy spectrum for the NMSGUT and used it to
evaluate threshold effects in the gauge evolution. Our experience
with fits is  that the Unification scale is generally  raised
above the one-loop values. We have also attempted surveys to find
viable  unification parameters($\Delta_{X,G,3}$) over  the (vast)
parameter space and these also support this intuition
\cite{nmsgut1,precthresh}. This increase can take $M_X$ to values
as large as $M_{Planck}$ while still remaining in the perturbative
domain. Thus gauge mediated baryon decay is unmeasurably small in
this theory. Together with $M_X$ all other masses, in particular
those of the three triplet types that mediate $d=5$ baryon decay,
also rise. Thus, \emph{prima facie}, not only $d=6$ but also $d=5$
proton decay may be controllable. However when we searched for
fits without imposing any constraint on B-violation rates the
rates    are faster than the current limits by about 6 orders of
magnitude. We find\cite{gutupend} that inclusion of GUT scale
threshold effects due to the {\bf{120}}-plet and searches of the
parameter space under a constraint to suppress B-violation is
necessary before acceptable B-violation rates are reached.

It should be kept in mind   that our results carry systematic
errors due to the omission of various effects such as right-handed
neutrino thresholds (important due to their light masses $\sim
10^8- 10^{12}$ GeV),   mass splitting among Susy partners at
$M_Z$,   non-diagonality in the supersymmetric threshold
corrections, loop corrections to scalar masses and neglect of
 renormalization effects from
$M_{Planck}$ (where the soft SUGRY-NUHM parameters should
presumably be specified) to $M_X^0$ which is the   high energy
matching scale      we have chosen for convenience. Accounting for
multiple and disparate  Susy threshold corrections also leads to
errors that are quite complex to account for and can be an
important source of theoretical uncertainty.
 The non-diagonal threshold effects at large $\tan\beta$ can   be
calculated from the same diagrams by inserting the appropriate
flavor structure and diagonalization. All these improvements are
required for later stages of our program. However
 these effects are  higher order
corrections to the large diagonal  modifications to
 $y_{d,s}$  and the cancellation that prevents
the bottom Yukawa from also being strongly renormalized by gluino
exchange. The non-diagonal effects in both the fermionic and
scalar sectors are driven by the small off diagonal Yukawa
couplings at $M_X$ that we determined by fitting the quark and
lepton mixings. Hence     they will  modify the diagonal effects
only marginally. Nevertheless their presence makes it clear that
one needs to be circumspect in claiming too high accuracies.

  An  increase in  $M_X$ could provide  resolution of a
   nagging difficulty\cite{trmin}  in  the
  MSGUT : the Landau pole in the gauge coupling evolution above
  $M_X$. When  $M_X$ is  closer to the Planck scale the
  coincidence  of the SO(10) Landau pole  with the Planck scale
  strengthens our speculation that the UV condensation to be
  expected in such a supersymmetric Asymptotically Strong(AS) theory
  \cite{trmin,tas}  acts as   a physical cutoff for the perturbative
  SO(10) theory and perhaps even as the scale of an  induced gravity that arises
  from this theory(supersymmetry cures the ambiguities that plagued the original
  induced gravity ideas\cite{david}) . We made a beginning in\cite{tas} by
  demonstrating,    using supersymmetric strong coupling heuristics\cite{seiberg},
   that in a toy   AS Susy GUT the condensation actually takes place and breaks the (toy)
  GUT symmetry, and that the vevs responsible are
  \emph{calculable}.  It is encouraging that the development of the
  theory  in regard to  apparently unrelated  features  has naturally brought us
  to the point where a number of  intractable fundamental features
  have  become pliable to a synthetic interpretation.

In this   paper  we have indicated a feasible route to pursue if
the New or Full Minimal Supersymmetric SO(10) GUT is to
 become vulnerable to falsification by the upcoming data from the
LHC and it's successors and the large water Cerenkov detectors now
being planned to further raise the limits on proton decay. This
route evades difficulties and confusion that had waylaid the
fermion fitting program in MSGUTs in the recent past. We have
shown that the way out is to accept the difficulties as indicative
of a particular type of sfermion spectra which will modify the
$d,s$ quark  Yukawa couplings significantly at the threshold
$M_S$. To our knowledge this is a novel proposal that ties the
fermion spectrum achievable by a Susy GUT tightly to its soft
breaking terms. Such a program is
 feasible and plausible precisely because of the constraints
 under which the NMSGUT functions and  the lack of freedom to add
 additional fermion mass operators   so abundantly available
 in SO(10) Susy theories based on low dimensional Higgs fields and
 non-renormalizable operators.

 If the colliders do not belie
their promises they may sketch the supersymmetric spectra for us.
The detection of even a single supersymmetric particle(likely a
chargino or a smuon in terms of the typical SUGRY-NUHM NMSGUT
spectra we find) will  anchor the Susy mass patterns emerging from
the RG analysis of the NMSGUT. With these spectra in hand it will
  have to face both edges of the scissor we have shown operative:
failure at $M_X$ to fit fermion spectra unless the sfermion  and
gaugino spectrum is such as to push the down quark type  fermion
Yukawa couplings above $M_S$ to values where they match with the
otherwise accurate and consistent fits we find at $M_X$ and
failure to fit the neutrino masses if the ${\bf{\oot}}$ couplings
are used to correct the second generation mass patterns.

In sum, the NMSGUT having inherited  the strengths of its parent
is revealing new virtues as well as new weaknesses and, while
threatening still to plunge into the yawning crevasse of
falsification, yet promises to carry the long winding  caravan of
Grand Unification not only across the Grand Desert that set its
first horizons but across threshold jungles beyond that first
horizon up into the rarefied heights where gauge forces and
gravity meld into their primordial pleromal\cite{tas} unity.

 \vspace{ .5 true cm}

\section*{Acknowledgments}
 \vspace{ .5 true cm}

The work of C.S.A. was supported by   grant No. SR/S2/HEP-11/2005
from the Department of Science and Technology of the Government of
India and that of S.K.G. by a University Grants Commission Senior
Research Fellowship. It is a pleasure for C.S.A.  to acknowledge
the hospitality of the High Energy Theory Group ICTP,Trieste and
in particular Goran  Senjanovic. C.S.A. thanks Goran  Senjanovic
for frank conversations on where MSGUTs stand today  as well as
Borut Bajc and Alejandra Melfo for discussions on supersymmetry
breaking.We thank Sunil Mittal for writing the first version of
the 2-loop renormalization group flow Mathematica  program that we
use as a cross check. C.S.A. acknowledges useful correspondence
with S. Antusch and W.A. Porod and informative discussions with K.
Babu and Ts. Enkhbat.   C.S.A. also thanks Ila Garg and
 Charanjit Kaur for help with the preparation of the
 final manuscript and especially the tables.
\vspace{ .5 true cm}

  \section*{Note added in proof}
 \vspace{ .3 true cm}
   On 13 December 2011, just before we
received the  proofs, the ATLAS-CMS 5$fb^{-1}$ Higgs mass results
indicating $M_{h^0}\sim 125 ~GeV$ were announced. It is amusing to
see that our second example fit coincides with this result. More
seriously such Higgs mass values indicate heavy sparticle spectra
and large $A_0$ values : as many of  the plethora of
``instantaneous papers'' released on December 14 and 15 have
noted.   Our spectra  are strongly of the large  $A_0$ type  but
stand out as falsifiable and distinct due to the normal sparticle
hierarchy with ultra heavy stops and various other distinctive
features  like light smuons. They are also distinct in that they
necessarily embrace $A_0,\mu$ values so large as to render the
MSSM vacuum  stable against decay to the CCB vacua that large
$A_0,\mu$ parameters imply [71]. We emphasize that these features
of the sparticle spectra are relatively independent of the details
by which the GUT achieves fits of the fermion spectra at $M_X^0$.
They are required by the need to lower $y_{d,s}$ while preserving
$y_b$.
% \section*{Appendix A : MSGUT couplings and vevs.}

\appendix
  \section{}
  \subsection {  MSGUT couplings and vevs.}
\Alph{section} \Alph{equation}
   We provide some further details concerning the
MSGUT in this appendix for the readers reference.  The
superpotential contains the mass parameters
 \bea
 m: {\bf{210}}^{\bf{2}} \qquad ; \qquad M : {\bf{126\cdot{\overline {126}}}}
 ;\qquad M_H : {\bf{10}}^{\bf{2}}
\eea and trilinear couplings
  \bea
 \lambda : {\bf{210}}^{\bf{3}} \qquad ; \qquad  \eta  :
 {\bf{210\cdot 126\cdot{\overline {126}}}}
 ;\qquad  \gamma \oplus {\bar\gamma}  : {\bf{10 \cdot 210}\cdot(126 \oplus
{\overline {126}}})
  \eea

In addition   one has two complex symmetric matrices
$h_{AB},f_{AB}$ of Yukawa couplings of the  $\mathbf{10,\oot}$
Higgs multiplets to the $\mathbf{16 .16} $ matter bilinears. The
$U(3)$ ambiguity due to $SO(10)$ `flavor' redefinitions can be
used to remove 9 of the 24 real parameters in $f,h$. In addition
rephasing of the remaining 4 fields
$\mathbf{\Phi,H,\Sigma,\overline\Sigma}$ removes 4 phases from the
14 parameters in $m,M,M_H,\lambda,\eta,\gamma,{\bar\gamma}$
leaving   25 superpotential parameters to begin with. Strictly
speaking, since a fine tuning to keep one pair of doublets light
is an intrinsic part of the MSGUT scenario, an additional
\emph{complex} parameter (say $M_H$ ) may be considered as fixed
so that there are actually 23 free superpotential parameters. In
addition the electroweak scale vev $v_W$, and $\tan \beta$ are
relevant external parameters for the light  fermion spectrum
determined by the GUT Yukawa structures. The overall superheavy
scale(identified with the real parameter $m_{210}$) is fixed by the
identification of the unification scale determined by the RG flow
with the mass of the gauge $X[3,2,\pm {4\over 3}]$ sub-multiplet.

The GUT scale vevs that break the gauge symmetry down to the SM
symmetry (see \cite{ag1} for complete details)
are{\cite{aulmoh,ckn}} \bea
 {\langle(15,1,1)\rangle}_{210}& :&
\langle{\phi_{abcd}}\rangle={a\over{2}}
\epsilon_{abcdef}\epsilon_{ef}\\
\hfil\break
\langle(15,1,3)\rangle_{210}~&:&~\langle\phi_{ab\ta\tb}\rangle=\omega
\epsilon_{ab}\epsilon_{\ta\tb}\quad\\
\quad\langle(1,1,1)\rangle_{210}~&:& ~\langle\phi_{ {\tilde
\alpha}{\tilde \beta} {\tilde \gamma}{\tilde \delta}}
\rangle=p\epsilon_{{\tilde \alpha} {\tilde \beta} {\tilde
\gamma}{\tilde \delta}}\quad\\
 \hfil\break
\langle(10,1,3)\rangle_{\oot} ~&:&
  \langle{\overline\Sigma}_{\hat{1}\hat{3}\hat{5}
\hat{8}\hat{0}}\rangle= \bar\sigma \quad\\
\langle({\overline{10}},1,3)\rangle_{126} ~&:&
\langle{\Sigma}_{\hat{2}\hat{4}\hat{6}\hat{7}\hat{9}}
\rangle=\sigma. \eea In the above $a,b..=1..6$ are SO(6) indices
and $\tilde{\alpha},\tilde{\beta}..=7..10$ are SO(4) indices and
hats indicate their pairwise complexification.

 The vanishing of the D-terms of the SO(10) gauge sector
 potential imposes only the condition $
 |\sigma|=|{\overline{\sigma}}| $.
The relevant system of (electrically neutral) superpotential
extremization equations $F_{a,p,\omega,\sigma,\bar\sigma}=0$   can
be reduced to a single cubic equation \cite{abmsv}
 for a variable $x= -\lambda\omega/m$, in terms of
 which the vevs $a,\omega,p,\sigma,
 {\overline\sigma}$ are specified  :
\be C(x,\xi)=8 x^3 - 15 x^2 + 14 x -3 +\xi (1-x)^2 =0\label{cubic}
\ee where $\xi ={{ \lambda M}\over {\eta m}} $. Then the
dimensionless vevs in units of (m/$\lambda$) are $\omt=-x$
\cite{abmsv} and \be \at={{ (x^2 +2 x -1)}\over (1-x)}\quad ;\quad
 \pt={{x(5 x^2-1)}\over {(1-x)^2}}\quad ; \quad
\sgt\sgbt={2\over \eta}{{\lambda x(1-3x)(1+x^2)}\over {(1-x)^2}}
\label{dlvevs}\ee
 This   exhibits the crucial importance of the
parameters $\xi,x$. Note that one can trade\cite{abmsv,bmsv,bmsv2}
the parameter $\xi$ for $x$ with advantage using equation
(\ref{cubic}) since $\xi$ is  uniquely fixed by $x$.  By a survey
of  the behavior of the theory as a function of the complex
parameter $x$ we  thus cover the behavior of the three different
solutions possible for each complex value of $\xi$.

Among the mass matrices   is the all important Higgs doublet mass
matrix \cite{ag1,ag2}  ${\cal H}$ which  can be diagonalized by a
bi-unitary transformation\cite{abmsv,bmsv,ag2}: from the  pairs of
Higgs doublets $h^{(i)},{\bar h}^{(i)}$ arising from the SO(10)
fields to a new set $H^{(i)},{\bar H}^{(i)}$ of fields in terms of
which the doublet mass terms  are diagonal. \bea {\overline U}^T
{\cal H}U &=&   Diag ( m_H^{(1)},m_H^{(2)},....)
 \nnu
 h^{(i)} &=& U_{ij} H^{(j)}  \qquad ;\qquad   {\bar h}^{(i)} = {\bar
U}_{ij} {\bar H}^{(j)}  \eea

 To keep one pair
of these doublets light one  tunes $M_H$ so that $Det{\cal H}=0$.
  In the effective theory at low energies the
GUT Higgs doublets  $h^{(i)},{\bar h}^{(i)}$  are present in the
massless doublets   $H^{(1)},{\bar H}^{(1)}$ in a proportion
determined by the  first columns of the matrices  $U,{\bar U}$ :
    \bea E < <M_X \qquad : \qquad  {  h}^{(i)} &\rightarrow&
      {  \alpha}_i {  H}^{(1)} \quad
  ; \quad
{ \alpha}_i = {  U}_{i1} \nnu
   {\bar h}^{(i)} &\rightarrow&  {\bar \alpha}_i {\bar H}^{(1)} \quad
  ; \quad
{\bar\alpha}_i = {\bar U}_{i1}
  \eea
The all important normalized 6-tuples $\alpha,\bar\alpha$ can be
easily determined\cite{abmsv,bmsv,gmblm,bmsv2,blmdm}  by solving
the zero mode conditions: ${\cal {H}} \alpha =0 ~ ;~  \bar\alpha^T
{\cal{H} }=0 $. Explicit formulae may be found in \cite{nmsgut1}.
We use a convention where the first component i.e
$\alpha_1,{\bar{\alpha}}_1$ is real.

\subsection{120-plet superpotential}

The Pati-Salam decomposition of the rest of the \textbf{120}-plet
terms terms in the superpotential is :  \bea
\frac{m_{\Theta}}{2(3!)}{\Theta}_{ijk}{\Theta}_{ijk} &=&
\frac{m_{\Theta}}{12}[6
 {\Theta}_{\mu\nu}^{(s)} {{\Theta}}^{\mu\nu}_{(s)}+ 6
{{\Theta}}_{\sigma}^{~\lambda\alpha\dot{\alpha}}{\Theta}_{\lambda\alpha\dot{\alpha}}^{~\sigma}\\
 & & + 3(\vec{\tilde{{\Theta}}}_{(a)(R)}^{\mu\nu}\cdot\vec{{\Theta}}_{\mu\nu(R)}^{(a)}+
 \vec{\tilde{{\Theta}}}_{(a)(L)}^{\mu\nu}\cdot\vec{{\Theta}}_{\mu\nu(L)}^{(a)})\\
 & & -6
 {\Theta}^{\alpha\dot{\alpha}}{\Theta}_{\alpha\dot{\alpha}}]\\
\nonumber \\
\nonumber \\
\nonumber \\
% \eea
%\bea
\frac{\rho}{4!} {\Theta}_{ijm}{\Theta}_{klm}\Phi_{ijkl}  &=&
\frac{\rho}{4!}[8i
 {\Theta}_{\mu\lambda}^{(s)}{{\Theta}}^{\nu\lambda}_{(s)}{\Phi}_{\nu}^{~\mu}+8i
 {{\Theta}}_{\sigma}^{~\mu\alpha\dot{\alpha}}{{\Theta}}_{\nu\alpha\dot{\alpha}}^
 {~\sigma}{\Phi}_{\mu}^{~\nu}\\
& & -
 4\sqrt{2}({\Theta}_{\mu\lambda}^{(s)}\vec{\tilde{{\Theta}}}_{(a)(R)}^{\mu\nu}\cdot\vec{{\Phi}}_{\nu(R)}^
 {~\lambda}+ {\Theta}_{\mu\lambda}^{(s)}\vec{\tilde{{\Theta}}}_{(a)(L)}^{\mu\nu}\cdot\vec{{\Phi}}_{\nu(L)}^
 {~\lambda}\\
 & & -
 {{\Theta}}^{\mu\lambda}_{(s)}\vec{{\Theta}}_{\mu\nu(R)}^{(a)}\cdot\vec{{\Phi}}_{\lambda(R)}^
 {~\nu}- {{\Theta}}^{\mu\lambda}_{(s)} \vec{{\Theta}}_{\mu\nu(L)}^{(a)}\cdot\vec{{\Phi}}_{\lambda(L)}
 ^{~\nu})\\
 & & +4\sqrt{2}({\Theta}^{\alpha\dot{\alpha}}{{\Theta}}_{\nu\alpha}^{~\mu\dot{\beta}}
 {\Phi}_{\mu\dot{\alpha}\dot{\beta}}^{~\nu}-{\Theta}^{\alpha\dot{\alpha}}
{{\Theta}}_{\nu\dot{\alpha}}^{~\mu\beta}{\Phi}_{\mu\alpha\beta}^{~\nu})\\
 & & +8({{\Theta}}_{\nu}^{~\mu\alpha\dot{\alpha}}{\Theta}_{\mu\lambda}^{(s)}{\Phi}^
 {\nu\lambda(s)}_{~~\alpha\dot{\alpha}}-
 {{\Theta}}_{\nu}^{~\mu\alpha\dot{\alpha}}{{\Theta}}^{\nu\lambda}_{(s)}\Phi_{\mu\lambda
 \alpha\dot{\alpha}}^{(s)})\\
 & & - 2(\tilde{{\Theta}}^{\mu\nu}_{(a)\dot{\alpha}\dot{\beta}}{{\Theta}}_{\nu\alpha}^
 {~\lambda\dot{\beta}}\Phi_{\mu\lambda(s)}^{~~~\alpha\dot{\alpha}}+
 \tilde{{\Theta}}_{(a)\alpha\beta}^{\mu\nu}{{\Theta}}_{\nu\dot{\alpha}}^{~\lambda\beta}
 \Phi_{\mu\lambda(s)}^{~~\alpha\dot{\alpha}}\\
 & & -{\Theta}_{\mu\nu\dot{\alpha}\dot{\beta}}^{(a)}{{\Theta}}_{\lambda\alpha}^{~\nu\dot{\beta}}
 {\Phi}^{\mu\lambda\alpha\dot{\alpha}}_{(s)}-{\Theta}_{\mu\nu\alpha\beta}^{(a)}
 {{\Theta}}_{\lambda\dot{\alpha}}^{~\nu\beta} {\Phi}^{\mu\lambda\alpha
 \dot{\alpha}}_{(s)})\\
 & & - 2 (\vec{\tilde{{\Theta}}}_{(a)(L)}^{\mu\nu}\cdot\vec{{\Theta}}_{\mu\nu(L)}^{(a)} -
 \vec{\tilde{{\Theta}}}_{(a)(R)}^{\mu\nu}\cdot\vec{{\Theta}}_{\mu\nu(R)}^{(a)})\Phi\\
 & & -4\sqrt{2}({{\Theta}}_{\lambda}^{~\mu\alpha\dot{\alpha}}{\Theta}_{(a)\dot{\alpha}
 \dot{\beta}}^{\nu\lambda}\Phi_{\mu\nu\alpha}^{(a)\dot{\beta}}-{{\Theta}}_{\lambda}^
 {~\mu\alpha\dot{\alpha}}{\Theta}_{(a)\alpha\beta}^{\nu\lambda}\Phi_{\mu\nu\dot{\alpha}}^
 {(a)\beta})\\
  & & - 2\sqrt{2}({\Theta}^{\alpha\dot{\alpha}}\tilde{{\Theta}}_{~(a)\dot{\alpha}\dot{\beta}}^
  {\mu\nu}\Phi_{\mu\nu\alpha}^{(a)\dot{\beta}}+ {\Theta}^{\alpha\dot{\alpha}}\tilde{{\Theta}}^
  {\mu\nu}_{(a)\alpha\beta}\Phi_{\mu\nu\dot{\alpha}}^{(a)\beta})\\
  & & +4\sqrt{2}({\Theta}_{\lambda}^{~\nu\alpha \dot{\alpha}}{\Theta}_{\mu\alpha}^{~\lambda\dot{\beta}}
  \Phi_{\nu\dot{\alpha}\dot{\beta}}^{~\mu}+{\Theta}_{\lambda}^{~\nu\alpha \dot{\alpha}}
  {\Theta}_{\mu\dot{\alpha}}^{~\lambda\beta}\Phi_{\nu\alpha\beta}^{~\mu})\\
  & & -2({{\Theta}}_{\nu\alpha}^{~\lambda\dot{\beta}}\tilde{{\Theta}}_{~~\dot{\alpha}
  \dot{\beta}}^{\mu\nu(a)}\Phi_{\mu\lambda(s)}^{~~\alpha\dot{\alpha}}+
 {{\Theta}}_{\nu\dot{\alpha}}^{~\lambda\beta}\tilde{{\Theta}}_{~~\alpha\beta}^{\mu\nu(a)}
  \Phi_{\mu\lambda(s)}^{~~\alpha\dot{\alpha}} \\
  &- & {{\Theta}}_{\lambda\dot{\alpha}}^{~\nu\beta}{\Theta}_{\mu\nu\alpha\beta}^{(a)}
  {\Phi}^{\mu\lambda\alpha\dot{\alpha}}_{(s)}-{{\Theta}}_{\lambda\alpha}^{~\nu\dot{\beta}}
  {\Theta}_{\mu\nu\dot{\alpha}\dot{\beta}}^{(a)}\Phi^{\mu\lambda\alpha\dot{\alpha}}_{(s)})\\
  & & -4\sqrt{2}(\vec{{\Theta}}_{\mu\nu (R)}^{(a)}\cdot(\vec{{\Theta}}_{(a)R}^{\nu\lambda}\times
  \vec{\Phi}_{\lambda (R)}^{~\mu})\\
  & &+ \vec{{\Theta}}_{\mu\nu (L)}^{(a)}\cdot(\vec{{\Theta}}_{(a)L}^
  {\nu\lambda}
  \times \vec{\Phi}_{\lambda (L)}^{~\mu}))]\eea
  \newpage
  \bea
\frac{\zeta}{2(3!)} {\Theta}_{ijk}\Sigma_{ijlmn} \Phi_{klmn}
 & =& \frac{\zeta}{2(3!)}[-6 \sqrt{2}i
 ({\Theta}_{\mu\lambda}^{(s)}\tilde{\Sigma}^{\mu\nu}_{(a)}{\Phi}_{\nu}^{~\lambda}-{{\Theta}}^
 {\mu\lambda}_{(s)}\Sigma_{\mu\nu}^{(a)}{\Phi}_{\lambda}^{~\nu})\\
   & &
   -12i({\Theta}_{\mu\lambda}^{(s)}{\Sigma}_{\nu}^{~\mu\alpha\dot{\alpha}}
   {\Phi}_{~~\alpha\dot{\alpha}}^{\nu\lambda(s)}+{{\Theta}}^{\nu\lambda}_{(s)}
 {\Sigma}_{\nu}^{~\mu\alpha\dot{\alpha}}\Phi_{\mu\lambda\alpha\dot{\alpha}}^{(s)})\\
 & & -
 12\sqrt{2}({\Theta}_{\mu\lambda}^{(s)}\vec{\Sigma}_{(s)(R)}^{\nu\lambda}\cdot
 \vec{{\Phi}}_{\nu(R)}^{~\mu}
-{{\Theta}}^{\nu\lambda}_{(s)}\vec{\Sigma}_{\mu\lambda(L)}^{(s)}\cdot
\vec{{\Phi}}_{\nu(L)}^{~\mu})\\
 & & + 6\sqrt{2}i({{\Theta}}^{\mu\lambda}_{(s)}{\Sigma}_{\lambda\alpha\dot{\alpha}}^
 {~\nu}\Phi_{\mu\nu(a)}^{~~\alpha\dot{\alpha}}- {\Theta}_{\mu\lambda}^{(s)}{\Sigma}_
 {\nu\alpha\dot{\alpha}}^{~\lambda}\tilde{\Phi}^{\mu\nu\alpha\dot{\alpha}}_{(a)})\\
 & & - 6\sqrt{2}({{\Theta}}_{\nu}^{~\lambda\alpha\dot{\alpha}}\tilde{\Sigma}^
 {\mu\nu}_{(a)}\Phi_{\mu\lambda\alpha\dot{\alpha}}^{(s)}+ {{\Theta}}_{\lambda}^
 {~\nu\alpha\dot{\alpha}} \Sigma_{\mu\nu}^{(a)}{\Phi}_{\alpha\dot{\alpha}}^
 {\mu\lambda(s)})\\
 & & -12{\sqrt{2}}i( {{\Theta}}_{\lambda}^{~\mu\alpha\dot{\alpha}}{\Sigma}_{\nu\dot{\alpha}}^
 {~\lambda\beta}{\Phi}_{\mu\alpha\beta}^{~\nu}+
 {{\Theta}}_{\lambda}^{~\mu\alpha\dot{\alpha}}{\Sigma}_{\mu\alpha}^
 {~\nu\dot{\beta}}{\Phi}_{\nu\dot{\alpha}\dot{\beta}}^{~\lambda})\\
  & & +6\sqrt{2}({{\Theta}}_{\nu\dot{\alpha}}^{~\lambda\beta}\Sigma_{\mu\lambda\alpha\beta}^
 {(s)}\tilde{\Phi}^{\mu\nu\alpha\dot{\alpha}}_{(a)} +{{\Theta}}_{\lambda\alpha}^
 {~\nu\dot{\beta}}\Sigma_{~~\dot{\alpha}\dot{\beta}}^{\mu\lambda(s)}\Phi_{\mu\nu(a)}^
 {~~\alpha\dot{\alpha}})\\
  & &
  +12i{{\Theta}}_{\nu}^{~\mu\alpha\dot{\alpha}}{\Sigma}_{\mu\alpha\dot{\alpha}}^
  {~\nu}\Phi\\
  & & +12i
  (\vec{\tilde{{\Theta}}}_{(a)(L)}^{\mu\nu}\cdot\vec{\Sigma}_{\mu\lambda(L)}^{(s)}{\Phi}_{\nu}^
  {~\lambda}
   +
 \vec{{\Theta}}_{\mu\nu(R)}^{(a)}\cdot\vec{\Sigma}^{\mu\lambda}_{(s)(R)}{\Phi}_{\lambda}^{~\nu}
  )\\
 & & -6i(\tilde{{\Theta}}^{\mu\nu}_{(a)\dot{\alpha}\dot{\beta}}{\Sigma}_{\nu\alpha}^
 {~\lambda\dot{\beta}}\Phi_{\mu\lambda(s)}^{~~\alpha\dot{\alpha}}- \tilde{{\Theta}}^
 {\mu\nu}_{(a)\alpha\beta}{\Sigma}_{\nu\dot{\alpha}}^{~\lambda\beta}\Phi_{\mu\lambda(s)}
 ^{~~\alpha\dot{\alpha}}\\
 & &
 -{\Theta}_{\mu\nu\dot{\alpha}\dot{\beta}}^{(a)}{\Sigma}^{~\nu\dot{\beta}}_{\lambda\alpha}
 {\Phi}^{\mu\lambda\alpha\dot{\alpha}}_{(s)} + {\Theta}_{\mu\nu\alpha\beta}^{(a)}{\Sigma}^
 {~\nu\beta}_{\lambda\dot{\alpha}}{\Phi}^{\mu\lambda\alpha\dot{\alpha}}_{(s)})\\
 & &
-12(\vec{{\Theta}}_{\mu\nu(L)}^{(a)}\vec{{\Phi}}_{\lambda(L)}^{~\mu}-
\vec{{\Theta}}_{\mu\nu(R)}^
{(a)}\vec{{\Phi}}_{\lambda(R)}^{~\mu})\Sigma^{\nu\lambda}_{(a)}\\
 & & +6({\Theta}_{\dot{\alpha}}^{\beta}\Sigma_{\mu\nu\alpha\beta}^{(s)}{\Phi}^
 {\mu\nu\alpha\dot{\alpha}}_{(s)}
  - {\Theta}_{\alpha}^{\dot{\beta}}\Sigma_{~~\dot{\alpha}\dot{\beta}}^{\mu\nu(s)}
  \Phi^{~~\alpha\dot{\alpha}}_{\mu\nu(s)})\\
  & &
  -6\sqrt{2}i({\Theta}_{\alpha}^{\dot{\beta}}{\Sigma}_{\nu}^{~\mu\alpha\dot{\alpha}}
  {\Phi}^{~\nu}_{\mu\dot{\alpha}\dot{\beta}}+{\Theta}_{\dot{\alpha}}^{\beta}{\Sigma}_
  {\nu}^{~\mu\alpha\dot{\alpha}}{\Phi}^{~\nu}_{\mu\alpha\beta})\\
& & +12
({{\Theta}}_{\nu}^{~\mu\alpha\dot{\alpha}}{\Sigma}_{\lambda\alpha\dot{\alpha}}^
{~\nu}{\Phi}_{\mu}^{~\lambda}-{{\Theta}}_{\nu}^{~\mu\alpha\dot{\alpha}}{\Sigma}_
{\mu\alpha\dot{\alpha}}^{~\lambda}{\Phi}_{\lambda}^{~\nu})\\
  & &
  +12({{\Theta}}_{\nu}^{~\mu\alpha\dot{\alpha}}\Sigma_{\mu\lambda\alpha\beta}^{(s)}
 {\Phi}_{(s)\dot{\alpha}}^{\nu\lambda\beta}
  -{{\Theta}}_{\nu}^{~\mu\alpha\dot{\alpha}}\Sigma^{\nu\lambda}_{(s)\dot{\alpha}
  \dot{\beta}}\Phi^{(s)\dot{\beta}}_{\mu\lambda\alpha})\\
  & & - 12 ({{\Theta}}_{\lambda}^{~\mu\alpha\dot{\alpha}}\Sigma_{\mu\nu}^{(a)}\Phi_
  {(a)\alpha\dot{\alpha}}^{\nu\lambda})\\
  & & +6i (\tilde{{\Theta}}^{\mu\nu}_{(a)\dot{\alpha}\dot{\beta}}{\Sigma}_{\nu}^
  {\lambda\alpha\dot{\alpha}}\Phi_{\mu\lambda\alpha}^{(s)\dot{\beta}}+
  \tilde{{\Theta}}^{\mu\nu}_{(a)\alpha\beta}{\Sigma}_{\nu}^{~\lambda\alpha\dot{\alpha}}
  \Phi_{\mu\lambda\dot{\alpha}}^{(s)\beta}\\
  & & +{\Theta}_{\mu\nu\dot{\alpha}\dot{\beta}}^{(a)}{\Sigma}_{\lambda}^{~\nu\alpha\dot{\alpha}}
  {\Phi}^{\mu\lambda\dot{\beta}}_{(s)\alpha}+{\Theta}_{\mu\nu\alpha\beta}^{(a)}{\Sigma}_
  {\lambda}^{~\nu\alpha\dot{\alpha}}{\Phi}^{\mu\lambda\beta}_{(s)\dot{\alpha}})\\
  & & + 12\sqrt{2}[\vec{\tilde{{\Theta}}}^{\mu\nu}_{(a)(L)}\cdot(\vec{{\Phi}}_{\nu (L)}^
  {~\lambda}\times
  \vec{\Sigma}_{\mu\lambda(L)}^{(s)})\\
 & & - \vec{{\Theta}}_{\mu\nu (R)}^{(a)}\cdot(\vec{{\Phi}}_{\lambda (R)}^{~\nu}\times
\vec{\Sigma}_{(R)}^{\mu\lambda(s)})]\\
  & & +
  6\sqrt{2}i
  ({\Theta}_{\mu\nu\dot{\alpha}\dot{\beta}}^{(a)}{\Sigma}_{\lambda\alpha}^{~\mu\dot{\beta}}
  \Phi^{\nu\lambda\alpha\dot{\alpha}}_{(a)}+{\Theta}_{\mu\nu\alpha\beta}^{(a)}{\Sigma}_
  {\lambda\dot{\alpha}}^{~\mu\beta}\Phi^{\nu\lambda\alpha\dot{\alpha}}_{(a)}
  )]
\eea

\bea
 \frac{\bar{\zeta}}{2(3!)}{\Theta}_{ijk} \bar{\Sigma}_{ijlmn}\Phi_{klmn}
& =& \frac{\bar{\zeta}}{2(3!)}[6  \sqrt{2}i
 ({\Theta}_{\mu\lambda}^{(s)}\tilde{\bar{\Sigma}}^{\mu\nu}_{(a)}{\Phi}_{\nu}^{~\lambda}-
 {{\Theta}}^{\mu\lambda}_{(s)}\bar{\Sigma}_{\mu\nu}^{(a)}{\Phi}_{\lambda}^{~\nu})\\
 & & -12i({\Theta}_{\mu\lambda}^{(s)}{\bar{\Sigma}}_{\nu}^{~\mu\alpha\dot{\alpha}}
{\Phi}_{~~\alpha\dot{\alpha}}^{\nu\lambda(s)}+
 {{\Theta}}^{\nu\lambda}_{(s)}{\bar{\Sigma}}_{\nu}^{~\mu\alpha\dot{\alpha}}
 \Phi_{\mu\lambda\alpha\dot{\alpha}}^{(s)})\\
 & & -
 12\sqrt{2}({\Theta}_{\mu\lambda}^{(s)}\vec{\bar{\Sigma}}_{(L)}^{\nu\lambda(s)}\cdot
  \vec{{\Phi}}_{\nu(L)}^{~\mu}
 - {{\Theta}}^{\nu\lambda}_{(s)}
\vec{\bar{\Sigma}}_{\mu\lambda(R)}^{(s)}\cdot\vec{{\Phi}}_{\nu(R)}^{~\mu}
 )\\
 & & - 6\sqrt{2}i({{\Theta}}^{\mu\lambda}_{(s)}{\bar{\Sigma}}_{\lambda\alpha\dot
 {\alpha}}^{~\nu}\Phi_{\mu\nu}^{(a)\alpha\dot{\alpha}}-
 {\Theta}_{\mu\lambda}^{(s)}
 {\bar{\Sigma}}_{\nu\alpha\dot{\alpha}}^{~\lambda}\tilde{\Phi}^{\mu\nu\alpha\dot
 {\alpha}}_{(a)})\\
 & & + 6\sqrt{2}({{\Theta}}_{\nu}^{~\lambda\alpha\dot{\alpha}}\tilde{\bar{\Sigma}}^
 {\mu\nu}_{(a)}\Phi_{\mu\lambda\alpha\dot{\alpha}}^{(s)}+ {{\Theta}}_{\lambda}^
 {~\nu\alpha\dot{\alpha}}\bar{\Sigma}_{\mu\nu}^{(a)}{\Phi}_{~~~\alpha\dot{\alpha}}^
 {\mu\lambda(s)})\\
 & & -12{\sqrt{2}}i({{\Theta}}_{\lambda}^{~\mu\alpha\dot{\alpha}}{\bar
 {\Sigma}}_{\nu\alpha}^{~\lambda\dot{\beta}}{\Phi}_{\mu\dot{\alpha}
 \dot{\beta}}^{~\nu}+ {{\Theta}}_{\lambda}^{~\mu\alpha\dot{\alpha}}
 {\bar{\Sigma}}_{\mu\dot{\alpha}}
 ^{~\nu\beta}{\Phi}_{\nu\alpha\beta}^{~\lambda})\\
  & & -6\sqrt{2}({{\Theta}}_{\nu\alpha}^{~\lambda\dot{\beta}}\bar{\Sigma}_{\mu\lambda
 \dot{\alpha}\dot{\beta}}^{(s)}\tilde{\Phi}^{\mu\nu\alpha\dot{\alpha}}_{(a)}
 +
 {{\Theta}}_{\lambda\dot{\alpha}}^{~\nu\beta}\bar{\Sigma}_{~~~\alpha\beta}^{\mu\lambda(s)}
 \Phi_{\mu\nu}^{(a)\alpha\dot{\alpha}})\\
  & &
  -12i{{\Theta}}_{\nu}^{~\mu\alpha\dot{\alpha}}{\bar{\Sigma}}_{\mu\alpha\dot{\alpha}}^
  {~\nu}\Phi\\
 & & +12i
  (\vec{\tilde{{\Theta}}}_{(R)}^{\mu\nu(a)}\cdot\vec{\Sigma}_{\mu\lambda(R)}^{(s)}{\Phi}_{\nu}^
  {~\lambda}+\vec{{\Theta}}_{\mu\nu(L)}^{(a)}\cdot\vec{\Sigma}^{\mu\lambda(s)}_{(L)}{\Phi}_
  {\lambda}^{~\nu}
  )\\
 & & +6i(\tilde{{\Theta}}^{\mu\nu(a)}_{~~\dot{\alpha}\dot{\beta}}{\bar{\Sigma}}_{\nu\alpha}^
 {~\lambda\dot{\beta}}\Phi_{\mu\lambda(a)}^{~~\alpha\dot{\alpha}} -\tilde{{\Theta}}^{\mu\nu(a)}_
 {~~\alpha\beta}{\bar{\Sigma}}_{\nu\dot{\alpha}}^{~\lambda\beta}\Phi_{\mu\lambda(a)}^
 {~~\alpha\dot{\alpha}}\\
 & &
 -{\Theta}_{\mu\nu\dot{\alpha}\dot{\beta}}^{(a)}{\bar{\Sigma}}^{~\nu\dot{\beta}}_
 {\lambda\alpha}{\Phi}^{\mu\lambda\alpha\dot{\alpha}}_{(a)}
 +{\Theta}_{\mu\nu\alpha\beta}^{(a)}
 {\bar{\Sigma}}^{~\nu\beta}_{\lambda\dot{\alpha}}{\Phi}^{\mu\lambda\alpha
 \dot{\alpha}}_{(a)})\\
 & &
 -12(\vec{{\Theta}}_{\mu\nu(L)}^{(a)}\cdot\vec{{\Phi}}_{\lambda(L)}^{~\mu}-\vec{{\Theta}}_{\mu\nu(R)}^{(a)}
 \cdot\vec{{\Phi}}_{\lambda(R)}^{~\mu})\bar{\Sigma}^{\nu\lambda}_{(a)}\\
 & &
 -6({\Theta}_{\alpha}^{~\dot{\beta}}\bar{\Sigma}_{\mu\nu\dot{\alpha}\dot{\beta}}^{(s)}
 {\Phi}^{\mu\nu\alpha\dot{\alpha}}_{(s)}
  -{\Theta}_{\dot{\alpha}}^{~\beta}\bar{\Sigma}_{~~\alpha\beta}^{\mu\nu(s)}
  \Phi^{~~\alpha\dot{\alpha}}_{\mu\nu(s)})\\
  & &
  -6\sqrt{2}i({\Theta}_{\alpha}^{~\dot{\beta}}{\bar{\Sigma}}_{\nu}^
  {~\mu\alpha\dot{\alpha}}{\Phi}^{~\nu}_{\mu\dot{\alpha}\dot{\beta}}+
  {\Theta}_{\dot{\alpha}}^{~\beta}{\bar{\Sigma}}_{\nu}^{~\mu\alpha\dot{\alpha}}
  {\Phi}^{~\nu}_{\mu\alpha\beta})\\
& & +12  ({{\Theta}}_{\nu}^{~\mu\alpha\dot{\alpha}}{\bar{\Sigma}}_
{\lambda\alpha\dot{\alpha}}^{~\nu}{\Phi}_{\mu}^{~\lambda}-{{\Theta}}_{\nu}^
{~\mu\alpha\dot{\alpha}}{\bar{\Sigma}}_{\mu\alpha\dot{\alpha}}^{~\lambda}
{\Phi}_{\lambda}^{~\nu})\\
  & &
  +12({{\Theta}}_{\nu}^{~\mu\alpha\dot{\alpha}}\bar{\Sigma}_{\mu\lambda
  \dot{\alpha}\dot{\beta}}^{(s)}{\Phi}_{~~~\alpha}^{\nu\lambda(s)\dot{\beta}}-
  {{\Theta}}_{\nu}^{~\mu\alpha\dot{\alpha}}\bar{\Sigma}^{\nu\lambda(s)}_
  {~~\alpha\beta}\Phi^{~~~\beta}_{\mu\lambda(s)\dot{\alpha}})\\
  & & -12  ({{\Theta}}_{\lambda}^{~\mu\alpha\dot{\alpha}}\bar{\Sigma}_{\mu\nu}^
  {(a)}\Phi_{(a)\alpha\dot{\alpha}}^{\nu\lambda})\\
  & & +6i (\tilde{{\Theta}}^{\mu\nu}_{(a)\dot{\alpha}\dot{\beta}}{\bar{\Sigma}}_
  {\nu}^{~\lambda\alpha\dot{\alpha}}\Phi_{\mu\lambda\alpha}^{(s)\dot{\beta}}+
  \tilde{{\Theta}}^{\mu\nu}_{(a)\alpha\beta}{\bar{\Sigma}}_{\nu}^{~\lambda\alpha
  \dot{\alpha}}\Phi_{\mu\lambda\dot{\alpha}}^{(s)\beta}\\
  & & +{\Theta}_{\mu\nu\dot{\alpha}\dot{\beta}}^{(a)}{\bar{\Sigma}}_{\lambda}^{~\nu\alpha
  \dot{\alpha}}{\Phi}^{\mu\lambda(s)\dot{\beta}}_{~~~\alpha}+
  {\Theta}_{\mu\nu\alpha\beta}^{(a)}{\bar{\Sigma}}_{\lambda}^{~\nu\alpha\dot{\alpha}}
  {\Phi}^{\mu\lambda(s)\beta}_{~~~\dot{\alpha}})\\
  & & + 12\sqrt{2} [\vec{\tilde{{\Theta}}}_{(a)(R)}^{\mu\nu}\cdot(\vec{{\Phi}}_{\nu (R)}^
  {~\lambda}\times
  \vec{\bar{\Sigma}}_{\mu\lambda(R)}^{(s)})\\
  & &- \vec{{\Theta}}_{\mu\nu (L)}^{(a)}\cdot(\vec{{\Phi}}_{\lambda (L)}^{~\nu}\times
\vec{\bar{\Sigma}}_{(s)(L)}^{\mu\lambda})]\\
  & & -6\sqrt{2} i (\Phi^{\nu\lambda\alpha\dot{\alpha}}_{(a)}{\bar{\Sigma}}_
  {\lambda\alpha}^{~\mu\dot{\beta}}{\Theta}_{\mu\nu\dot{\alpha}\dot{\beta}}^{(a)}+
  \Phi^{\nu\lambda\alpha\dot{\alpha}}_{(a)}{\bar{\Sigma}}_{\lambda\dot{\alpha}}^
  {~\mu\beta}{\Theta}_{\mu\nu\alpha\beta}^{(a)})]
\eea

 \vspace{ .3 true cm}

  \section{  : Tables of NMSGUT   mass matrices  }
Here superfield  mass   matrix rows are labeled by barred irreps
and columns by unbarred.

  {\bf{(i)}} The 26 letters of the English alphabet are just sufficient to label the 26 distinct MSSM
   multiplet types that occur in the (N)MSGUT. Calligraphic letters denote the superfield mass matrices of the corresponding
   alphabetically named MSSM multiplet type.The masses of 13 Unmixed fields
   are given in Table 2.

\begin{table}
$$
\begin{array}{l|l|l}
{\rm Field }[SU(3),SU(2),Y] &  PS  \qquad  Fields  & {\rm \qquad Mass}  \\
 \hline
 &&\\
 A[1,1,4],{} \bar A[1,1,-4] &{{\s^{44}_{(R+)}}\over
\sq}, {{\os_{44(R-)}}\over \sq}&
 2( M + \eta (p +3a +  6 \omega )) \\
 M[6,1,{8\over 3}],{}{\ovl M} [(\bar 6,1, -{8\over 3}] &
  (\Sigb^{'(R+)}_{\bm\bn(R+)},{}
 \s^{'\bm\bn}_{(R-)})_{\bm\leq\bn} &
 2 (M + \eta (p -a + 2 \omega )) \\
N[6,1,-{4\over 3}],{}\bar N [(\bar 6,1, {4\over 3}] &
 (\Sigb_{\bm\bn}^{'(R-)},
  \s^{'\bm\bn}_{(R+)} )_{\bm\leq\bn} &
 2 (M + \eta (p -a-2\omega )) \\
 O[1,3,-2],{}\bar O [(1,3, +2] &
 {{{\vec\s}_{44(L)}}\over \sq},{}
 {{{{\vec{\Sigb}}^{44}_{(L)}}}\over \sq} &
 2 (M + \eta (3a-p)) \\
 &&\\
 W[6,3,{2\over 3}],{}{\overline W} [({\bar 6},3, -{2\over 3}] &
 {{{\vec\s'}_{\bm\bn(L)}}} ,
 {\vec\Sigb}^{\bm \bn}_{(L)}  &
 2 (M - \eta (a+p)) \\
I[3,1,{10\over 3}],{}\bar I [(\bar 3,1,- {10\over 3}] &
\phi_{~\bn(R+)}^4,{}
 \phi_{4(R-)}^{~\bn} &
 -2 (m + \lambda (p+a+4\omega)) \\
S[1,3,0] & \vec\phi^{(15)}_{(L)} & 2(m+\lambda(2a-p))\\
Q[8,3,0]& {\vec\phi}_{\bm(L)}^{~\bn}&
 2 (m - \lambda (a +p)) \\
U[3,3,{4\over 3}],{} \bar U[ \bar 3,3,-{4\over 3}] &
{\vec\phi}_{\bm(L)}^{~4},{} {\vec\phi}_{4(L)}^{~\bm}&
 -2 (m - \lambda (p-a)) \\
 &&\\
V[1,2,-3],{} \bar V[ 1,2,3] & {{{\phi}_{44\alpha\dot
2}}\over\sq},{} {{\phi^{44}_{\alpha\dot 1}}\over \sq}&
 2 (m  + 3 \lambda (a + \omega)) \\
B[6,2,{5\over 3}],{}\bar B [(\bar 6,2, -{5\over 3}] &
 (\phi_{\bm\bn\alpha\dot 1}',
 \phi^{'\bm\bn}_{\alpha\dot 2} )_{\bm\leq\bn} &
 -2 (m + \lambda (\omega -a )) \\
Y[6,2,-{1\over 3}],{}\bar Y [(\bar 6,2, {1\over 3}] &
 (\phi_{\bm\bn\alpha\dot 2}',
\phi^{'\bm\bn}_{\alpha\dot 1})_{\bm\leq\bn} &
 2 (m - \lambda (a+\omega )) \\Z[8,1,2],{} \bar Z[ 8,1,-2] & {\phi}_{~\bm(R+)}^{\bn}
{\phi}_{\bm(R-)}^{~\bn}&
 2 (m + \lambda (p-a)) \\
\end{array}
$$
\label{unmix}\caption{     Masses   of the unmixed states in terms
of the superheavy vevs . The $SU(2)_L$ contraction
 order is always $\bar F^{\alpha} F_{\alpha} $.  The absolute value
   of the expressions in the column ``Mass" is understood.
 For sextets of $SU(3)$ the 6 unit norm fields are denoted by a prime
 : $\Sigma'_{\bar\mu\bar\nu}= \Sigma_{\bar\mu\bar\nu},
   \bar\mu >\bar\nu ,\Sigma'_{\bar\mu\bar\mu}=\Sigma_{\bar\mu\bar\mu}/{\sqrt{2}} $
 and similarly for $\bar 6$.  }
\end{table}

\newpage
\vspace{ .3 cm}
 {\bf{ ii)\hspace{ 1.0 cm}  Mixed states}}\hfil\break

\vspace{ .3 cm}

{\bf{a)}}~~$[8,2,-1](\hspace{2mm}\bar{C}_1,\bar{C}_2,\bar{C}_3\hspace{2mm})\bigoplus
[8,2,1](\hspace{2mm}C_1,C_2,C_3\hspace{2mm})\equiv (\bar\Sigma_{
~\dot{2}}^{A \alpha}, \Sigma_{ ~\dot{2}}^{A\alpha},{\Theta}_{
~\dot{2}}^{A\alpha}\hspace{2mm}) \bigoplus\\ (\Sigma_{~ \alpha
\dot{1}}^{A},\bar\Sigma_{~ \alpha \dot{1}}^{A},{\Theta}_{ ~\alpha
\dot{1}}^{A})(A=1.....8)$ \bea {\cal{C}}=  \left(
\begin{array}{ccc}
2(-M + \eta (a + \omega))& 0 & -i(\omega-p)\bar\zeta \\
0 & 2(-M + \eta (a + \omega))& -i(\omega+ p)\zeta\\
i(\omega - p)\zeta& i(\omega + p)\bar\zeta& -m_\Theta
+\frac{\rho}{3}a
\end{array}\right) \nonumber \eea

{\bf{b)}}~~$[\bar{3},2,-\frac{7}{3}](\hspace{2mm}\bar{D}_1,
\bar{D}_2,\bar{D}_3\hspace{2mm})\oplus
[3,2,\frac{7}{3}](\hspace{2mm}D_1,D_2,D_3\hspace{2mm})\equiv
(\Sigma^{\bar\nu \alpha }_{4\dot{2}},
\bar\Sigma_{4\dot{2}}^{\bar\nu \alpha},{\Theta}_{4
\dot{2}}^{\bar\nu\alpha}\hspace{2mm}) \oplus\\
(\bar\Sigma_{\bar\nu \alpha \dot{1}}^{4},\Sigma_{\bar\nu \alpha
\dot{1}}^{4},{\Theta}_{\bar\nu\alpha \dot{1}}^{4})$ \bea
{\cal{D}}=
  \left( \begin{array}{ccc}
2(M + \eta (a + \omega))& 0 & (i \omega +ip - 2i a )\zeta \\
0 & 2(M + \eta (a + 3\omega))& ( -3i\omega -ip - 2i a)\bar\zeta\\
-(i \omega +i p - 2ia)\bar\zeta & (3i\omega + i p + 2i a )\zeta&
m_\Theta +\frac{\rho}{3}(a + 2\omega)
\end{array}\right)\nonumber \eea

{\bf{c)}}~~   $[\bar 3,2,-{1\over 3}](\bar E_1, \bar E_2,\bar
E_3,\bar E_4,\bar E_5,\bar E_6) \oplus [3,2,{1\over
3}](E_1,E_2,E_3,E_4,E_5,E_6)$\hfil\break $.\qquad\qquad \equiv
(\Sigma_{4 \dot 1}^{\bar\mu\alpha}, \Sigb_{4\dot 1}^{\bm \alpha},
\phi^{\bm 4\alpha}_{(s)\dot 2} , \phi^{(a) \bm 4\alpha}_{\dot
2},\lambda^{\bm 4\alpha}_{\dot 2},\Theta_{4
\dot{1}}^{\bar\sigma\alpha}) \oplus  (\bar\Sigma_{\bar\mu \alpha
\dot{2}}^{4}\s_{\bm\alpha\dot 2}^4,\phi_{\bm 4\alpha\dot 1}^{(s)},
\phi_{\bm 4\alpha\dot 1}^{(a)},\lambda_{\bm\alpha\dot
1},\Theta_{\bar\sigma\alpha}^{4 \dot{1}}) $ \bea {\cal{E}}=
{\scriptsize \left(
\begin{array}{cccccc}-2(M+\e(a-\om))&0&0&0&0&(i \omega -ip + 2ia )\zeta\\ 0&-2(M+\e(a-3\om))&
-2\sq i\e\sss&2i\e\sss&ig\sq\ssb^*& (-3 i\omega +ip + 2ia )\bar\zeta\\
0&2i\sq\e\ssb&-2(m+\la(a-\om))&-2\sq\la\om&2g(a^*-\om^*)& -\sqrt{2}\bar{\zeta}\bar{\sigma}\\
0&-2i\e\ssb&-2\sq\la\om&-2(m-\la\om)&\sq g(\om^*-p^*)& \bar{\sigma}\bar{\zeta}\\
0&-ig\sq\sss^*&2g(a^*-\om^*)&g\sq(\om^*-p^*)&0&0\\
( -i\omega +ip - 2ia )\bar\zeta&(3i\omega -ip - 2ia )\zeta&
-\sqrt{2} \zeta \sigma & \sigma\zeta&0& -(m_\Theta
+\frac{\rho}{3}a - \frac{2}{3}\rho \omega)\\
\end{array}\right) } \nonumber\eea
{\bf{d)}}~~$[1,1,-2](\hspace{2mm}\bar{F}_1,\bar{F}_2,\bar{F}_3,\bar{F}_4\hspace{2mm})\oplus
[1,1,2](\hspace{2mm}F_1,F_2,F_3,F_4\hspace{2mm})\equiv
(\frac{\bar\Sigma_{44 (R0)}}{\sqrt{2}}, \Phi_{(R-)
}^{(15)},\lambda_{(R-) },\frac{\Theta_{44}}{{\sqrt{2}}}\hspace{2mm}) \oplus\\
(\frac{\Sigma_{(R0)
}^{44}}{\sqrt{2}},\Phi_{(R+)}^{(15)},\lambda_{(R+)},\frac{\Theta^{44}}{\sqrt{2}})$
\bea {\cal{F}}=
  \left( \begin{array}{cccc}
2(M + \eta (p + 3a))& -2i\sqrt{3}\eta\sigma &
-g\sqrt{2}\bar{\sigma}^*&
-6i\bar{\zeta}\omega \\
2i\sqrt{3}\eta \bar{\sigma}& 2(m + \lambda (p + 2a))& \sqrt{24}ig
\omega^*&
\sqrt{{3}}\bar{\zeta}\bar{\sigma}\\
-g\sqrt{2}\sigma^*& -\sqrt{24}ig\omega^*& 0&0\\
 6i\zeta\omega& \sqrt{3}\zeta\sigma&0& m_\Theta + a \rho
\end{array}\right) \nonumber \eea

%\newpage
{\bf{e)}}~~$ [1,1,0] (G_1,G_2,G_3,G_4,G_5,G_6) \equiv
(\phi,\phi^{(15)},\phi^{(15)}_{(R0)},{{\s^{44}_{(R-)}}\over \sq},
{{\Sigb_{44((R+)}}\over \sq}, {{{\sq \lambda^{(R0)} -
{\sqrt{3}}\lambda^{(15)}}\over {\sqrt{5}}}})$

\bea {\cal{G}}= 2\left({\begin{array}{cccccc} m&0 &
\sqs\la\om & {{i\e\ssb}\over \sq}&{-i\e\sss\over \sq}&0\\
0& m + 2 \la a & 2\sq\la\om& i\e\ssb\sqtt &-i\e\sss\sqtt&0\\
\sqs\la\om&2\sq\la\om&m+\la(p+2a)& -i\e\sqt\ssb & i\sqt\e\sss&0\\
{{i\e\ssb}\over\sq}& i\e\ssb\sqtt&-i\e\sqt\ssb&0&
M+\e(p+3a -6\om)&{{\sqf g\sss^*}\over  2 }\\
{{-i\e\sss}\over\sq}& -i\e\sss\sqtt&i\e\sqt\sss&
M+\e(p+3a -6\om)&0&{{\sqf g \ssb^*}\over 2}\\
0&0&0&{{\sqf g\sss^*}\over 2}&{{\sqf g\ssb^*}\over 2}&0
\end{array}}\right)
\nonumber\eea

{\bf{f)}} ~~
 $[1,2,-1](\bar{h}_1,\bar{h}_2,\bar{h}_3,\bar{h}_4,\bar{h}_5,\bar{h}_6)\oplus
[1,2,1](h_1,h_2,h_3,h_4,h_5,h_6)\equiv(H^{\alpha }_{\dot{2}},
\bar\Sigma_{\dot{2}}^{(15)\alpha},\\\Sigma_{\dot{2}}^{(15)\alpha},\frac
{\Phi_{44}^{\dot{2}\alpha}}{\sqrt{2}},\Theta^{\alpha}_{\dot{2}},\Theta_{\dot{2}}^
{(15)\alpha}\hspace{2mm}) \oplus (H_{\alpha
\dot{1}},\bar\Sigma_{\alpha \dot{1}}^{(15)},\Sigma_{\alpha
\dot{1}}^{(15)},\frac{\Phi_{\alpha}^{44\dot{1}}}{\sqrt{2}},\Theta_{\alpha\dot{1}}
,\Theta_{\alpha\dot{1}}^{(15)})$\bea {\cal{H}}= {\scriptsize
  \left( \begin{array}{cccccc}
-M_H & \bar{\gamma}\sqrt{3}(\omega-a) & -\gamma\sqrt{3}(\omega +
a)&
-\bar{\gamma}\bar{\sigma}&kp & -\sqrt{3}ik\omega \\
 -\bar{\gamma}\sqrt{3}(\omega+ a)& 0 & -(2M + 4\eta(a+ \omega))&0 &
 -\sqrt{3}\bar{\zeta}\omega & i(p+2\omega)\bar{\zeta}\\
\gamma\sqrt{3}(\omega-a) & -(2M + 4\eta(a- \omega))&0 & -2\eta
\bar{\sigma}\sqrt{3}& \sqrt{3}\zeta\omega& -i(p-2\omega)\zeta\\
-\sigma\gamma & -2\eta\sigma\sqrt{3}&0 & -2m + 6\lambda(\omega-a)&
\zeta\sigma & \sqrt{3}i\zeta\sigma\\
pk& \sqrt{3}\bar{\zeta}\omega& -\sqrt{3}\omega\zeta&
\bar{\zeta}\bar{\sigma}& -m_{\Theta}&
\frac{\rho}{\sqrt{3}}i\omega\\
\sqrt{3}ik\omega&i(p-2\omega)\bar{\zeta}&
 -i(p+2\omega)\zeta& -\sqrt{3}i\bar{\zeta}\bar{\sigma}&
  -\frac{\rho}{\sqrt{3}}i\omega& -m_\Theta - \frac{2\rho}{3}a\\
\end{array}\right) } \nonumber \eea
The above matrix is to be diagonalized after imposing the fine
tuning condition $Det {\cal H} =0$ to keep one pair of doublets
light.\\

{\bf{g)}}~~ $[\bar 3,1,-{4\over 3}](\bar J_1,\bar J_2,\bar
J_3,\bar J_4,\bar J_5) \oplus [3,1,{4\over
3}](J_1,J_2,J_3,J_4,J_5)$ \hfil\break $.\qquad\qquad\equiv
(\s^{\bm4}_{(R-)},\phi_4^{\bm},
\phi_4^{~\bm(R0)},\lambda_4^{~\bm},\Theta^{\bar{\mu}4}_{(R-)})
\oplus (\Sigb_{\bm4(R+)},\phi_{~\bm}^4,
\phi_{\bm(R0)}^{~4},\lambda_{\bm}^4,\Theta_{\bar\mu 4}^{(R+)})$

\bea {\cal{J}}= \left({\begin{array}{ccccc} 2(M+\e(a+p-2\om))&
-2\e\ssb&2\sq \e\ssb&-ig\sq\sss^*& 2\zeta(a - 2 \om)\\
2\e\sss&-2(m+\la a)&-2\sq\la\om&-2ig\sq a^*& -\sigma \zeta\\
-2\sq\e\sss&-2\sq\la\om&-2(m+\la(a+p))&-4i g\om^*&  \sqrt{2} \sigma \zeta\\
-ig\sq\ssb^*&2\sq i g a^*&4i g\om^*&0&0\\
2 \bar\zeta(a- 2\om) &\bar{\sigma}\bar\zeta& -\sqrt{2}
\bar{\sigma} \bar{\zeta}&0& m_{\Theta} +\frac{\rho}{3}(p-2\omega)
 \end{array}}\right)
\nonumber\eea\\

{\bf{h)}}~~$[\bar{3},1,\frac{8}{3}](\hspace{1mm}\bar{K}_1,\bar{K}_2\hspace{1mm})\oplus
[3,1,-\frac{8}{3}](\hspace{2mm}K_1,K_2,\hspace{2mm})\equiv
(\Sigma^{\bar\mu 4}_{(R+)},\Theta^{\bar\mu 4}_{(R+)}) \oplus
(\bar\Sigma_{\bar\mu 4(R-)},\Theta_{\bar\mu 4(R-)})$ \bea
{\cal{K}}=
  \left( \begin{array}{cc}
2(M + \eta (a + p+ 2\omega))& 2\zeta(a +2\om) \\
2\bar\zeta(a + 2\om) & m_\Theta + \frac{\rho}{3}(p+2\omega)\\
\end{array}\right) \nonumber\eea

%\newpage
{\bf{i)}}~~$[\bar{6},1,-\frac{2}{3}](\hspace{1mm}\bar{L}_1,\bar{L}_2\hspace{1mm})\oplus
[6,1,\frac{2}{3}](\hspace{2mm}L_1,L_2,\hspace{2mm})\equiv
(\Sigma'^{\bar\mu \bar\nu(s)}_{(R0)},
~~\Theta'^{\bar\mu\bar\nu(s)})
\oplus\\
(\bar\Sigma'_{\bar\mu\bar\nu(s)(R0)},~~
\Theta'_{\bar\mu\bar\nu(s)})$
 \bea {\cal{L}}= \left( \begin{array}{cc}
2(M + \eta (p-a))& - 2i\zeta\omega\\
2i\bar{\zeta} \omega & m_\Theta - \frac{\rho}{3}a\\
\end{array}\right) \nonumber\eea

{\bf{j)}}~~$[\bar{3},3,\frac{2}{3}](\hspace{1mm}\bar{P}_1,\bar{P}_2\hspace{1mm})\oplus
[3,3,-\frac{2}{3}](\hspace{2mm}P_1,P_2,\hspace{2mm})\equiv
(\vec{\bar{\Sigma}}^{\bar\mu 4}_{(L)},\vec{\Theta}^{\bar\mu
4}_{(L)}) \oplus (\vec{\Sigma}_{\bar\mu
4(L)},\vec{\Theta}_{\bar\mu 4(L)})$ \bea {\cal{P}}=
 \left( \begin{array}{cc}
2(M + \eta (a-p))& 2a\bar{\zeta}\\
2a\zeta & m_\Theta - \frac{\rho}{3}p \\
\end{array}\right)\nonumber\eea

 {\bf{k)}}~~$ [8,1,0](R_1,R_2)\equiv (\hat\phi_{\bm}^{~\bn},\hat\phi_{\bm
(R0)}^{~\bn})  $

 \bea
{\cal{R}} = 2 \left({\begin{array}{cc} (m-\lambda a ) &
-\sqrt{2}\lambda\omega \\ -\sqrt{2}\lambda\omega & m+\lambda( p-a)
\end{array}}\right)
\nonumber\eea\\

{\bf{l)}}~~$[\bar{3},1,\frac{2}{3}](\bar{t}_1,\bar{t}_2,\bar{t}_3,\bar{t}_4,\bar{t}_5,
\bar{t}_6,\bar{t}_7)\oplus
[3,1,-\frac{2}{3}](t_1,t_2,t_3,t_4,t_5,t_6,t_7)\equiv(H^{\bar\mu 4
},
\bar\Sigma_{(a)}^{\bar\mu4},\\\Sigma_{(a)}^{\bar\mu4},\Sigma_{(R0)}^{\bar\mu4},
\Phi_{4(R+)}^{\bar\mu},\Theta^{\bar\mu
4(s)},\Theta^{\bar\mu4}_{(R0)}) \oplus
(H_{\bar\mu4},\bar\Sigma_{\bar\mu 4(a)},\Sigma_{\bar\mu
4(a)},\bar\Sigma_{\bar\mu
4(R0)},\Phi_{\bar\mu(R-)}^{4},\Theta_{\bar\mu 4
(s)},\Theta_{\bar\mu 4(R0)})$ \bea {\cal{T}}={\scriptsize
  \left( \begin{array}{ccccccc}
M_H & \bar{\gamma}(a+p) & \gamma(p-a)& 2\sqrt{2}i\omega
\bar{\gamma}& i
\bar{\sigma}\bar{\gamma} &\sqrt{2}ka& \sqrt{2}ik\omega \\
 \bar{\gamma}(p-a)& 0 & 2M &0 & 0 & \sqrt{2}a\bar\zeta& \sqrt{2}i\omega \bar{\zeta}\\
\gamma(p+a) &2M &0 & 4\sqrt{2}i\omega\eta & 2i\eta\bar{\sigma}&
-\sqrt{2}  a \zeta&
\sqrt{2}i\omega \zeta \\
-2\sqrt{2}i\omega\gamma & -4\sqrt{2}i\omega\eta&0 & 2M + 2\eta
p+2\eta a& -2\sqrt{2}
\eta \bar{\sigma} & 2i\omega \zeta& 2\zeta a\\
i\sigma\gamma&2i\eta\sigma&0&
2\sqrt{2}\eta\sigma&-2m-2\lambda(a+p-4\omega)&
\sqrt{2}i\sigma\zeta& -\sqrt{2}\zeta\sigma\\
\sqrt{2}ka & -\sqrt{2}  a\bar\zeta& \sqrt{2}a\zeta& -2i
\bar{\zeta}\omega&
\sqrt{2}i\bar{\zeta}\bar{\sigma}& m_\Theta + \frac{\rho}{3}a& -\frac{2i}{3} \rho \omega\\
 -\sqrt{2}i k \omega& -\sqrt{2}i\omega \bar{\zeta}& -\sqrt{2}i\omega \zeta& 2\zeta a
 & \sqrt{2}\bar{\sigma}\bar{\zeta}& \frac{2i}{3} \rho \omega& m_\Theta +\frac{\rho}{3}p\\
\end{array}\right) } \nonumber \eea

{\bf{m)}}~~$ [3,2,{5\over 3}](\bar X_1,\bar X_2,\bar X_3) \oplus
[3,2,-{5\over 3}](X_1,X_2,X_3)\hfil\break .\qquad\qquad \equiv
(\phi^{(s)\bm4}_{\alpha\dot 1} , \phi^{(a)\bm4}_{\alpha\dot 1}
,\lambda^{\bm4}_{\alpha\dot 1}) \oplus(\phi_{\bm4\alpha\dot
2}^{(s)}, \phi_{\bm4\alpha\dot 2}^{(a)},\lambda_{\bm4\alpha\dot
2})
  $

\bea {\cal{X}}= \left({\begin{array}{ccc} 2(m+\la(a+\om))&
-2\sq \la \om &-2g(a^*+\om^*)\\
-2\sq \la \om &2(m+\la \om)& {\sq}g(\om^* +p^*)\\
 -2 g(a^* +\om^*) &\sq g(\om^* + p^*)&0
 \end{array}}\right)
\nonumber\end{eqnarray}

\section{ Susy Threshold Corrections}
  A complete evaluation of 1-loop radiative corrections for the MSSM in the absence of generation mixing is available in
  \cite{piercebagger} which  is an invaluable and essential reference for our work.  The
approximations(to their own exact 1-loop results) of
\cite{piercebagger} are however used in our  actual calculation
with the modification that for the down type ($T_{3L}=-1/2$)
fermions we also add the Bino correction given in \cite{antusch}
to the chargino and gluino corrections considered in the
approximation   advocated in \cite{piercebagger}; in any case
difference is   small. The two give closely similar  results. As
in \cite{piercebagger}, we use the gluon and squark-gluino
correction to correct the top quark Yukawa from that given by the
pole mass and apply the squark-gluino corrections to the running
up and charm quark Yukawa couplings since  these corrections are
not insignificant either( see Tables 5,11).

 We next summarize the approximations to the  1-loop threshold corrections
computed in \cite{piercebagger}.  The  corrections are computed in
the \emph{absence} of generation mixing. The dominant corrections
to the top quark Yukawa are the gluon and squark-gluino loops. To
estimate the leading corrections to the top quark Yukawa $y_t$,
they set the electroweak couplings and the Yukawas  to zero :  $g
= g' = 0$ and $y_t = y_b = 0$   and simplify the resulting
expressions by setting the external momenta $p^2 = 0$ because
$m_t$ is much smaller than a typical squark or gluino mass. In
that limit, the physical top quark mass is given by
\begin{equation}
m_t\ =\ \hat m_t(Q) \left[\, 1 +\,{\Delta m_t \over
\hat{m}_t}\,\right] \label{mt phys}\ ,
\end{equation}
where ${\hat m}_t$ is the MSSM tree level mass and $\Delta m_t$ is
the one-loop correction due to (heavy) sparticles computed in the
effective MSSM. The one-loop correction ${\Delta m_t \over
\hat{m}_t}$ receives two important contributions. The first is the
finite  gluon correction
\begin{eqnarray}
\left({\Delta m_t \over m_t}\right)^{tg}\   &=& \
{g^2_3\over12\pi^2}\ \left[\,3
 \ln\left({M_Z^2\over m_t^2}\right) + 5\,\right] \label{mt corr}\ .
\end{eqnarray}
The  renormalization scale  is always chosen to be $M_Z$.  The
second correction comes from the top squark($\tilde
t$)/gluino($\tilde g$) loops,
\begin{eqnarray}
\left({\Delta m_t\over m_t}\right)^{\tilde t\tilde g}\ &=& \ - \
{g^2_3\over12\pi^2}\ \Bigg\{ \,B_{1}(0,m_{\tilde g},m_{\tilde
t_1}) + B_{1}(0,m_{\tilde g},m_{\tilde t_2})\, \nonumber\\ &&\ -\
\sin (2\theta_t) \ \left({m_{\tilde g} \over m_t}\right) \
\Bigg[\,B_{0}(0,m_{\tilde g},m_{\tilde t_1}) - B_{0}(0,m_{\tilde
g},m_{\tilde t_2})\,\Bigg] \Bigg\} \label{mt gluino}
\end{eqnarray}
where $\theta_t$ is the top-squark LR mixing angle (for any
sfermion mixing angle $\theta_{ f}\,:\, \tan 2 \theta_{\tilde f}=
2(M^2_{LR}/(M^2_{LL}-M^2_{RR} ))_{ f} $ and the approximate forms
of the loop functions are :
\begin{equation}
B_{0}(0,m_1,m_2)\ =\ - \ln\left(M^2\over Q^2\right) + 1 +
{m^2\over m^2-M^2}\ln\left(M^2\over m^2\right)\ , \label{b0(0)}
\end{equation}
\begin{equation}
B_{1}(0,m_1,m_2)\ =\ {1\over2}\left[\, -\ \ln\left(M^2\over
Q^2\right) + {1\over2} + {1\over 1-x} + {\ln  x \over (1-x)^2} -
\theta(1-x)\ln x \,\right]\ , \label{b1(0)}
\end{equation}
with $M=\max(m_1,m_2),$ $m=\min(m_1,m_2),$ and $x=m_2^2/m_1^2$.

For the bottom quark   it is appropriate to use the running mass
at $Q=M_Z$ and thus to absorb the gluon corrections in the running
bottom quark mass. Taking the bottom quark pole mass $m_b=4.9$
GeV, and $\alpha_s(M_Z)=0.118$, they find the standard-model
\mbox{\footnotesize$\overline{\rm DR}~$} value $\hat m_b(M_Z)^{\rm
SM}=2.92$ GeV. See \cite{xzz} for a recent review of accurate RG
corrected fermion Yukawas.

 As in the case of the top quark we then interpret the relation
 between the SM and MSSM masses(Yukawas) as :
\begin{equation}
\hat m_b(M_Z)^{\rm SM} \ =\   \hat m_b(M_Z)  \ \left[\, 1 +\left(
{\Delta m_b \over \hat{m}_b}\right)^{\rm massive}\,\right]\ .
\end{equation}
instead of the (perturbatively) equivalent form quoted in
\cite{piercebagger}
\begin{equation}
\hat m_b(M_Z)\ =\ \hat m_b(M_Z)^{\rm SM}\ \left[\, 1 - \left(
{\Delta m_b \over m_b}\right)^{\rm massive}\,\right]\ .
\end{equation}
This is appropriate since the radiative corrections should be
calculated consistently in terms of the MSSM couplings.
 Ignoring  the small
$W$, $Z$, Higgs, and neutralino contributions  leaves the
squark/gluino and squark/chargino loops :
\begin{equation}
\left({\Delta m_b\over m_b}\right)^{\rm massive}\ = \
\left({\Delta m_b\over m_b}\right)^{\tilde b\tilde g}\ + \
\left({\Delta m_b\over m_b}\right)^{\tilde t\tilde\chi^+} \
.\label{mb app}
\end{equation}
The squark/gluino contribution is again given by (\ref{mt
gluino}), with the substitution $t\rightarrow b$. To approximate
the squark/chargino contribution, they set $g = g' =
 y_b =  y_t = 0$, except for terms that are enhanced by the
Higgsino mass parameter $\mu$ or by $\tan\beta$. The chargino
masses are set to the Wino mass $M_2$ and $\mu$, respectively
(this is an excellent approximation in all the parameter sets we
will actually find  since $\mu >>M_2,m_t$).  In this case
squark/chargino loops give rise to the following terms for the
down quarks(with similar formulae also for down type charged
leptons)
\begin{eqnarray}
&&\left({\Delta m_b\over m_b}\right)^{\tilde t\tilde\chi^+}\ = \ {
y_t^2\over16\pi^2}\ \mu\ {A^0_t\tan\beta-\mu \over m_{{\tilde
t}_2}^2-m_{{\tilde t}_1}^2} \ \Bigg[\, B_0(0,\mu,m_{{\tilde
t}_1})-B_0(0,\mu,m_{{\tilde t}_2})\, \Bigg] \nonumber \\\qquad &&-
\ {g^2\over16\pi^2}\,\Bigg\{ {\mu M_2\tan\beta \over
\mu^2-M_2^2}\, \Bigg[\, c_t^2 B_0(0,M_2,m_{{\tilde t}_1}) + s_t^2
B_0(0,M_2,m_{{\tilde t}_2}) \Bigg]\ \nonumber \\ &&+\
(\mu\leftrightarrow M_2) \ \Bigg\}\ ,
\end{eqnarray}
where $B_{0}(0,m_1,m_2)$ is defined in (\ref{b0(0)}), and $c_t\
(s_t)$ is $\cos\theta_t\ (\sin\theta_t)$  and we have changed the
sign of $\mu$ since our convention  is opposite to that of
\cite{piercebagger}.

The same formulae,  \emph{mutatis mutandis},  can be used for the
down and strange quarks and, dropping the gluino correction  and
neutrino-Dirac coupling, for the $T_{3L}=-1/2$ leptons. In a
slight improvement we also add the Bino corrections for the
$T_{3L}=-1/2$ fermions taken from the formulae quoted in
\cite{antusch} : \bea \Delta m_i/(m_i \tan\beta)   =   \frac{1}{16
\pi^2} \left[ \frac{{g'}^2}{6} \frac{M_1}{\mu} \left(
H_2(v_{\tilde{Q}_i}, x_1) +  2 H_2(v_{\tilde{d}_i}, x_1) \right) +
\frac{{g'}^2}{9} \frac{\mu}{M_1}
H_2(w_{\tilde{Q}_i},w_{\tilde{d}_i}) \right]\eea where
$u_{\tilde{f}} = m_{\tilde{f}}^2/M_3^2$, $v_{\tilde{f}} =
m_{\tilde{f}}^2/\mu^2$, $w_{\tilde{f}} = m_{\tilde{f}}^2/M_1^2$,
$x_1 = M_1^2/\mu^2$ and $x_2 = M_2^2/\mu^2$ for $i=d,s,b$.   All
mass parameters are assumed to be real. $H_2$ is defined as
\begin{equation}
H_2(x,y) = \frac{x \ln x}{(1-x)(x-y)} + \frac{y \ln
y}{(1-y)(y-x)}.
\end{equation}
  $H_2$ is negative for positive $x$ and $y$ and $|H_2|$
is maximal, if its arguments are minimal, and vice versa.

The main point to note is that due to large $\tan\beta$, for the b
quark, a large (MSSM Yukawa  lowering)  gluino correction can be
canceled and even overcome  by a large $A_0 y_t$ driven chargino
correction to a degree that it alleviates the known
tension\cite{king} between the  value of ${\hat m}^{SM}_b$ from
experiment and the value ${\hat m}^{MSSM}_b$   required by SO(10)
grand unification. The same is not true for the down and strange
quark due to their small Yukawas, so that for them the gluino
correction serves just to lower the SM  Yukawa couplings as needed
to match the MSSM/GUT values. Conversely for the charged leptons
there are no gluino corrections. The structure of the threshold
corrections is thus just what  is needed to alleviate   the
problems that arise in SO(10) Yukawa unification including those
specific to the NMSGUT.    It is is easy to check  that the down
and strange Yukawas can be significantly lowered only by the
gluino correction with a large positive  value of the ratio
$\mu/M_3$. There is   a preference for lower values of squark
masses for the first two generations. On the other hand for the
third generation one has the opposite problem : the desired value
of ${\hat y}_b^{MSSM}(M_Z)$   corresponds to ${\hat
m}^{SM}_b(M_Z)$ at least as large as $3.15$ GeV   whereas the
actual central value is about $2.9\pm .2\,\, GeV$. Thus one needs
to countervail against the gluino correction  and a large  value
of $A_t$ can accomplish this. In this paper we have confined
ourselves to searches where $\mu $ is real and positive, thus
$A_0$ emerges negative and large. Although we allow the initial
gaugino mass parameter at $M_X$ to be negative (and negative
$m_{1/2}$ values are indeed  chosen by the search algorithm: see
Tables 3,9) the low energy values are  positive. Moreover note
that the large values of $A_0$ used cause the gaugino mass ratios
at $M_Z$ to vary considerably from the approximately 1:2:7 (i.e.
$1:g_2^2/g_1^2:g_3^2/g_1^2$) ratio expected on the basis of one-loop
RG flow down to $M_Z$ from  universal gaugino mass boundary
condition we use  at $M_X$.   The range of possibilities for these
ratios that we actually encounter in our work -given that
$A_0,m_{\tilde {f}}$ are always large - is so different from that
encountered  at small values of these parameters  that the oft
advocated abandonment of the universal gaugino mass condition at
$M_X$ to achieve such diversity seems ill advised till it has
indeed been verified that suitable large  values of $A_0$ etc will
not serve equally well.   Finally we note that the   third
generation sfermion masses are driven to large values by the large
negative values of the Higgs mass-squared parameters which occur
(already at one loop) multiplied with the Yukawa couplings squared
in the RGE for sfermion masses squared. This enabling feature is
discussed more completely in \cite{gutupend}(Appendix \textbf{B}).

For completeness, we also mention the 1-loop electroweak symmetry
breaking conditions  that we imposed:
\begin{eqnarray}
\mu^2 &=& {1\over2}\Biggl[\, \tan2\beta\biggl(\overline
m_{H_2}^2\tan\beta-\overline m_{H_1}^2\cot\beta\biggr) \ -\ M_Z^2\
  \Biggr]\  \nonumber\\ B
&=& {1\over 2}\biggl(\tan 2\beta \, (\overline m_{H_2}^2-\overline
m_{H_1}^2) \ -\ M_Z^2 \sin 2 \beta\biggr)
\label{ourmuB}\end{eqnarray}
where $H_2\equiv H, H_1\equiv{\overline H},  \overline m_{H_1}^2 =
m_{H_1}^2 - t_1/v_1 , \ \overline m_{H_2}^2 = m_{H_2}^2 - t_2/v_2
$  and $v_{1,2},t_{1,2}$ are vevs and tadpoles(calculated using a
subroutine taken from the SPheno Susy spectrum code\cite{porod}
based on the formulae of\cite{piercebagger}) of the effective
potential. Note that the Higgs scalar masses we quote are also
output from SPHENO\cite{porod} subroutine that however uses the
method of \cite{loophiggs}. Thus the pseudo-scalar mass $M_A$ is
computed from the one-loop corrected value of $B$ in
eqn.(\ref{ourmuB}) above using the tree level relation ($M_A^2=2
B/\sin2\beta$) while the Higgs scalar masses are computed using
the effective potential masses. The 1-loop pole masses for these
particles require also computation of the
self-energies\cite{piercebagger} which are not included here. Note
that we use the interpolation formula suggested in
\cite{piercebagger} for the electroweak vev in the MSSM:\be
v=[248.6 +0.9\, Ln\,{\frac{M_{\tilde q}}{M_Z}}]\, GeV\quad ;\quad
M_{\tilde q}^2 = 4 m_{\frac{1}{2}}^2 + m_{\tilde f}^2  \ee

\clearpage

%\section{Example Fits}
 %\usepackage{epsfig}
\begin{table}
 $$
 \begin{array}{|c|c|c|c|}
 \hline
  {\bf{Appendix \, D} }  &{\bf{Two}} &{\bf{ Example}} & {\bf{fits}}   \\
  \hline
 {\rm Parameter }&Value &{\rm  Field }&\hspace{10mm} Masses\\
 &&{\rm}[SU(3),SU(2),Y]&\hspace{10mm}( Units\,\,of 10^{16} Gev)\\ \hline
       \chi_{X}&  0.0338           &A[1,1,4]&     19.73 \\ \chi_{Z}&
    0.0149
                &B[6,2,{5/3}]&            0.0224\\
           h_{11}/10^{-6}& -0.3965         &C[8,2,1]&{      2.59,     18.50,     18.98 }\\
           h_{22}/10^{-4}&  9.9684    &D[3,2,{7/ 3}]&{      0.57,      5.84,      8.41 }\\
                   h_{33}& -0.3872     &E[3,2,{1/3}]&{      0.03,      1.03,      1.03 }\\
 f_{11}/10^{-6}&
 -0.2749-  0.1421i
                      &&{     1.729,      9.94,     11.97 }\\
 f_{12}/10^{-6}&
  2.0663+  0.4203i
          &F[1,1,2]&      0.34,      0.34
 \\f_{13}/10^{-5}&
  1.0541-  0.5949i
                  &&      1.95,     10.08  \\
 f_{22}/10^{-5}&
 -0.9652-  0.6818i
              &G[1,1,0]&{     0.004,      0.04,      0.05 }\\
 f_{23}/10^{-4}&
  0.6791-  1.3076i
                      &&{     0.052,      1.23,      1.25 }\\
 f_{33}/10^{-3}&
  2.6506-  1.3235i
              &h[1,2,1]&{     0.234,      2.52,      3.38 }\\
 g_{12}/10^{-3}&
  0.5392+  0.1579i
                 &&{      4.83,      9.75 }\\
 g_{13}/10^{-3}&
  2.5557-  4.6432i
     &I[3,1,{10/3}]&      0.07\\
 g_{23}/10^{-2}&
  1.0256+  3.9768i
          &J[3,1,{4/3}]&{     0.057,      0.75,      0.75 }\\
 \lambda/10^{-2}&
 -1.9313-  0.2227i
                 &&{      3.34,     13.47 }\\
 \eta&
 -5.2157-  3.1585i
   &K[3,1, {8/ 3}]&{      3.31,     17.44 }\\
 \rho&
 -0.7161-  1.5279i
    &L[6,1,{2/ 3}]&{      3.06,     25.18 }\\
 k&
 -0.6362-  0.0019i
     &M[6,1,{8/ 3}]&     26.49\\
 \zeta&
 -0.9531-  0.3806i
     &N[6,1,{4/ 3}]&     24.23\\
 \bar\zeta &
  1.4806+  1.2424i
          &O[1,3,2]&     24.62\\
       m/10^{16}GeV& 0.00306    &P[3,3,{2/ 3}]&{      2.37,     21.08 }\\
     m_\Theta/10^{16}GeV&  -2.805e^{-iArg(\lambda)}     &Q[8,3,0]&     0.054\\
             \gamma& 3.18293        &R[8,1, 0]&{      0.02,      0.06 }\\
              \bar\gamma&  0.5553     &S[1,3,0]&    0.0761\\
 x&
  0.6944+  0.5020i
         &t[3,1,{2/ 3}]&{      0.04,      0.78,      2.70,      3.33        }\\\Delta_X&     -0.30 &&{     12.75,     13.63,     93.77 }\\
              \Delta_{G}&   5.350           &U[3,3,{4/3}]&     0.064\\
 \Delta\alpha_{3}(M_{Z})&  -0.016               &V[1,2,3]&     0.055\\
    \{M^{\nu^c}/10^{11}GeV\}&{  0.0108,    0.33,  107.72    }&W[6,3,{2/ 3}]&             23.51  \\
  \{M^{\nu}_{ II}/10^{-7}eV\}&  0.0109,      0.33,            108.83               &X[3,2,{5/ 3}]&     0.019,     1.189,     1.189\\
                  M_\nu(meV)&{  2.2845,    7.62,   41.82    }&Y[6,2, {1/3}]&              0.03  \\
    \{\rm{Evals[f]}\}/ 10^{-5} eV&{ 0.02972,   0.903,  297.00         }&Z[8,1,2]&              0.06  \\
 \hline\hline
 \mbox{Soft parameters}&{\rm m_{\frac{1}{2}}}=
          -445.837
 &{\rm m_{0}}=
          7018.005
 &{\rm A_{0}}=
         -2.8605 \times 10^{   5}
 \\
 \mbox{at $M_{X}^0$}&\mu=
          2.2652 \times 10^{   5}
 &{\rm B}=
         -3.5698 \times 10^{  10}
  &{\rm tan{\beta}}=           50.0000\\
 &{\rm M^2_{\bar H}}=
         -3.6296 \times 10^{  10}
 &{\rm M^2_{  H} }=
         -3.4956 \times 10^{  10}
 &
 {\rm R_{\frac{b\tau}{s\mu}}}=
  0.9530
  \\
 Max(|L_{ABCD}|,|R_{ABCD}|)&
          2.3854 \times 10^{ -18}
  {\,\rm{GeV^{-1}}}&& \\
 \hline\end{array}
 $$
 \label{table a} \caption{\small{Case 1 : Column 1 contains values   of the NMSGUT-SUGRY-NUHM  parameters at $M_X$
  derived from an  accurate fit to all 18 fermion data and compatible with RG constraints.
 Unification parameters and mass spectrum of superheavy and superlight fields are  also given.
 The values of $\mu(M_X^0),B(M_X^0)$ are determined by RG evolution from $M_Z$ to $M_X^0$
 of the values determined by the EWRSB conditions.}}\end{table}
 \begin{table}
 $$
 \begin{array}{|c|c|c|c|c|}
 \hline
 &&&&\\
 {\rm  Parameter }&Target =\bar O_i &Uncert.= \delta_i    &Achieved= O_i &Pull =(O_i-\bar O_i)/\delta_i\\
 \hline
    y_u/10^{-6}&  2.044938&  0.781166&  2.045688&  0.000960\\
    y_c/10^{-3}&  0.996830&  0.164477&  0.997657&  0.005028\\
            y_t&  0.378790&  0.015152&  0.378779& -0.000722\\
    y_d/10^{-5}&  7.128042&  4.155648&  7.073763& -0.013061\\
    y_s/10^{-3}&  1.351200&  0.637766&  1.368823&  0.027632\\
            y_b&  0.540954&  0.280755&  0.542439&  0.005288\\
    y_e/10^{-4}&  1.276823&  0.191523&  1.276139& -0.003569\\
  y_\mu/10^{-2}&  2.589599&  0.388440&  2.589488& -0.000285\\
         y_\tau&  0.566903&  0.107712&  0.565812& -0.010128\\
             \sin\theta^q_{12}&    0.2210&  0.001600&    0.2210&            0.0008\\
     \sin\theta^q_{13}/10^{-4}&   28.6321&  5.000000&   28.6231&           -0.0018\\
     \sin\theta^q_{23}/10^{-3}&   33.6894&  1.300000&   33.6848&           -0.0036\\
                      \delta^q&   60.0208& 14.000000&   59.9598&           -0.0044\\
    (m^2_{12})/10^{-5}(eV)^{2}&    5.2894&  0.560672&    5.2895&            0.0002\\
    (m^2_{23})/10^{-3}(eV)^{2}&    1.6906&  0.338115&    1.6906&            0.0001\\
           \sin^2\theta^L_{12}&    0.2934&  0.058683&    0.2934&            0.0000\\
           \sin^2\theta^L_{23}&    0.4585&  0.137561&    0.4585&           -0.0003\\
           \sin^2\theta^L_{13}&    0.0249&  0.019000&    0.0249&            0.0005\\
% \hline
 %  Eigenvalues(\Delta_{\bar u})&   0.015898&   0.034527&   0.035837&\\
%   Eigenvalues(\Delta_{\bar d})&   0.010434&   0.041268&   0.042342&\\
% Eigenvalues(\Delta_{\bar \nu})&   0.012491&   0.063265&   0.064544&\\
 %  Eigenvalues(\Delta_{\bar e})&   0.011460&   0.042634&   0.045031&\\
 %       Eigenvalues(\Delta_{Q})&   0.013873&   0.026937&   0.028614&\\
  %      Eigenvalues(\Delta_{L})&   0.001385&   0.041835&   0.044312&\\
  %   \Delta_{\bar H},\Delta_{H}&        2.958530   &        1.438419    &{}&\\
 \hline
 \alpha_1 &
  0.3453-  0.0000i
 & {\bar \alpha}_1 &
  0.4938-  0.0000i
 &\\
 \alpha_2&
  0.1119-  0.0124i
 & {\bar \alpha}_2 &
  0.3341+  0.0788i
 &\\
 \alpha_3 &
  0.0721-  0.0878i
 & {\bar \alpha}_3 &
  0.0730-  0.1041i
 &\\
 \alpha_4 &
 -0.6519+  0.6401i
 & {\bar \alpha}_4 &
  0.2867+  0.4862i
 &\\
 \alpha_5 &
 -0.0217-  0.0724i
 & {\bar \alpha}_5 &
 -0.0646-  0.1725i
 &\\
 \alpha_6 &
 -0.1213-  0.0089i
 & {\bar \alpha}_6 &
 -0.5050-  0.1211i
 &\\
  \hline
 \end{array}
 $$
 \label{table b} \caption{\small{Example 1:    with $\chi_X=\sqrt{ \sum_{i=1}^{17}
 (O_i-\bar O_i)^2/\delta_i^2}=  0.0338 $. Target values,  at $M_X^0$ of the fermion Yukawa
 couplings and mixing parameters, together with the estimated uncertainties, achieved values and pulls.
 The eigenvalues of the wavefunction renormalization increment  matrices $\Delta_i$ for fermion lines and
 the factors for Higgs lines are given, assuming the external Higgs is 10-plet dominated.
 The Higgs fractions $\alpha_i,{\bar{\alpha_i}}$ which control the MSSM fermion Yukawa couplings  are also
 given. Notice the dominance of the first components $\alpha_1,{\bar{\alpha_1}}$
 consistently with  the assumption made. Right-handed neutrino threshold  effects   have been ignored.
  We have truncated numbers for display although all calculations are done at double
 precision.}}
 \end{table}
 \begin{table}
 $$
 \begin{array}{|c|c|c|c|}
 \hline &&&\\ {\rm  Parameter }&SM(M_Z) & m^{GUT}(M_Z) & m^{MSSM}=(m+\Delta m)^{GUT}(M_Z) \\
 \hline
    m_d/10^{-3}&   2.90000&   0.61793&   2.87054\\
    m_s/10^{-3}&  55.00000&  11.95726&  55.58066\\
            m_b&   3.00000&   3.24165&   2.99969\\
    m_e/10^{-3}&   0.48657&   0.47227&   0.48588\\
         m_\mu &   0.10272&   0.09579&   0.10267\\
         m_\tau&   1.74624&   1.73972&   1.74281\\
    m_u/10^{-3}&   1.27000&   1.08293&   1.27036\\
            m_c&   0.61900&   0.52813&   0.61955\\
            m_t& 172.50000& 146.75487& 172.43486\\
 \hline
 \end{array}
 $$
 \label{table c}
 \caption{\small{Example 1: Values of standard model
 fermion masses in GeV at $M_Z$ compared with the masses obtained from
 values of GUT derived  Yukawa couplings  run down from $M_X^0$ to
 $M_Z$  both before and after threshold corrections.
  Fit with $\chi_Z=\sqrt{ \sum_{i=1}^{9} (m_i^{MSSM}- m_i^{SM})^2/ (m_i^{MSSM})^2} =
0.0149$.}}
 \end{table}
 \begin{table}
 $$
 \begin{array}{|c|c|c|c|}
 \hline
 {\rm  Parameter}  &Value&  {\rm  Parameter}&Value \\
 \hline
                       M_{1}&            223.01&   M_{{\tilde {\bar {u}}_1}}&           9225.58\\
                       M_{2}&            661.94&   M_{{\tilde {\bar {u}}_2}}&           9222.53\\
                       M_{3}&            827.66&   M_{{\tilde {\bar {u}}_3}}&          29944.82\\
     M_{{\tilde {\bar l}_1}}&           2890.72&               A^{0(l)}_{11}&        -172061.46\\
     M_{{\tilde {\bar l}_2}}&            230.51&               A^{0(l)}_{22}&        -171854.14\\
     M_{{\tilde {\bar l}_3}}&          19852.55&               A^{0(l)}_{33}&        -103991.62\\
        M_{{\tilde {L}_{1}}}&          12250.20&               A^{0(u)}_{11}&        -209999.77\\
        M_{{\tilde {L}_{2}}}&          12080.49&               A^{0(u)}_{22}&        -209998.40\\
        M_{{\tilde {L}_{3}}}&          18971.40&               A^{0(u)}_{33}&        -106625.21\\
     M_{{\tilde {\bar d}_1}}&           4654.53&               A^{0(d)}_{11}&        -172245.57\\
     M_{{\tilde {\bar d}_2}}&           4651.04&               A^{0(d)}_{22}&        -172244.04\\
     M_{{\tilde {\bar d}_3}}&          66504.44&               A^{0(d)}_{33}&         -59367.97\\
          M_{{\tilde {Q}_1}}&           9877.51&                   \tan\beta&             50.00\\
          M_{{\tilde {Q}_2}}&           9875.28&                    \mu(M_Z)&         174668.73\\
          M_{{\tilde {Q}_3}}&          51813.13&                      B(M_Z)&
          5.6389 \times 10^{   9}
 \\
 M_{\bar {H}}^2&
         -2.9298 \times 10^{  10}
 &M_{H}^2&
         -3.3648 \times 10^{  10}
 \\
 \hline
 \end{array}
 $$
 \label{table d} \caption{ \small {Example 1: Values (GeV) in  of the soft Susy parameters  at $M_Z$
 (evolved from the soft SUGRY-NUHM parameters at $M_X^0$).
 The  values of soft Susy parameters  at $M_Z$
 determine the Susy threshold corrections to the fermionYukawas.
 The matching of run down fermion Yukawas in the MSSM to the SM   parameters
 determines  soft SUGRY parameters at $M_X$. Note the  heavier third
 sgeneration.  The values of $\mu(M_Z)$ and the corresponding soft
 Susy parameter $B(M_Z)=m_A^2 {\sin 2 \beta }/2$ are determined by
 imposing electroweak symmetry breaking conditions. $m_A$ is the
 mass of the CP odd scalar in the Doublet Higgs. The sign of
 $\mu$ is assumed positive. }}
 \end{table}
 \begin{table}
 $$
 \begin{array}{|c|c|}
 \hline {\mbox {Field } }&Mass(GeV)\\
 \hline
                M_{\tilde{G}}&            827.66\\
               M_{\chi^{\pm}}&            661.94,         174668.77\\
       M_{\chi^{0}}&            223.01,            661.94,         174668.76    ,         174668.76\\
              M_{\tilde{\nu}}&         12250.021,         12080.306,         18971.280\\
                M_{\tilde{e}}&           2891.07,          12250.30,            224.01   ,          12080.79,          18815.46,          20000.51  \\
                M_{\tilde{u}}&           9225.51,           9877.36,           9222.40   ,           9875.18,          29942.71,          51814.72  \\
                M_{\tilde{d}}&           4654.61,           9877.70,           4651.10   ,           9875.48,          51808.66,          66507.96  \\
                        M_{A}&         531090.49\\
                  M_{H^{\pm}}&         531090.49\\
                    M_{H^{0}}&         531090.47\\
                    M_{h^{0}}&            114.58\\
 \hline
 \end{array}
 $$
 \label{table e}\caption{\small{Example 1:Spectra of supersymmetric partners calculated ignoring generation mixing effects.
 Inclusion of such effects   changes the spectra only marginally. Due to the large
 values of $\mu,B,A_0$. The LSP and light chargino are  essentially pure Bino and Wino($\tilde W_\pm $).
   The light  gauginos and  light Higgs  $h^0$, are accompanied by a light smuon and  sometimes  selectron.
 The rest of the sfermions have multi-TeV masses. The mini-split supersymmetry spectrum and
 large $\mu,A_0$ parameters help avoid problems with FCNC and CCB/UFB instability\cite{kuslangseg,gutupend}.
 The sfermion masses  are ordered by generation not magnitude. This is useful in understanding the spectrum
  calculated including generation mixing effects
  }}\end{table}
 \begin{table}
 $$
 \begin{array}{|c|c|}
 \hline {\mbox {Field } }&Mass(GeV)\\
 \hline
                M_{\tilde{G}}&            821.02\\
               M_{\chi^{\pm}}&            660.62,         174947.06\\
       M_{\chi^{0}}&            222.65,            660.62,         174947.05    ,         174947.05\\
              M_{\tilde{\nu}}&          12075.63,          12245.38,         18876.120\\
                M_{\tilde{e}}&            324.73,           2900.24,          12076.12   ,          12245.66,          18706.96,          19837.44  \\
                M_{\tilde{u}}&           9240.12,           9243.22,           9860.87   ,           9892.76,          29570.81,          51588.50  \\
                M_{\tilde{d}}&           4693.25,           4696.77,           9861.19   ,           9893.10,          51582.41,          66310.78  \\
                        M_{A}&         532268.90\\
                  M_{H^{\pm}}&         532268.91\\
                    M_{H^{0}}&         532268.89\\
                    M_{h^{0}}&            113.41\\
 \hline
 \end{array}
 $$
 \label{table f}\caption{\small{Example 1: Spectra of supersymmetric partners calculated including  generation mixing effects.
 Inclusion of such effects   changes the spectra only marginally. Due to the large
 values of $\mu,B,A_0$. The LSP and light chargino are  essentially pure Bino and Wino($\tilde W_\pm $).
  Note that the ordering of the eigenvalues in this table follows their magnitudes, comparison
 comparison with the previous table is necessary to identify the sfermions}}\end{table}

 \begin{table}
 $$
 \begin{array}{|c|c|c|c|}
 \hline
 {\rm Parameter }&Value &{\rm  Field }&\hspace{10mm} Masses\\
 &&{\rm}[SU(3),SU(2),Y]&\hspace{10mm}( Units\,\,of 10^{16} Gev)\\ \hline
       \chi_{X}&  0.0528           &A[1,1,4]&   3502.30 \\ \chi_{Z}&
    0.0227
                &B[6,2,{5/3}]&            0.4053\\
           h_{11}/10^{-6}& -2.6106         &C[8,2,1]&{    162.52,   1146.22,   1385.40 }\\
           h_{22}/10^{-4}& -8.1592    &D[3,2,{7/ 3}]&{    139.23,   2043.80,   2350.42 }\\
                   h_{33}&  0.3114     &E[3,2,{1/3}]&{      0.96,    150.01,    150.01 }\\
 f_{11}/10^{-6}&
 -0.0630-  0.0424i
                      &&{   317.336,    670.73,   1575.37 }\\
 f_{12}/10^{-6}&
 -0.0357-  2.0877i
          &F[1,1,2]&     79.95,     79.95
 \\f_{13}/10^{-5}&
 -0.6349-  0.1413i
                  &&    663.71,   1826.48  \\
 f_{22}/10^{-5}&
-10.5738+  6.5114i
              &G[1,1,0]&{     0.074,      0.69,      2.34 }\\
 f_{23}/10^{-4}&
  0.5095-  0.3396i
                      &&{     2.342,     41.69,     42.08 }\\
 f_{33}/10^{-3}&
 -0.4395-  0.0514i
              &h[1,2,1]&{     3.924,    291.17,    397.51 }\\
 g_{12}/10^{-3}&
 -8.5548-  3.4900i
                 &&{   1580.89,   2524.44 }\\
 g_{13}/10^{-3}&
 -5.3631- 18.5130i
     &I[3,1,{10/3}]&      0.82\\
 g_{23}/10^{-2}&
 61.1617+  6.3854i
          &J[3,1,{4/3}]&{     0.822,     57.80,    124.04 }\\
 \lambda/10^{-2}&
 -0.0798+  0.2049i
                 &&{    124.04,   1550.83 }\\
 \eta&
 -3.8375-  2.4483i
   &K[3,1, {8/ 3}]&{     54.56,   2400.54 }\\
 \rho&
  4.3429+  0.4901i
    &L[6,1,{2/ 3}]&{    164.11,   2693.32 }\\
 k&
  0.0163+  0.0160i
     &M[6,1,{8/ 3}]&   2934.27\\
 \zeta&
  2.1043-  0.5613i
     &N[6,1,{4/ 3}]&   2549.57\\
 \bar\zeta &
  2.0991-  0.7985i
          &O[1,3,2]&   3774.62\\
       m/10^{16}GeV& 0.05463    &P[3,3,{2/ 3}]&{    521.70,   2527.43 }\\
     m_\Theta/10^{16}GeV&-268.776e^{-iArg(\lambda)}     &Q[8,3,0]&     0.396\\
             \gamma& 1.51935        &R[8,1, 0]&{      0.45,      1.23 }\\
              \bar\gamma&  0.8968     &S[1,3,0]&    1.3123\\
 x&
  0.5241+  1.1792i
         &t[3,1,{2/ 3}]&{      1.36,     28.99,     95.99,    300.01        }\\\Delta_X&      2.15 &&{   1166.70,   1359.73,   5170.19 }\\
              \Delta_{G}& -17.689           &U[3,3,{4/3}]&     0.996\\
 \Delta\alpha_{3}(M_{Z})&  -0.007               &V[1,2,3]&     1.066\\
    \{M^{\nu^c}/10^{11}GeV\}&{  0.0014,  189.78,  746.17    }&W[6,3,{2/ 3}]&           1288.01  \\
  \{M^{\nu}_{ II}/10^{-11  }eV\}&  0.0004,     50.34,            197.92               &X[3,2,{5/ 3}]&     0.425,   158.136,   158.136\\
                  M_\nu(meV)&{  1.0169,    7.42,   42.31    }&Y[6,2, {1/3}]&              0.50  \\
    \{\rm{Evals[f]}\}/ 10^{-5} eV&{ 0.00008,  11.530,   45.33         }&Z[8,1,2]&              1.21  \\
 \hline\hline
 \mbox{Soft parameters}&{\rm m_{\frac{1}{2}}}=
          -558.176
 &{\rm m_{0}}=
          5739.456
 &{\rm A_{0}}=
         -2.8914 \times 10^{   5}
 \\
 \mbox{at $M_{X}^0$}&\mu=
          2.3744 \times 10^{   5}
 &{\rm B}=
         -3.9561 \times 10^{  10}
  &{\rm tan{\beta}}=           52.0000\\
 &{\rm M^2_{\bar H}}=
         -3.6652 \times 10^{  10}
 &{\rm M^2_{  H} }=
         -3.5552 \times 10^{  10}
 &
 {\rm R_{\frac{b\tau}{s\mu}}}=
  1.0655
  \\
 Max(|L_{ABCD}|,|R_{ABCD}|)&
          1.1270 \times 10^{ -17}
  {\,\rm{GeV^{-1}}}&& \\
 \hline\end{array}
 $$
 \label{table a} \caption{\small{Example 2 : Column 1 contains values   of the NMSGUT-SUGRY-NUHM  parameters at $M_X$
  derived from an  accurate fit to all 18 fermion data and compatible with RG constraints.
 Unification parameters and mass spectrum of superheavy and superlight fields are  also given.
 The values of $\mu(M_X^0),B(M_X^0)$ are determined by RG evolution from $M_Z$ to $M_X^0$
 of the values determined by the EWRSB conditions.}}\end{table}
 \begin{table}
 $$
 \begin{array}{|c|c|c|c|c|}
 \hline
 &&&&\\
 {\rm  Parameter }&Target =\bar O_i &Uncert.= \delta_i    &Achieved= O_i &Pull =(O_i-\bar O_i)/\delta_i\\
 \hline
    y_u/10^{-6}&  2.056309&  0.785510&  2.057005&  0.000886\\
    y_c/10^{-3}&  1.002379&  0.165393&  1.003313&  0.005643\\
            y_t&  0.383757&  0.015350&  0.383780&  0.001494\\
    y_d/10^{-5}&  6.031134&  3.516151&  5.966872& -0.018276\\
    y_s/10^{-3}&  1.142585&  0.539300&  1.164037&  0.039777\\
            y_b&  0.601133&  0.311988&  0.602353&  0.003909\\
    y_e/10^{-4}&  1.384241&  0.207636&  1.382867& -0.006614\\
  y_\mu/10^{-2}&  2.795186&  0.419278&  2.795532&  0.000827\\
         y_\tau&  0.634068&  0.120473&  0.630898& -0.026319\\
             \sin\theta^q_{12}&    0.2210&  0.001600&    0.2210&            0.0008\\
     \sin\theta^q_{13}/10^{-4}&   28.2913&  5.000000&   28.2975&            0.0012\\
     \sin\theta^q_{23}/10^{-3}&   33.2889&  1.300000&   33.2836&           -0.0041\\
                      \delta^q&   60.0208& 14.000000&   60.1344&            0.0081\\
    (m^2_{12})/10^{-5}(eV)^{2}&    5.4009&  0.572498&    5.4011&            0.0002\\
    (m^2_{23})/10^{-3}(eV)^{2}&    1.7352&  0.347036&    1.7351&           -0.0003\\
           \sin^2\theta^L_{12}&    0.2925&  0.058501&    0.2925&            0.0002\\
           \sin^2\theta^L_{23}&    0.4522&  0.135654&    0.4523&            0.0012\\
           \sin^2\theta^L_{13}&    0.0244&  0.019000&    0.0243&           -0.0015\\
% \hline
%  Eigenvalues(\Delta_{\bar u})&   0.096394&   4.208172&   4.263364&\\
%  Eigenvalues(\Delta_{\bar d})&   0.095511&   4.369212&   4.422943&\\
%Eigenvalues(\Delta_{\bar \nu})&   0.117462&   4.840710&   4.915154&\\
 % Eigenvalues(\Delta_{\bar e})&   0.120111&   4.357388&   4.439407&\\
 %      Eigenvalues(\Delta_{Q})&   0.085978&   5.054916&   5.101870&\\
 %      Eigenvalues(\Delta_{L})&   0.108810&   5.349132&   5.398868&\\
 %   \Delta_{\bar H},\Delta_{H}&        6.995974   &        2.816129    &{}&\\
 \hline
 \alpha_1 &
  0.4356+  0.0000i
 & {\bar \alpha}_1 &
  0.6832+  0.0000i
 &\\
 \alpha_2&
  0.0152-  0.0348i
 & {\bar \alpha}_2 &
  0.0294-  0.0607i
 &\\
 \alpha_3 &
  0.0074-  0.0199i
 & {\bar \alpha}_3 &
  0.0079-  0.0331i
 &\\
 \alpha_4 &
 -0.6267-  0.6436i
 & {\bar \alpha}_4 &
 -0.5709-  0.4473i
 &\\
 \alpha_5 &
  0.0078-  0.0160i
 & {\bar \alpha}_5 &
 -0.0265+  0.0002i
 &\\
 \alpha_6 &
 -0.0303-  0.0112i
 & {\bar \alpha}_6 &
 -0.0039+  0.0296i
 &\\
  \hline
 \end{array}
 $$
 \label{table b} \caption{\small{Example 2 :with $\chi_X=\sqrt{ \sum_{i=1}^{17}
 (O_i-\bar O_i)^2/\delta_i^2}= 0.0528 $. Target values,  at $M_X^0$ of the fermion Yukawa
 couplings and mixing parameters, together with the estimated uncertainties, achieved values and pulls.
 The eigenvalues of the wavefunction renormalization increment  matrices $\Delta_i$ for fermion lines and
 the factors for Higgs lines are given, assuming the external Higgs is 10-plet dominated.
 The Higgs fractions $\alpha_i,{\bar{\alpha_i}}$ which control the MSSM fermion Yukawa couplings  are also
 given. Notice the dominance of the first components $\alpha_1,{\bar{\alpha_1}}$
 consistently with  the assumption made. Right-handed neutrino threshold  effects   have been ignored.
  We have truncated numbers for display although all calculations are done at double
 precision.}}
 \end{table}
 \begin{table}
 $$
 \begin{array}{|c|c|c|c|}
 \hline &&&\\ {\rm  Parameter }&SM(M_Z) & m^{GUT}(M_Z) & m^{MSSM}=(m+\Delta m)^{GUT}(M_Z) \\
 \hline
    m_d/10^{-3}&   2.90000&   0.47862&   2.85870\\
    m_s/10^{-3}&  55.00000&   9.33699&  55.82819\\
            m_b&   2.90000&   3.19078&   2.90012\\
    m_e/10^{-3}&   0.48657&   0.46999&   0.48432\\
         m_\mu &   0.10272&   0.09496&   0.10258\\
         m_\tau&   1.74624&   1.73389&   1.73348\\
    m_u/10^{-3}&   1.27000&   1.08656&   1.27132\\
            m_c&   0.61900&   0.52997&   0.62010\\
            m_t& 172.50000& 146.48712& 172.21615\\
 \hline
 \end{array}
 $$
 \label{table c}
 \caption{\small{Example 2 : Values of standard model
 fermion masses in GeV at $M_Z$ compared with the masses obtained from
 values of GUT derived  Yukawa couplings  run down from $M_X^0$ to
 $M_Z$  both before and after threshold corrections.
  Fit with $\chi_Z=\sqrt{ \sum_{i=1}^{9} (m_i^{MSSM}- m_i^{SM})^2/ (m_i^{MSSM})^2} =
0.0227$.}}
 \end{table}
 \begin{table}
 $$
 \begin{array}{|c|c|c|c|}
 \hline
 {\rm  Parameter}  &Value&  {\rm  Parameter}&Value \\
 \hline
                       M_{1}&            198.83&   M_{{\tilde {\bar {u}}_1}}&           7246.47\\
                       M_{2}&            621.41&   M_{{\tilde {\bar {u}}_2}}&           7242.44\\
                       M_{3}&            592.21&   M_{{\tilde {\bar {u}}_3}}&          30148.11\\
     M_{{\tilde {\bar l}_1}}&           2895.27&               A^{0(l)}_{11}&        -167770.16\\
     M_{{\tilde {\bar l}_2}}&            219.15&               A^{0(l)}_{22}&        -167544.51\\
     M_{{\tilde {\bar l}_3}}&          31721.05&               A^{0(l)}_{33}&         -92942.20\\
        M_{{\tilde {L}_{1}}}&          11211.89&               A^{0(u)}_{11}&        -212183.50\\
        M_{{\tilde {L}_{2}}}&          11025.49&               A^{0(u)}_{22}&        -212182.25\\
        M_{{\tilde {L}_{3}}}&          25383.06&               A^{0(u)}_{33}&        -106623.25\\
     M_{{\tilde {\bar d}_1}}&           2280.17&               A^{0(d)}_{11}&        -167689.77\\
     M_{{\tilde {\bar d}_2}}&           2275.99&               A^{0(d)}_{22}&        -167688.64\\
     M_{{\tilde {\bar d}_3}}&          72463.40&               A^{0(d)}_{33}&         -49795.07\\
          M_{{\tilde {Q}_1}}&           8854.12&                   \tan\beta&             52.00\\
          M_{{\tilde {Q}_2}}&           8851.95&                    \mu(M_Z)&         174787.15\\
          M_{{\tilde {Q}_3}}&          55736.03&                      B(M_Z)&
          5.4479 \times 10^{   9}
 \\
 M_{\bar {H}}^2&
         -2.8079 \times 10^{  10}
 &M_{H}^2&
         -3.4167 \times 10^{  10}
 \\
 \hline
 \end{array}
 $$
 \label{table d} \caption{ \small {Example 2 : Values (GeV) in  of the soft Susy parameters  at $M_Z$
 (evolved from the soft SUGRY-NUHM parameters at $M_X^0$).
 The  values of soft Susy parameters  at $M_Z$
 determine the Susy threshold corrections to the fermion Yukawas.
 The matching of run down fermion Yukawas in the MSSM to the SM   parameters
 determines  soft SUGRY parameters at $M_X$. Note the  heavier third
 sgeneration.  The values of $\mu(M_Z)$ and the corresponding soft
 Susy parameter $B(M_Z)=m_A^2 {\sin 2 \beta }/2$ are determined by
 imposing electroweak symmetry breaking conditions. $m_A$ is the
 mass of the CP odd scalar in the in the Doublet Higgs. The sign of
 $\mu$ is assumed positive. }}
 \end{table}
 \begin{table}
 $$
 \begin{array}{|c|c|}
 \hline {\mbox {Field } }&Mass(GeV)\\
 \hline
                M_{\tilde{G}}&            592.21\\
               M_{\chi^{\pm}}&            621.41,         174787.19\\
       M_{\chi^{0}}&            198.83,            621.41,         174787.18    ,         174787.18\\
              M_{\tilde{\nu}}&         11211.692,         11025.294,         25382.974\\
                M_{\tilde{e}}&           2895.63,          11212.00,            208.97   ,          11025.89,          25369.23,          31732.18  \\
                M_{\tilde{u}}&           7246.37,           8853.95,           7242.31   ,           8851.81,          30146.48,          55737.26  \\
                M_{\tilde{d}}&           2280.32,           8854.33,           2276.11   ,           8852.17,          55732.48,          72466.16  \\
                        M_{A}&         532347.51\\
                  M_{H^{\pm}}&         532347.52\\
                    M_{H^{0}}&         532347.50\\
                    M_{h^{0}}&            124.53\\
 \hline
 \end{array}
 $$
 \label{table e}\caption{\small{Example 2: Spectra of supersymmetric partners calculated ignoring generation mixing effects.
 Inclusion of such effects   changes the spectra only marginally. Due to the large
 values of $\mu,B,A_0$. The LSP and light chargino are  essentially pure Bino and Wino($\tilde W_\pm $).
   The light  gauginos and  light Higgs  $h^0$, are accompanied by a light smuon and  sometimes  selectron.
 The rest of the sfermions have multi-TeV masses. The mini-split supersymmetry spectrum and
 large $\mu,A_0$ parameters help avoid problems with FCNC and CCB/UFB instability\cite{kuslangseg,gutupend}.
 The sfermion masses  are ordered by generation not magnitude. This is useful in understanding the spectrum
  calculated including generation mixing effects
  }}\end{table}
 \begin{table}
 $$
 \begin{array}{|c|c|}
 \hline {\mbox {Field } }&Mass(GeV)\\
 \hline
                M_{\tilde{G}}&            586.63\\
               M_{\chi^{\pm}}&            620.15,         175052.83\\
       M_{\chi^{0}}&            198.46,            620.15,         175052.81    ,         175052.81\\
              M_{\tilde{\nu}}&          11020.47,          11206.92,         25306.768\\
                M_{\tilde{e}}&            301.27,           2903.91,          11021.07   ,          11207.23,          25292.91,          31613.59  \\
                M_{\tilde{u}}&           7268.78,           7272.81,           8837.25   ,           8874.19,          29777.84,          55531.91  \\
                M_{\tilde{d}}&           2371.66,           2375.70,           8837.65   ,           8874.57,          55527.11,          72292.12  \\
                        M_{A}&         534359.68\\
                  M_{H^{\pm}}&         534359.69\\
                    M_{H^{0}}&         534359.67\\
                    M_{h^{0}}&            123.59\\
 \hline
 \end{array}
 $$
 \label{table f}\caption{\small{Example 2 : Spectra of supersymmetric partners calculated including  generation mixing effects.
 Inclusion of such effects   changes the spectra only marginally. Due to the large
 values of $\mu,B,A_0$. The LSP and light chargino are  essentially pure Bino and Wino($\tilde W_\pm $).
  Note that the ordering of the eigenvalues in this table follows their magnitudes, comparison
 comparison with the previous table is necessary to identify the sfermions}}\end{table}

\end{document}